\documentclass[12pt,epsf]{article}
\usepackage{amsmath}
\usepackage{amssymb}
\usepackage{bm}
\usepackage[dvipdfmx]{graphicx}

\catcode `@=11 \@addtoreset{equation}{section} \catcode `@=12
\begin{document}
\setlength{\oddsidemargin}{0cm}
\setlength{\baselineskip}{7mm}

\begin{titlepage}

\begin{center}
  {\LARGE
Toward the construction of the general multi-cut solutions in Chern-Simons Matrix Models }
\end{center}
\vspace{0.2cm}
\baselineskip 18pt 
\renewcommand{\thefootnote}{\fnsymbol{footnote}}

\begin{center}
Takeshi {\sc Morita}$^{a,b}$\footnote{%
E-mail address: morita.takeshi@shizuoka.ac.jp
}, 
Kento S{\sc ugiyama}$^{b}$\footnote{
E-mail address: sugiyama.kento.15@shizuoka.ac.jp}

\renewcommand{\thefootnote}{\arabic{footnote}}
\setcounter{footnote}{0}

\vspace{0.4cm}

{\it
a. Department of Physics,
Shizuoka University \\
836 Ohya, Suruga-ku, Shizuoka 422-8529, Japan 
\vspace{0.2cm}
\\
b. Graduate School of Science and Technology, Shizuoka University\\
836 Ohya, Suruga-ku, Shizuoka 422-8529, Japan
}

\end{center}

%\newpage

\vspace{1.5cm}

\begin{abstract}
In our previous work \cite{Morita:2017oev}, we pointed out that various multi-cut solutions exist in the Chern-Simons (CS) matrix models at large-$N$ due to a curious structure of the saddle point equations.
In the ABJM matrix model, these multi-cut solutions might be regarded as  the condensations of the D2-brane instantons.
However many of these multi-cut solutions including the ones corresponding to the condensations of the D2-brane instantons were obtained numerically only.
In the current work, we propose an ansatz for the multi-cut solutions which may allow us to derive the analytic expressions for all these solutions.
As a demonstration, we derive several novel analytic solutions in the pure CS matrix model and the ABJM matrix model.
We also develop the argument for the connection to the instantons.

\end{abstract}
%\\

%\\

\end{titlepage}

\tableofcontents

\section{Introduction}

The $1/N$ expansion \cite{Brezin1978} is a quite powerful technique in matrix models, and it makes us possible to analyze the models in the non-perturbative regime.
Not only that, in string theories, this expansion may correspond to the perturbative expansion of the string coupling  \cite{Brezin:1990rb, Douglas:1989ve, Gross:1989vs, Gross:1989aw, Banks:1996vh, Ishibashi:1996xs,Dijkgraaf:1997vv, Berenstein:2002jq}, and it might play important role to reveal quantum gravity.
Particularly, in the last decade, the analysis of the large-$N$ Chern-Simons (CS) matrix models has been developed quite remarkably. 
(See \cite{Hatsuda:2015gca, Marino:2016new} for reviews.)
The CS matrix models are obtained via the localization of the three dimensional supersymmetric CS matter theories on a sphere \cite{Pestun:2007rz, Kapustin:2009kz, Kapustin:2010xq, Jafferis:2010un, Hama:2010av}, which describe the low energy dynamics of the superstring theories and M-theory, and, through these developments, various non-perturbative aspects of the string theories have been revealed including the derivation of the $N^{3/2}$ factor \cite{Drukker:2010nc} of the free energy in the $N$ M2-brane theory \cite{Klebanov:1996un, Morita:2013wla, Morita:2014ypa}. 
These results provide us quite strong evidences for the AdS/CFT correspondence \cite{Maldacena:1997re, Itzhaki:1998dd, Aharony:2008ug}.

 In this article, we mainly investigate the $U(N)$ pure CS matrix model \cite{Kapustin:2009kz, Marino:2002fk, Aganagic:2002wv} among the various CS matrix models, since other models can be regarded as the variations of this model and we can expect that the application to these other models might be straightforward.

The partition function of the pure CS matrix model is given by
\begin{equation}
Z(k,N)
=\frac{1}{N!} \int \prod_{i=1}^N \frac{du_i}{2\pi} e^{-\frac{N}{4\pi i \lambda} \sum_i u^2_i} \prod_{i<j}^N \Bigl[ 2 \sinh{\frac{u_i-u_j}{2}} \Bigr]^2.
\label{partition-CS}
\end{equation}
Here $k$ is the CS level and $\lambda := N/k$ is the 't~Hooft coupling, and we will consider the 't~Hooft limit ($N \rightarrow \infty$, $\lambda$: fixed) of this model.
This partition function resembles the Gaussian Hermitian matrix model.
The difference appears only in the Vandermonde determinant, but this simple difference provides quite rich structures in the CS matrix models.

It is known that we can compute this partition function exactly at arbitrary $\lambda$ and $N$ \cite{Kapustin:2009kz, Tierz:2002jj}.
However, the 't~Hooft expansion of this model shows non-trivial properties and it is still valuable to investigate them \cite{Pasquetti:2009jg, Hatsuda:2015owa, Honda:2016vmv, Honda:2017qdb, Chattopadhyay:2017ckc}.
This is similar to the situations of the Gaussian matrix model and the Gross-Witten-Wadia model \cite{PhysRevD.21.446,Wadia:2012fr} which show non-trivial behaviors at large-$N$ \cite{Buividovich:2015oju, Alvarez:2016rmo, Okuyama:2017pil}, although we can calculate the partition functions exactly.

When we take the 't~Hooft limit, we can employ the saddle point approximation. 
The saddle point equation of the partition function (\ref{partition-CS}) with respect to $u_i$ is given by
\begin{equation}
u_i=\frac{2\pi i \lambda}{N} \sum_{j \neq i}^N \coth{\frac{u_i-u_j}{2}}, \qquad (i=1,\cdots, N).
\label{eom-CS}
\end{equation}
The exact solution of this equation at finite $\lambda$ which is characterized by a single cut of the eigenvalue distribution is known \cite{Aganagic:2002wv, Pasquetti:2009jg, Halmagyi:2003ze}.
This solution would be thermodynamically stable, since the free energy agrees with that of the $N \to \infty$ limit of the exact finite $N$ result \cite{Kapustin:2009kz, Tierz:2002jj}.

\begin{figure}
\begin{tabular}{cc}
\begin{minipage}{0.5\hsize}
\begin{center}
        \includegraphics[scale=0.2]{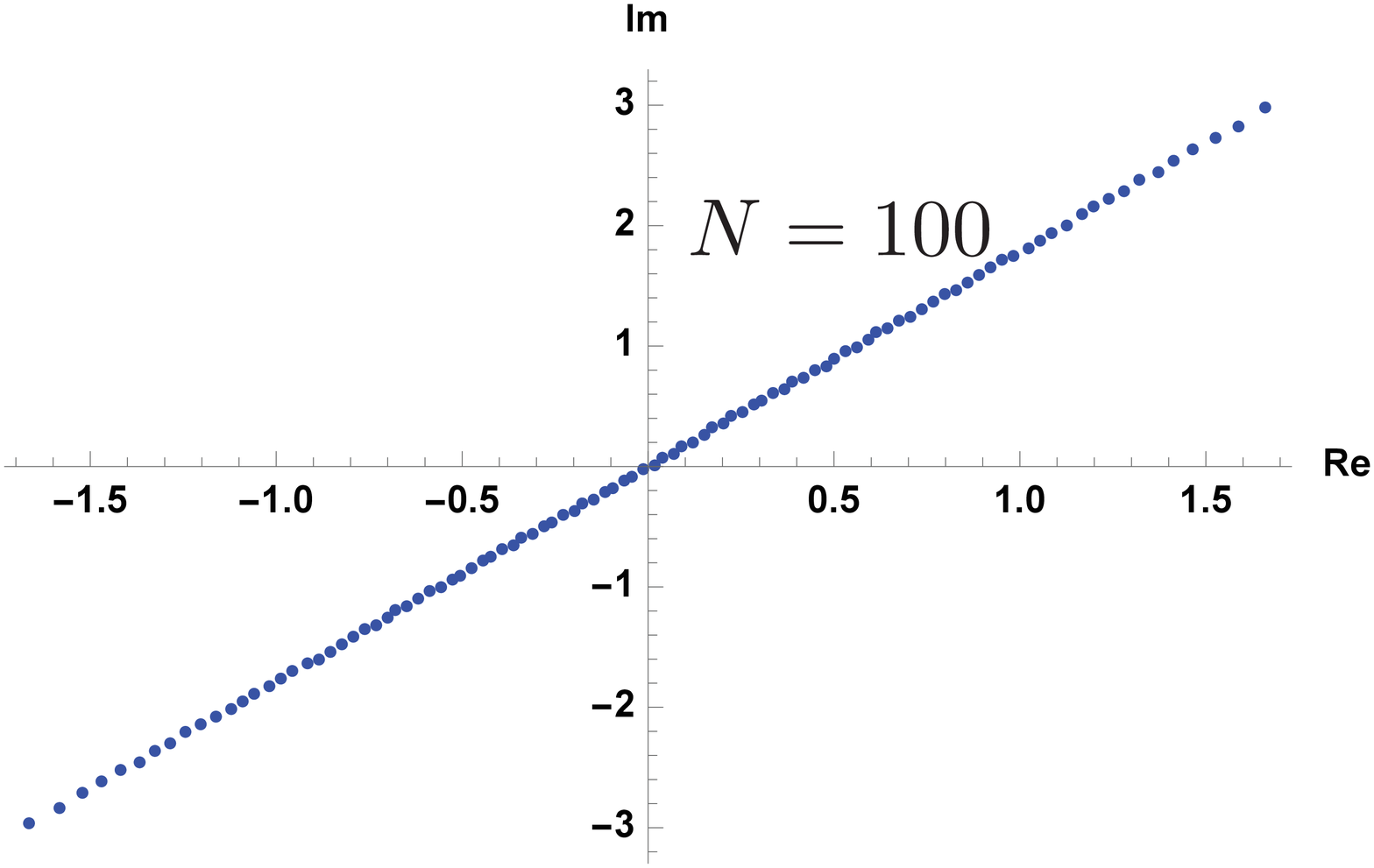}\\
    One-cut solution
\end{center}
\end{minipage}
\begin{minipage}{0.5\hsize}
\begin{center}
        \includegraphics[scale=0.2]{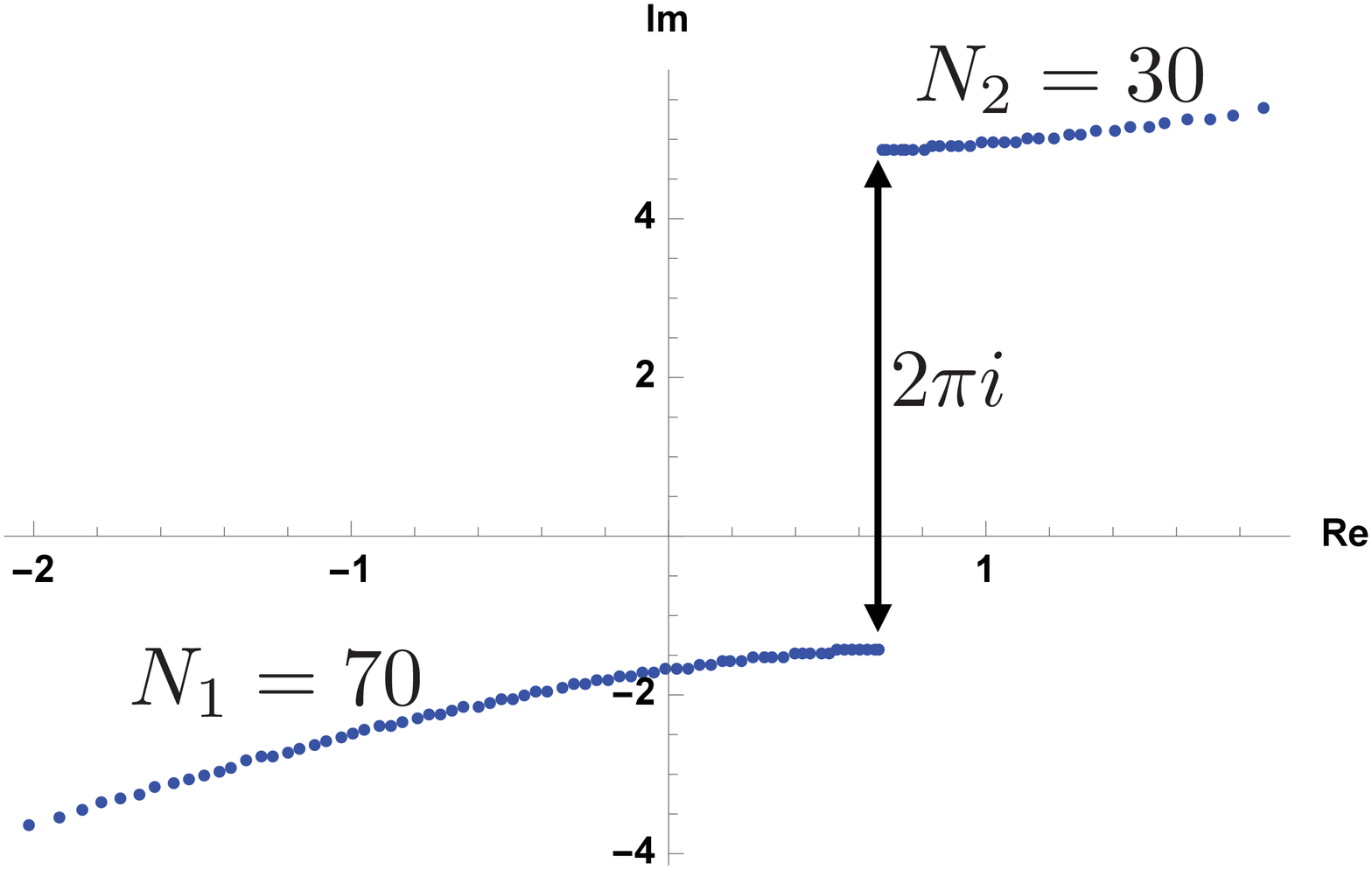}\\
    Stepwise two-cut solution
\end{center}
\end{minipage}
\\
\begin{minipage}{0.5\hsize}
\begin{center}
        \includegraphics[scale=0.2]{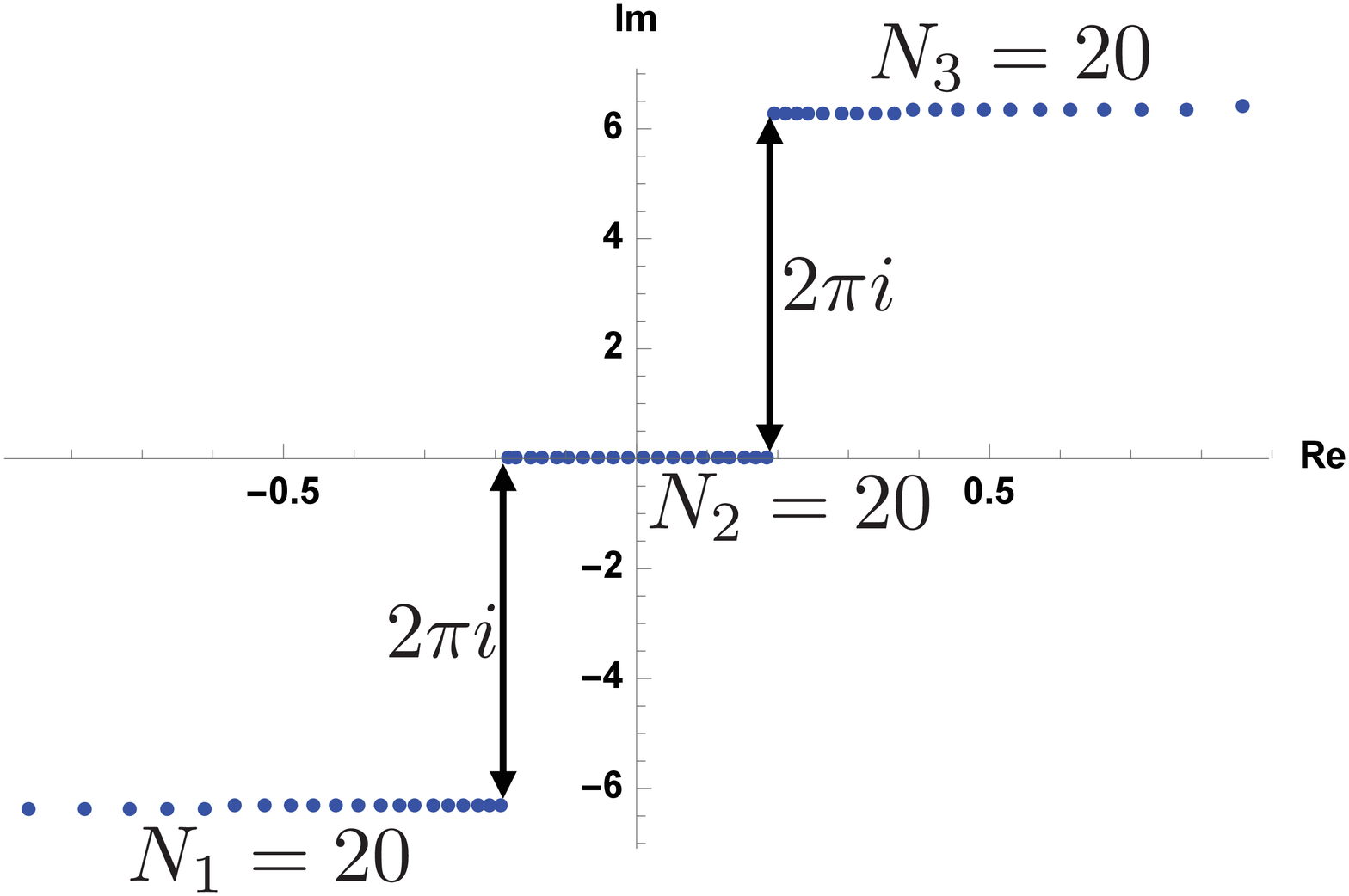}\\
    Stepwise multi-cut solution
\end{center}
\end{minipage}
\begin{minipage}{0.5\hsize}
\begin{center}
        \includegraphics[scale=0.2]{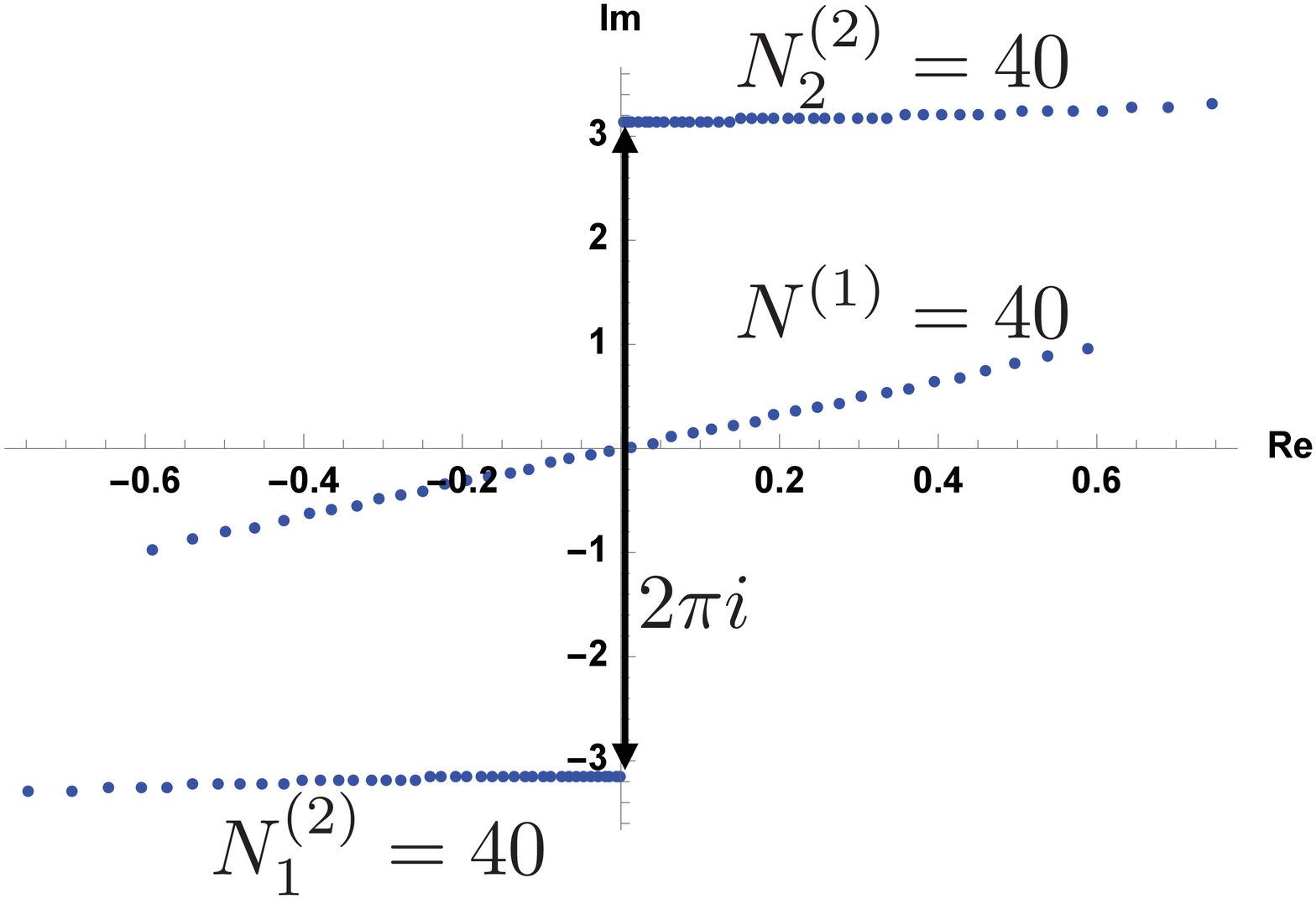}\\
Composition of the one-cut and stepwise two-cut solution 
\end{center}
\end{minipage}
\end{tabular}
       \caption{
   Eigenvalue distributions of  the pure CS matrix model. 
       We numerically solve the saddle point equation (\ref{eom-CS}) via the Newton method.     }
        \label{fig-numerical-CS}
\end{figure}

Then a question is whether this one-cut solution is unique or not.
Surprisingly it turned out that an infinite number of solutions are allowed in the saddle point equation (\ref{eom-CS}), which are characterized by the various multi-cuts \cite{Morita:2017oev, Morita:2011cs}\footnote{One important question is whether these multi-cut solutions contribute to the path-integral.
We do not consider this issue in this article.}.
See Figure \ref{fig-numerical-CS}.
These solutions were first found by solving the saddle point equations numerically through the Newton method \cite{Morita:2017oev, Morita:2011cs}, which have been employed in \cite{Herzog:2010hf, Niarchos:2011sn, Minwalla:2011ma}.
Later analytic expressions for some of the multi-cut solutions have been found by Ref.~\cite{Morita:2017oev}.

The purpose of this article is to develop the studies of Ref.~\cite{Morita:2017oev}, and provides analytic methods to treat all of these multi-cut solutions.
We will show that an integral formula (\ref{composite-CS}) for the resolvent related to the method of Migdal \cite{Migdal:1984gj} is quite useful.
By solving this integral formula either analytically or numerically, we will demonstrate that the eigenvalue distributions of the multi-cut solutions obtained through the Newton method shown in Figure \ref{fig-numerical-CS} can be reproduced.
All types of the Newton method solutions as far as we find may be explained by our formula, and we presume that our method might  be applicable to derive all possible solutions of the saddle point equation (\ref{eom-CS}). (Hence the formula (\ref{composite-CS}) is the main result of this article.)

The obtained analytic results tell us curious properties of the multi-cut solutions. 
We will see that there are two types of the multi-cuts in the pure CS matrix model.
One is the cuts which are separated by a multiple of $2 \pi i$. 
We refer to such cuts as ``stepwise multi-cuts" in this article. (See Figure \ref{fig-numerical-CS}.)
Another type of the multi-cuts is the composition of the stepwise multi-cut and one-cut (or another stepwise multi-cuts).
We refer to them as ``composite type".
We will show that, as the number of the composite type cuts increases, the genus of the resolvent increases similar to the multi-cut solutions in the ordinary matrix models. 
(Hence we need the higher genus generalizations of elliptic functions to describe the composite type multi-cuts.)
On the other hand, the stepwise multi-cuts do not change the genus\footnote{Although the resolvent of the stepwise multi-cut solution is not described by the higher genus generalizations of elliptic functions, the free energy is suppressed by $1/N^2$ as usual \cite{Morita:2017oev}.}, while they cause additional logarithmic singularities at the end points of each step in the resolvent. 
These properties might capture the geometrical natures of the multi-cut solutions.

We also discuss that our methods will work in other CS matrix models, and we propose a similar integral formula  (\ref{general-ABJM}) for the resolvent of the ABJM matrix model as an example. 
By using this formula, we will derive novel analytic solutions of the saddle point equation of the ABJM matrix model.

Generally the various multi-cut solutions in a matrix model may describe the different vacua of the system, and
these vacua would affect the perturbative vacuum through the instanton effects \cite{David:1990sk, David:1992za}.
Indeed the connection between the multi-cut solutions  and the D2-brane instantons in the ABJM matrix model \cite{Drukker:2011zy, Grassi:2014cla} was conjectured in Ref.~\cite{Morita:2017oev}.
We will develop this discussion and show a quantitative evidence for this connection.
Besides we comment on the relation to the membrane instanton in the pure CS matrix model \cite{Pasquetti:2009jg,Hatsuda:2015owa}.\\

The organization of this article is as follows.
In section \ref{sec-pCS}, we show the derivation of the multi-cut solutions in the pure CS matrix model.
In section \ref{sec-ABJM}, we argue the multi-cut solutions in the ABJM matrix model.
We also consider the connection to the D2-brane instantons.
We conclude in section \ref{sec-discussion} with some future directions.
In appendix \ref{app-holomorphy}, we introduce the derivation of the multi-cut solution via holomorphy in the pure CS matrix model.
This derivation is more powerful than the integral formula (\ref{composite-CS}) in certain situations.
In appendix \ref{app-negative-n}, we discuss the issue of ``negative steps".

%%%%%%%%%%%%%%%%%%%%%%%%%%%%%%%%%%%%%%%%%%%%%%%%%%%%%%%%%
\section{Multi-cut solutions in the pure CS matrix model}
\label{sec-pCS}
%%%%%%%%%%%%%%%%%%%%%%%%%%%%%%%%%%%%%%%%%%%%%%%%%%%%%%%%%

\begin{figure}
\begin{center}
        \includegraphics[scale=0.25]{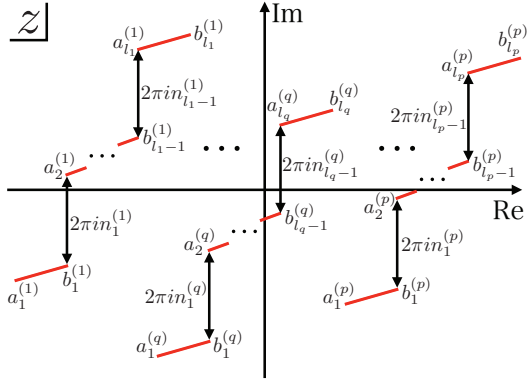}\\
\end{center}
       \caption{Eigenvalue distribution of the general solution (\ref{composite-CS}) in the pure CS matrix model on the $z$-plane.}
        \label{fig-image-CSgeneral}
\end{figure}

In this section, we will propose the integral formula (\ref{composite-CS}) which provides us a method to derive possible general solutions of the saddle point equation (\ref{eom-CS}) including the various multi-cut solutions shown in Figure \ref{fig-numerical-CS}.
Since the general solution will be characterized by a bit complicated multi-cuts sketched in Figure \ref{fig-image-CSgeneral}, we will first explain the derivations of several simpler multi-cut solutions which will give us insights about the general solution.

In order to derive the multi-cut solutions, we will employ the resolvent.
It is convenient to introduce new variables $U_i := \exp{\left( u_i \right)}$ and rewrite
 the saddle point equation (\ref{eom-CS}) as\footnote{If we use a new variable $\tilde{U}_i:=U_i e^{-2\pi i \lambda}$, \eqref{eom-CS2} becomes the saddle point equation of the Stieltjes-Wigert matrix model \cite{Tierz:2002jj}: $ \frac{1}{\tilde{U}_i} \log \tilde{U}_i = \frac{4\pi i \lambda}{N}  \sum_{j \neq i}^N \frac{1}{\tilde{U}_i-\tilde{U}_j} $. 
The advantage of the $U_i$ variable \cite{Suyama:2016nap} is that it makes equations symmetric under $U \to 1/U$ corresponding to the symmetry $u \to -u$ in the original variable \eqref{partition-CS}.
}
\begin{equation}
\log{U_i}=\frac{2\pi i \lambda}{N} \sum_{j \neq i}^N \frac{U_i+U_j}{U_i-U_j}, \qquad (i=1,\cdots, N).
\label{eom-CS2}
\end{equation}
Following \cite{Marino:2011nm, Suyama:2016nap}, we define the eigenvalue density $\rho(Z)$ and resolvent $v(Z)$ 
\begin{equation}
\rho(Z) := \frac{1}{N} \sum_{i=1}^N \delta (Z-U_i),
\qquad
v(Z) := \int_\mathcal{C} dW \rho(W) \frac{Z+W}{Z-W} ,
\label{resolvent-CS}
\end{equation}
where  $\mathcal{C}$ is the support of $\rho(Z)$.
Then the saddle point equation (\ref{eom-CS2}) becomes
\begin{equation}
V'(Z)=\lim_{\epsilon \rightarrow 0} \left[ v(Z+i\epsilon)+v(Z-i\epsilon) \right],
\qquad
(Z \in \mathcal{C}),
\qquad
V'(Z):=\frac{1}{\pi i\lambda} \log{Z}.
\label{eom-CS3}
\end{equation}
Besides, the resolvent satisfies the boundary conditions
\begin{equation}
\lim_{Z \rightarrow \infty} v(Z) = 1, \qquad \lim_{Z \rightarrow 0}v(Z) = -1,
\label{boundary-CS}
\end{equation}
through the definition (\ref{resolvent-CS}).
By using the resolvent, the eigenvalue density is described as
\begin{equation}
\rho(Z)=-\frac{1}{4\pi iZ} \lim_{\epsilon \rightarrow 0} \left[ v(Z+i\epsilon)-v(Z-i\epsilon) \right],
\qquad
(Z \in \mathcal{C}).
\label{rho-CS}
\end{equation}
In the following subsections, we will explore the solutions of the equation (\ref{eom-CS3}) which obey the boundary conditions (\ref{boundary-CS}).

%%%%%%%%%%%%%%%%%%%%%%%%%%%%%%%%%%%%%%%%%%%%%%%%%%%%%%%%%
\subsection{One-cut solution}
%%%%%%%%%%%%%%%%%%%%%%%%%%%%%%%%%%%%%%%%%%%%%%%%%%%%%%%%%

We review the derivation of the resolvent describing the one-cut solution shown in Figure \ref{fig-numerical-CS} (top-left) \cite{Aganagic:2002wv, Pasquetti:2009jg, Halmagyi:2003ze}.
There are various derivations of this solution, and we employ the integral method of Migdal \cite{Migdal:1984gj} which is useful for finding the general solution later.

\begin{figure}
\begin{tabular}{cc}
\begin{minipage}{0.5\hsize}
\begin{center}
        \includegraphics[scale=0.25]{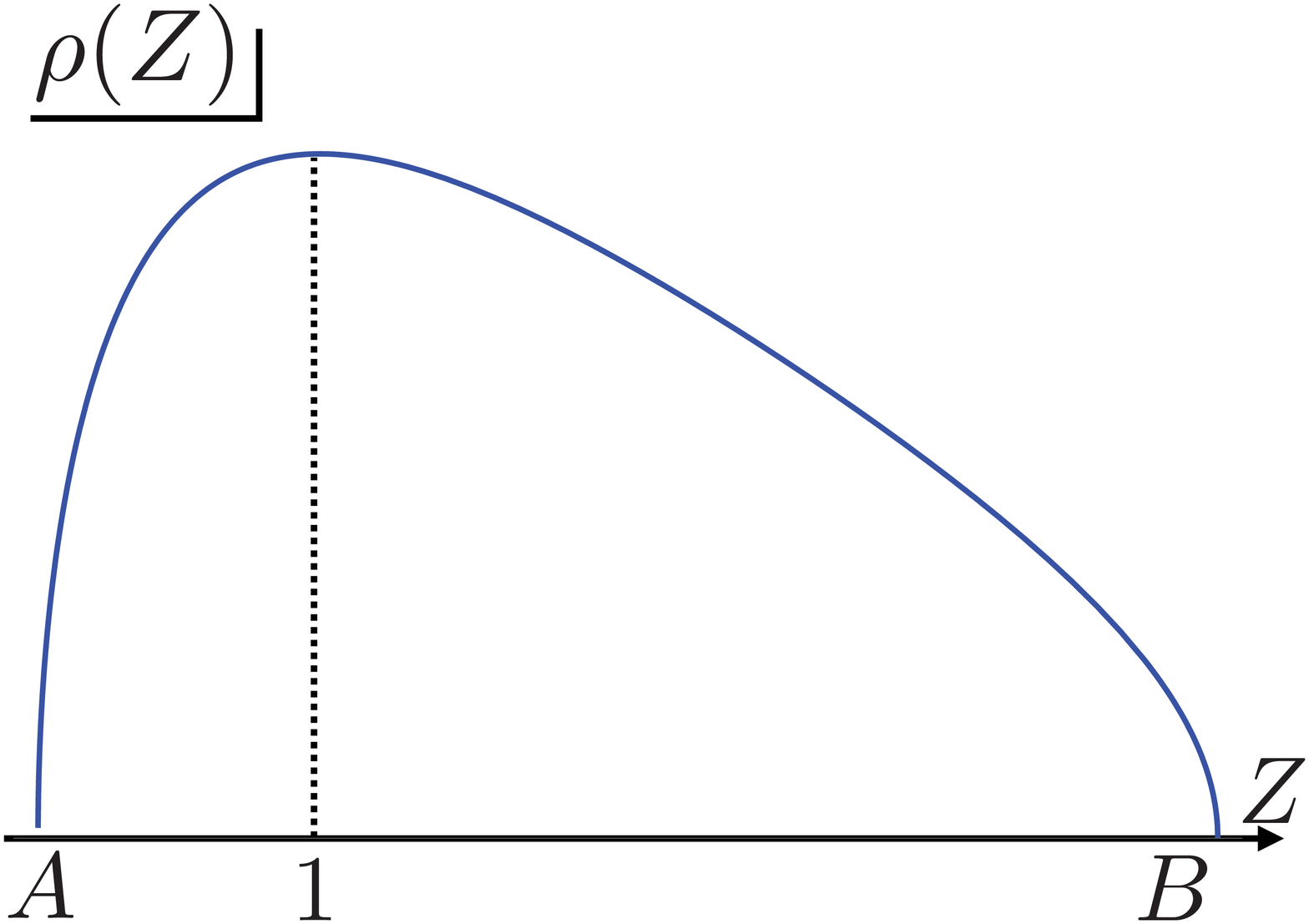}
   \end{center}
\end{minipage}
\begin{minipage}{0.5\hsize}
\begin{center}
        \includegraphics[scale=0.25]{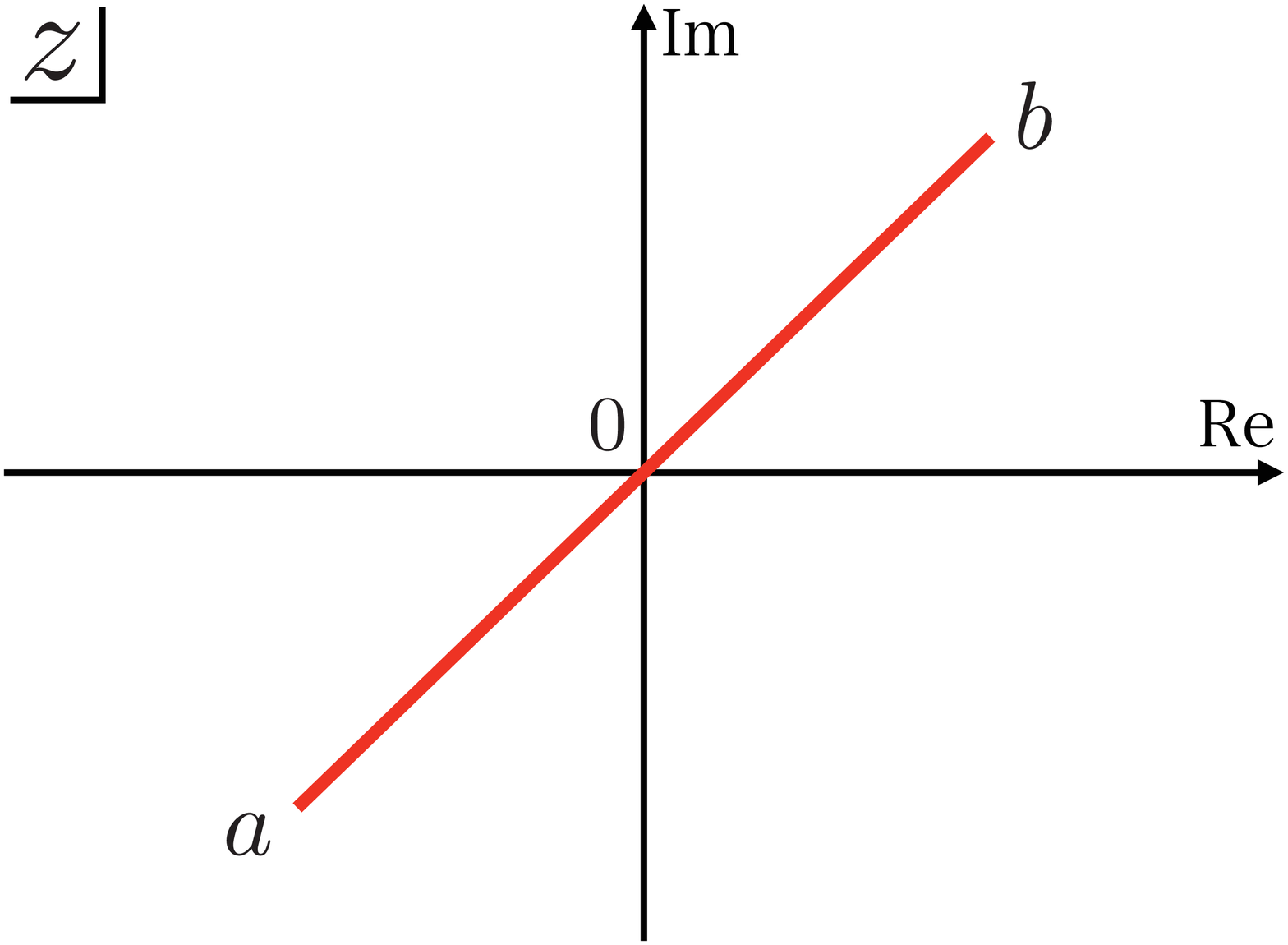}
\end{center}
\end{minipage}
\end{tabular}
       \caption{Schematic plot of the the eigenvalue density $\rho(Z)$ of the one-cut solution (\ref{density-CS}) and the eigenvalue distribution on the $z$-plane.
      $A,B$ and $a,b$ are given in (\ref{endpt-CS}).
       }
        \label{fig:CS1}
\end{figure}

The potential $V'(Z)$ (\ref{eom-CS3}) has the unique extreme at $Z=1$ (or $z=0$ where $z:=\log Z$), and the eigenvalues tend to be around there.
Hence we assume that $\rho(Z)$ has a single support on the interval $[A,B]$ near $Z=1$, where $A$ and $B$ ($|A| <|B|$) will be fixed soon.
We apply the ansatz \cite{Migdal:1984gj} for the solution of the saddle point equation (\ref{eom-CS3}) \cite{Halmagyi:2003ze}, 
\begin{equation}
v(Z)=\oint_{C_1} \frac{dW}{4\pi i} \frac{V'(W)}{Z-W} \sqrt{\frac{(Z-A)(Z-B)}{(W-A)(W-B)}}.
\label{Migdal-CS}
\end{equation}
Here the contour $C_1$ encircles the support $[A,B]$ counterclockwise\footnote{We employ the script $C$ for the closed  contours and $\mathcal{C}$ for the supports in this article. }.
By performing this integral\footnote{To perform this integral, we deform the contour $C_1$ so that it encloses the pole at $W=Z$ and the branch cut $W \in [-\infty, 0]$ of $\log W$ \cite{Kazakov:1995ae}.}, we obtain
\begin{align}
&v(Z)=\frac{1}{\pi i\lambda} \log{\left( \frac{f(Z)-\sqrt{f^2(Z)-4Z} }{2} \right)}, \nonumber \\
&f(Z)=f_0+f_1Z, \quad
f_0=\frac{2\sqrt{AB}}{\sqrt{A}+\sqrt{B}},
\quad
f_1=\frac{2}{\sqrt{A}+\sqrt{B}}.
\label{one-cut-CS}
\end{align}
Then, through the boundary conditions (\ref{boundary-CS}), $A$ and $B$ are determined as
\begin{equation}
A= \exp \left(-2\, {\rm arccosh} \left( e^{\pi i\lambda} \right) \right),
\qquad
B= \exp \left( 2\, {\rm arccosh} \left( e^{\pi i\lambda} \right) \right)=1/A.
\label{CS-AB}
\end{equation}
The eigenvalue density is obtained through (\ref{rho-CS}),
\begin{equation}
\rho(Z)=\frac{1}{4\pi^2 \lambda Z} \log{\left( \frac{Z+\sqrt{AB}-i \sqrt{(Z-A)(Z-B)}}{Z+\sqrt{AB}+i \sqrt{(Z-A)(Z-B)}} \right)},
\qquad
(Z \in [A,B]).
\label{density-CS}
\end{equation}
We sketch the profile of this density in Figure \ref{fig:CS1}.
In order to compare the obtained result with the numerical result shown in Figure \ref{fig-numerical-CS}, we rewrite our results by using the variable $z=\log Z$ which corresponds to $u_i$ in (\ref{eom-CS}).
Correspondingly, $A$ and $B$ are mapped to
\begin{equation}
b = \log{B}=2\, {\rm arccosh} \left( e^{\pi i\lambda} \right),
\qquad
a= \log{A}=-2\, {\rm arccosh} \left( e^{\pi i\lambda} \right),
\label{endpt-CS}
\end{equation}
and they satisfy $a=-b$. (This is expected, since the system is symmetric under $z \to -z$.)
See Figure \ref{fig:CS1}. 
This solution describes the numerically obtained one-cut solution shown in Figure \ref{fig-numerical-CS}.

Note that the resolvent in the $z$ variable has the branch cuts on $z \in [a+2\pi in , b+2\pi in]$, $(n \in \mathbb{Z})$, although the eigenvalues are distributed on $z \in [a , b]$ only.
These additional infinite number of the cuts are related to the periodicity  $u_i \to u_i + 2\pi i$ of the right hand side of the saddle point equation (\ref{eom-CS}), and the equation of motion (\ref{eom-CS3}) is not satisfied there. 
In this article, we refer to the solutions in which $k$ mobs of the eigenvalues exist in the $z$ plane as ``$k$-cut solution", and do not count these additional cuts as ``cuts''.

%%%%%%%%%%%%%%%%%%%%%%%%%%%%%%%%%%%%%%%%%%%%%%%%%%%%%%%%%
\subsection{Stepwise two-cut solution}
\label{sec-2-cut-CS}
%%%%%%%%%%%%%%%%%%%%%%%%%%%%%%%%%%%%%%%%%%%%%%%%%%%%%%%%%

We consider the derivation of the stepwise two-cut solution \cite{Morita:2017oev} plotted in Figure \ref{fig-numerical-CS} (top-right).

Since the potential $V'(Z)$ (\ref{eom-CS3}) has only the single extreme at $Z=1$, it might be difficult to imagine that the saddle point equation (\ref{eom-CS3}) allows such a two-cut solution.
The key is the periodicity $u_i \to u_i+2 \pi i$ of the right hand side of the saddle point equation (\ref{eom-CS}).
Thanks to this periodicity, strong interactions between the eigenvalues arise if they are separated by $2\pi i$, and these interactions make the various solutions shown in Figure \ref{fig-numerical-CS} possible.

Before considering the $N=\infty$ case, we study the $N=2$ case as an example \cite{Morita:2017oev, Morita:2011cs}.
In this case, the saddle point equation (\ref{eom-CS}) become
\begin{equation}
\frac{u_1}{2\pi i \lambda}=\frac{1}{2} \coth{\frac{u_1-u_2}{2}},  \qquad
\frac{u_2}{2\pi i \lambda}=-\frac{1}{2} \coth{\frac{u_1-u_2}{2}}.
\label{N=2}
\end{equation}
By summing these two equations, we find $u_1=-u_2$, and the equations reduce to
\begin{equation}
\frac{u_1}{2\pi i \lambda}=\frac{1}{2} \coth u_1.
\label{N=2'}
\end{equation}
This equation indeed allows infinite number of solutions.
At weak coupling $|\lambda| \ll 1$, we can perturbatively obtain the solutions,
\begin{eqnarray}
u_1= \pm \sqrt{\pi i \lambda} + \cdots ,\qquad
u_1= \pi i n + \frac{\lambda}{n}  + \cdots , 
\label{N=2-multi}
\end{eqnarray}
where $n$ is a non-zero integer.
The first solution would correspond to the one-cut solution (\ref{density-CS}) at large-$N$, while the second one indicates the existence of a new class of the solutions.
Particularly the second solution satisfies $u_1-u_2= 2 \pi in + O(\lambda)$, and they are separated by $2\pi in$.
Thus, the periodicity of the right hand side of the saddle point equation (\ref{eom-CS}) causes the various solutions as we expected.

\begin{figure}
\begin{tabular}{cc}
\begin{minipage}{0.5\hsize}
\begin{center}
        \includegraphics[scale=0.25]{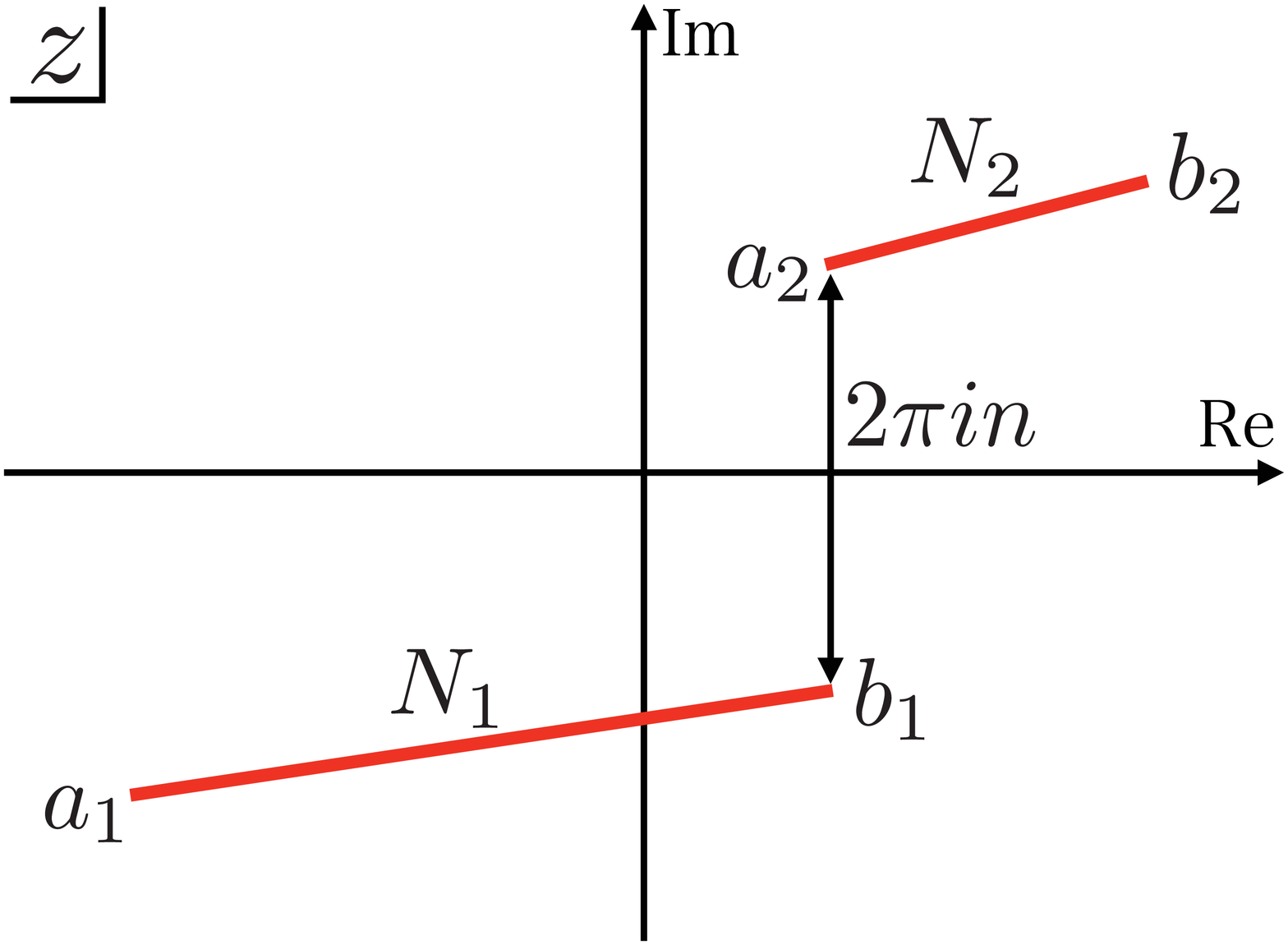}
\end{center}
\end{minipage}
\begin{minipage}{0.5\hsize}
\begin{center}
        \includegraphics[scale=0.25]{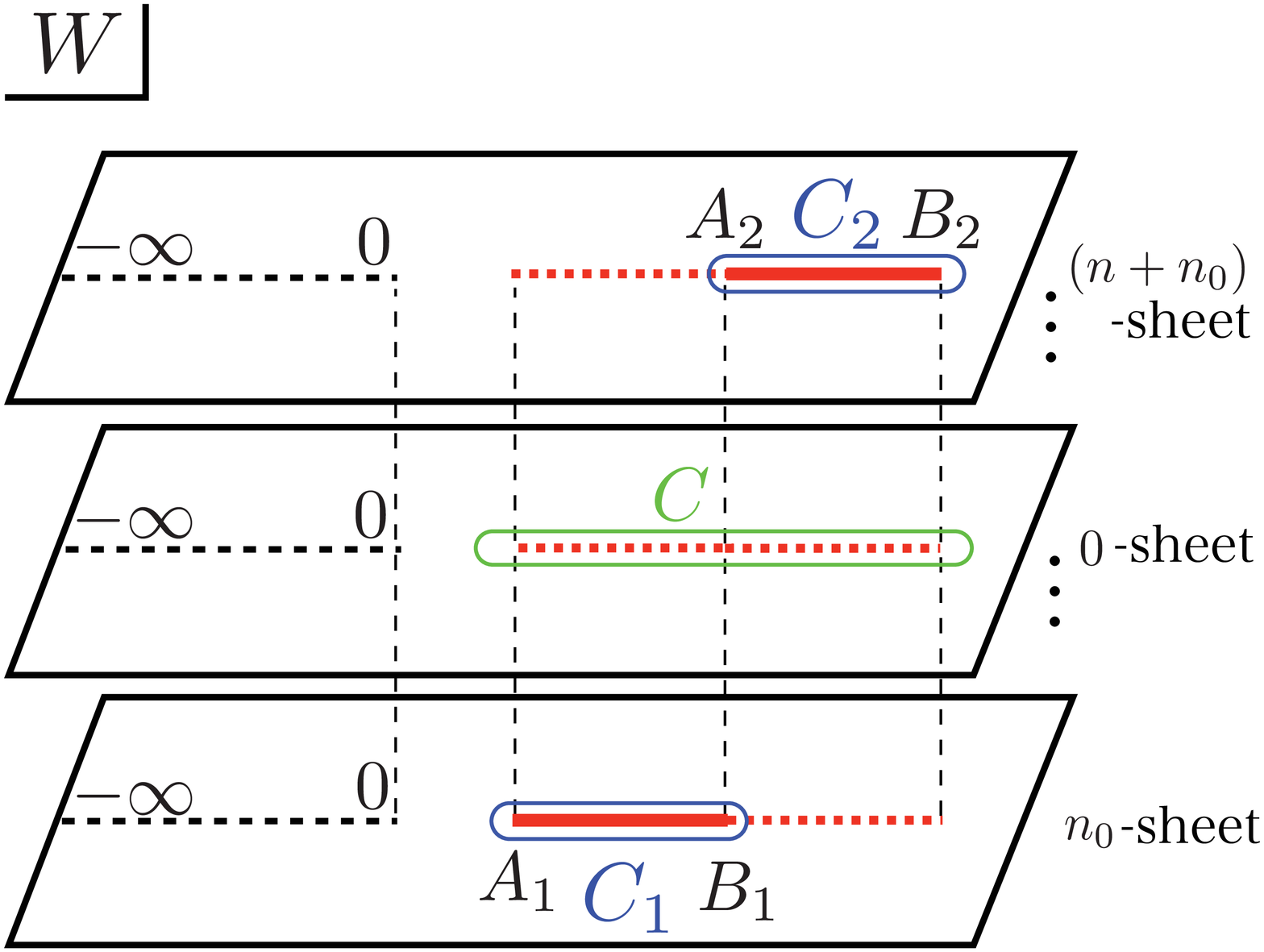}
\end{center}
\end{minipage}
\end{tabular}
       \caption{(Left) Sketch of the stepwise two-cut solution in the pure CS matrix model.
       The red lines describe the eigenvalue distributions.
(Right) Integral contours in the integral of the resolvent. 
     The doted lines denote the branch cuts of the integrand.
       The blue lines are the integral contour $C_1$ and $C_2$ in (\ref{Migdal-CS2}).
The green line denotes the contour $C$ in (\ref{Migdal-CS2''}).
 }
        \label{fig-two-CS}
\end{figure}

Let us move on the $N=\infty$ case.
To find the two-cut solution corresponding to the numerical result shown in Figure \ref{fig-numerical-CS},
we assume the two branch cuts $[a_i,b_i]$ where ${\rm Re}(a_i) \le {\rm Re} (b_i) $ and  ${\rm Im}(a_i) \le {\rm Im} (b_i) $ $(i=1,2)$ on the $z$-plane ($z= \log Z$) satisfying\footnote{\label{ftnt-tilt}We assume the condition ${\rm Re}(a_i) \le {\rm Re} (b_i) $ and  ${\rm Im}(a_i) \le {\rm Im} (b_i) $. This is because the potential $V'(z)= \frac{1}{\pi i \lambda} z$ forces the eigenvalues to compose such a configuration when $\lambda$ is real and positive.
We can see it from the results of the Newton method.} 
\begin{equation}
a_2=b_1+2\pi in.
\label{CS2ansatz}
\end{equation}
Here $n$ is a positive integer. (We will argue why we restrict $n$ positive in Appendix \ref{app-negative-n}.) 
We also assume that the first cut and the second cut consist of $N_1$ and $N_2(=N-N_1)$ eigenvalues, respectively. 
See Figure \ref{fig-two-CS}.
Since the branch cuts are always separated by the fixed number $2\pi i n$, we call this solution ``stepwise two-cut solution".

On the $Z$-plane, these cuts are mapped to $A_i=e^{a_i}$ and $B_i=e^{b_i}$ and they satisfy
\begin{align}
A_2=e^{2\pi in} B_1,
\label{CS-step}
\end{align}
through (\ref{CS2ansatz}). 
We assign a new symbol $D_1:=B_1$ for this point, since the properties of this point are different from $A_1$ and $B_2$ as we will see soon.
Note that, because of the branch cut of $\log Z$ in $V'(Z)$ (\ref{eom-CS3}), $A_2$ and $B_1$ stand different points on the Riemann surface.
See the right sketch of Figure \ref{fig-two-CS}.

By regarding the locations of these branch cuts, we propose the ansatz for the resolvent of the stepwise two-cut solution
\begin{equation}
v(Z)=\oint_{C_1 \cup \, C_2} \frac{dW}{4\pi i} \frac{V'(W)}{Z-W} \sqrt{\frac{(Z-A_1)(Z-B_2)}{(W-A_1)(W-B_2)}},
\qquad
V'(Z)=\frac{1}{\pi i\lambda} \log{Z}.
\label{Migdal-CS2}
\end{equation}
Here the integral contour $C_1$ and $C_2$ encircle the branch cut $[A_1,B_1]$ and $[A_2,B_2]$ counterclockwise, respectively, and they are on the different sheets as shown in Figure \ref{fig-two-CS}.
It is not difficult to show that this ansatz satisfies the saddle point equation (\ref{eom-CS3}) on the cut $[A_1,B_1]$ and $[A_2,B_2]$ \footnote{Our ansatz (\ref{Migdal-CS2}) is similar to the ansatz for the $m$-cut solution of Hermitian matrix models \cite{Migdal:1984gj} 
\begin{equation}
w(z)=\sum_{i=1}^{m} \oint_{C_i} \frac{dw}{4\pi i} \frac{V'(w)}{z-w} \prod_{i=1}^m \sqrt{\frac{(z-a_i)(z-b_i)}{(w-a_i)(w-b_i)}},
\label{Migdal-hermite}
\end{equation}
where $a_i$ and $b_i$ denote the end points of the $i$-th branch cuts ($i=1,\cdots, m$).
The difference is that  the end point $B_1$ and $A_2$ do not appear in the inside of the square root in our ansatz (\ref{Migdal-CS2}).
Since $B_1$ and $A_2$ are the same point on the different sheets, even though they do not appear in the square root, $v(Z)$ satisfies the saddle point equation (\ref{eom-CS3}) on the cuts.
 }.

\begin{figure}
\begin{tabular}{cc}
\begin{minipage}{0.5\hsize}
\begin{center}
        \includegraphics[scale=0.25]{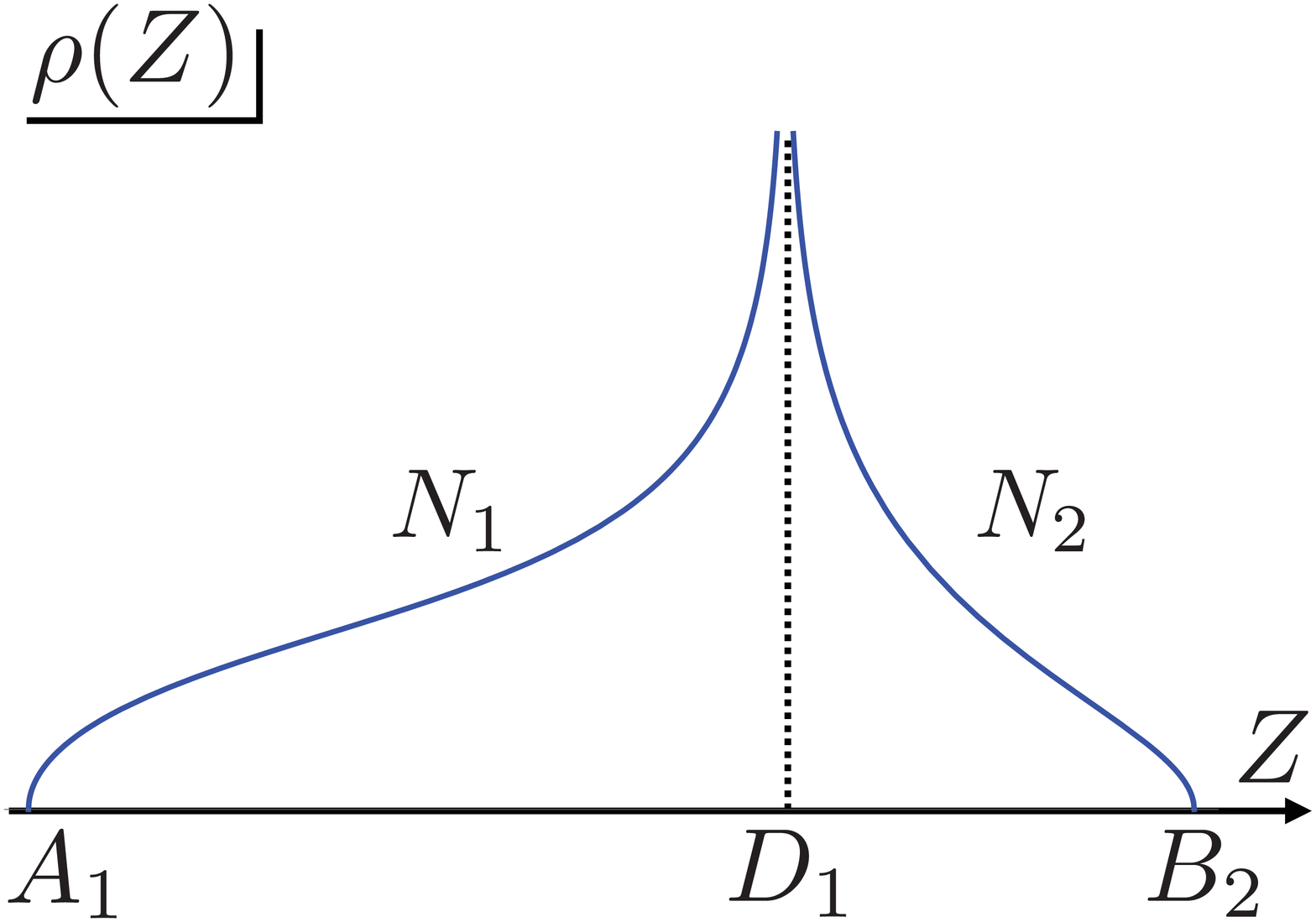}\\  
\end{center}
\end{minipage}
\begin{minipage}{0.5\hsize}
\begin{center}
        \includegraphics[scale=0.25]{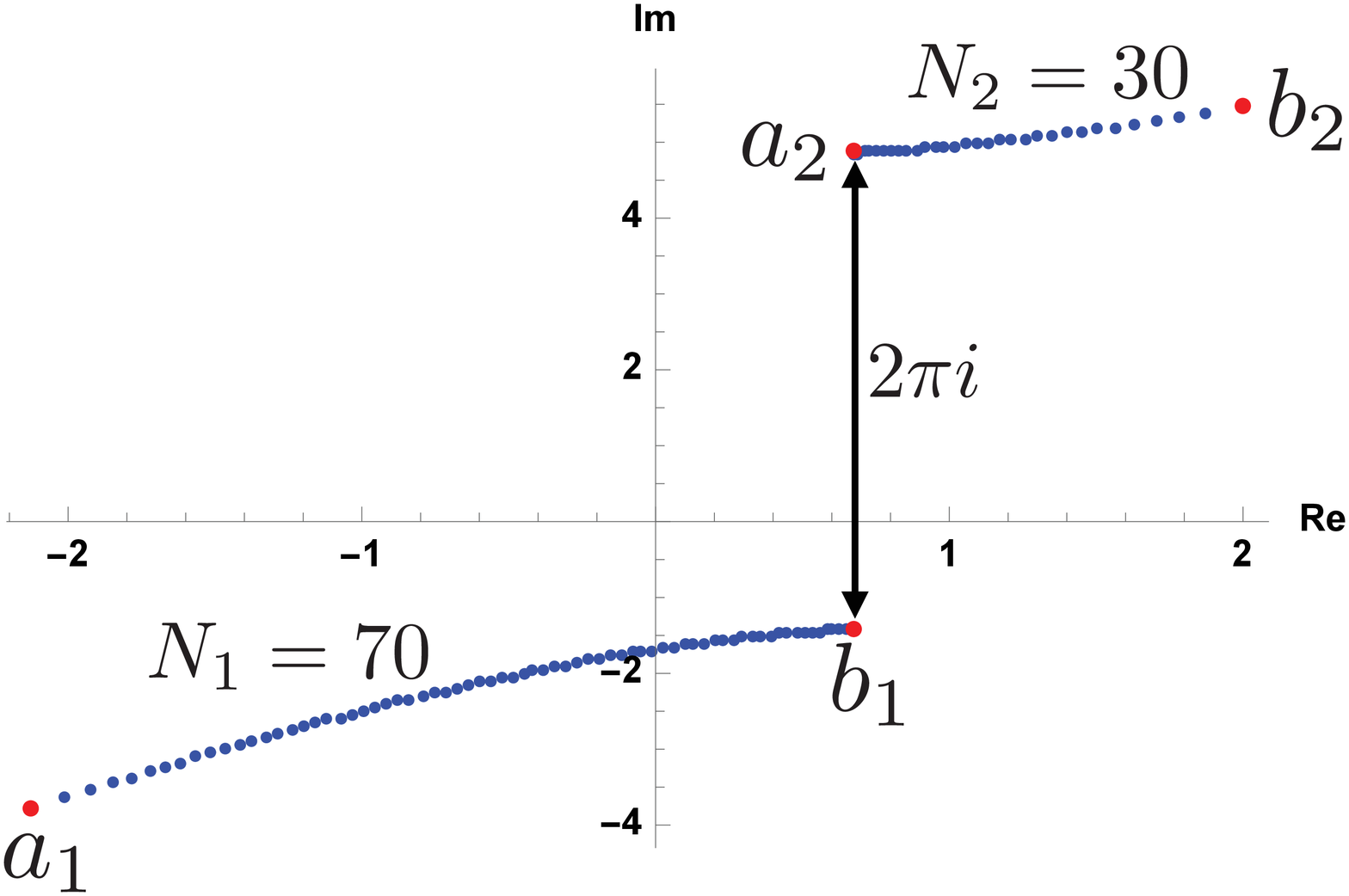}\\
\end{center}
\end{minipage}
\end{tabular}
       \caption{(Left) Schematic plot of the eigenvalue density for the stepwise two-cut solution (\ref{rho-CS2}) at a weak coupling.
Here we have projected the cut $[A_2,B_2]$ to the same sheet to $[A_1,B_1]$.
A logarithmic singularity appears at $Z=D_1(=B_1)$.
(Right)
Comparison of the Newton method (blue dots) and the result through (\ref{two-cut-CS}).
For the Newton method, we take $N_1=70$, $N_2=30$, $n=1$ and $\lambda=0.5$.
These two results agree very well.
}
        \label{fig-density-two-CS}
\end{figure}

Now we evaluate the integral in (\ref{Migdal-CS2}).
Since the integrand involves $\log W$ in $V'(W)$, we need to take care of the branch cut.
We assume that $C_1$ is on the $n_0$-th sheet\footnote{If we sum up the saddle point equation (\ref{eom-CS}), we obtain $\sum_{i=1}^N u_i=0$. This implies that the center-of-mass of the eigenvalues is at the origin. Thus the first cut  $\mathcal{C}_1$ may be on a negative sheet while the second cut  $\mathcal{C}_2$ may be on a positive sheet: $n_0 \le 0$ and $n+n_0 \ge 0$. }. (It implies $C_2$ is on the  $n+n_0$-th sheet through the ansatz (\ref{CS2ansatz}).)
Then we can evaluate the integral (\ref{Migdal-CS2}) as
\begin{align}
v(Z)=&\oint_{C} \frac{dW}{4\pi i} \frac{1}{\pi i \lambda} \frac{\log W}{Z-W}  \sqrt{\frac{(Z-A_1)(Z-B_2)}{(W-A_1)(W-B_2)}} + 
\oint_{C_1} \frac{dW}{4\pi i} \frac{2n_0}{ \lambda} \frac{1}{Z-W}  \sqrt{\frac{(Z-A_1)(Z-B_2)}{(W-A_1)(W-B_2)}}
\nonumber \\
&
+ 
\oint_{C_2} \frac{dW}{4\pi i} \frac{2(n+n_0)}{ \lambda} \frac{1}{Z-W}  \sqrt{\frac{(Z-A_1)(Z-B_2)}{(W-A_1)(W-B_2)}}
\label{Migdal-CS2''}.
\end{align}
Here the contour $C$ encircles the branch cut $[A_1, B_2]$ on the 0-th sheet.
See Figure \ref{fig-two-CS}.
The first integral is identical to (\ref{Migdal-CS}) and the second and third integrals have been done in \cite{Morita:2017oev}, and we obtain
\footnote{The second term can be written as
\begin{align}
\log{\left( \frac{q(Z)+\sqrt{q^2(Z)-4}}{2} \right)}
=2i \arctan{\left( \sqrt{\frac{Z-A_1}{Z-B_2}} \sqrt{\frac{B_2-D_1}{D_1-A_1}} \right)}.
\label{two-cut-arctan}
\end{align}
}
\begin{align}
v(Z)=&\frac{1}{\pi i\lambda} \log{\left( \frac{f(Z)-\sqrt{f^2(Z)-4Z}}{2} \right)}
+\frac{n}{\pi i\lambda} \log{\left( \frac{q(Z)+\sqrt{q^2(Z)-4}}{2} \right)}+\frac{n_0}{\lambda}, 
\nonumber \\
&f(Z)=f_0+f_1 Z, \qquad\quad
f_0=\frac{2\sqrt{A_1B_2}}{\sqrt{A_1}+\sqrt{B_2}},
\quad
f_1=\frac{2}{\sqrt{A_1}+\sqrt{B_2}}, \nonumber
\\
&q(Z)=\frac{q_1Z-q_0D_1}{Z-D_1},
\qquad
q_1=\frac{2(2D_1-A_1-B_2)}{B_2-A_1},
\quad
q_0=\frac{2(D_1B_2+D_1A_1-2A_1B_2)}{D_1(B_2-A_1)}.
\label{two-cut-CS}
\end{align}
This result agrees with that of Ref.~\cite{Morita:2017oev} which employs a different method\footnote{Our result (\ref{two-cut-CS}) differs from the resolvent (64) of our previous work \cite{Morita:2017oev} by a constant term. This is because Ref. \cite{Morita:2017oev} used a different variable $Z$ which was defined on page 17 of \cite{Morita:2017oev}. }. 
(In Appendix \ref{app-holomorphy}, we show how the resolvent (\ref{two-cut-CS}) satisfies the saddle point equation (\ref{eom-CS3}).
There, we also argue another derivation of this solution via holomorphy.)
From (\ref{rho-CS}), the eigenvalue density becomes
\begin{align}
\rho(Z) = & \frac{1}{4\pi^2 \lambda Z} \log{\left( \frac{Z+\sqrt{A_1B_2}-i \sqrt{(Z-A_1)(Z-B_2)}}{Z+\sqrt{A_1B_2}+i \sqrt{(Z-A_1)(Z-B_2)}} \right)} \\
&+ \frac{n}{\pi^2 \lambda Z}
\left \{
\begin{array}{l}
{\rm arctanh} \left( \sqrt{\dfrac{Z-A_1}{B_2-Z}} \sqrt{\dfrac{B_2-D_1}{D_1-A_1}} \right),
\qquad
(Z \in [A_1,B_1]),
\\
-{\rm arctanh} \left( \sqrt{\dfrac{B_2-Z}{Z-A_1}} \sqrt{\dfrac{D_1-A_1}{B_2-D_1}} \right),
\qquad
(Z \in [A_2,B_2]).
\end{array}
\right.
\label{rho-CS2}
\end{align}
The profile of this density at a small $\lambda$ is shown in Figure \ref{fig-density-two-CS}.
Particularly a logarithmic divergence at $Z=D_1$ arises from the second term due to the pole of $q(Z)$, although the integral of $\rho(Z)$ is finite \cite{Morita:2017oev}.
The existence of the divergence is quite contrast to the one-cut solution shown in Figure \ref{fig:CS1}.

Finally we have to fix the values of the undetermined constant $A_1,B_1(=D_1)$ and $B_2$.
We impose the two boundary conditions  at $Z=0$ and $Z=\infty$  (\ref{boundary-CS}) and the additional condition\footnote{If $N_1=N_2$, the solution becomes symmetric under $Z \to 1/Z$ and it makes the calculation much simpler.
We obtain $B_2=1/A_1$ and $A_2=1/B_1=e^{\pi i n}$, and we need to consider only one of  the boundary condition (\ref{boundary-CS}). } \begin{equation}
\frac{N_1}{N}=\int^{B_1}_{A_1} \rho(Z) dZ
=\frac{1}{4\pi i} \oint_{C_1} \frac{v(Z)}{Z} dZ,
\label{cycle-CS2}
\end{equation}
which demands that the $N_1$ eigenvalues are on the first cut $[A_1,B_1]$.
Thus $A_1,B_1$ and $B_2$ should be determined as the solution of these three equations and they are given as functions of the input parameters: $\lambda$, $n$ and $N_1/N$.
($n_0$ is determined when $A_1$ is fixed.)

However finding the solution for the general input parameters is difficult.
Also it is hard to answer whether the solution exists or not for the given parameters, and, even if it exists, whether it is unique or not.
In addition, even if we found a solution, if the eigenvalue density $\rho(Z)$ is not positive, the solution is not allowed.
For example, the one-cut solution (\ref{one-cut-CS}) is allowed only when $-1 \le \lambda \le 1$, if $\lambda$ is real \cite{Morita:2011cs}.

Some solvable cases were explored in \cite{Morita:2017oev}.
For example, at weak coupling $|\lambda| \ll 1$, 
the solution is uniquely given by 
\begin{align}
a_1 = &~ b_1-\frac{2\pi \lambda}{n} \tan{\left( \frac{\pi}{2} \frac{N_1}{N} \right)}+O(\lambda^2),  \qquad
 b_2 = a_2+\frac{2\pi \lambda}{n} \tan{\left( \frac{\pi}{2} \frac{N_2}{N} \right)}
+O(\lambda^2),
 \nonumber \\
b_1  = &~ d_1=a_2 -2\pi i n= - \frac{2\pi i n N_2}{N}
+\frac{\lambda \pi }{2n} 
\left( \tan{\left( \frac{\pi}{2} \frac{N_1}{N} \right)}-\tan{\left( \frac{\pi}{2} \frac{N_2}{N} \right)} \right) +O(\lambda^2).
\label{weak-ab}
\end{align}
Here $d_1:=\log D_1$.
In this case, the cuts are parallel to the real axis if $\lambda$ is real.

In the case of a finite $\lambda$, we can find the solution if $N_1=N_2$ and $n=1,2,3$ and 4.
In the case of $n=1$, the solution is given by
\begin{align}
B_2=1/A_1= e^{\pi i}\left( -i (e^{\pi i \lambda}-1)+\sqrt{2 e^{\pi i \lambda}-e^{2\pi i \lambda}}\right)^2, \qquad A_2=e^{2\pi i }B_1= e^{\pi i }.
\end{align}

Also we can find the solutions by solving (\ref{boundary-CS}) and (\ref{cycle-CS2}) numerically.
For example, when we take $\{\lambda,n,N_1/N \}=\{0.5,1,0.7\}$, we obtain a solution as shown in Figure \ref{fig-density-two-CS}\footnote{We use FindRoot in Mathematica in our numerical computation.
Then we find various solutions depending on the initial condition of FindRoot. 
We choose the solution which is consistent with the result of the Newton method.
It is unclear whether the other solutions are all meaningful, since the equations involve several multivalued functions which may lead to wrong numerical results.
Also some solutions might correspond to the eigenvalue density involving negative values which
are not allowed physically.}.
The result agrees with the numerical result derived through the Newton method in which we solve the saddle point equation (\ref{eom-CS}) at finite $N$ directly \cite{Herzog:2010hf, Niarchos:2011sn, Minwalla:2011ma}\footnote{Although we can derive the end points of the eigenvalue distribution, obtaining the distribution curve on the complex plane is technically difficult unless the coupling $\lambda$ is small as in (\ref{weak-ab}).}.

%%%%%%%%%%%%%%%%%%%%%%%%%%%%%%%%%%%%%%%%%%%%%%%%%%%%%%%%%
\subsection{Stepwise multi-cut solution}
%%%%%%%%%%%%%%%%%%%%%%%%%%%%%%%%%%%%%%%%%%%%%%%%%%%%%%%%%

\begin{figure}
\begin{tabular}{cc}
\begin{minipage}{0.5\hsize}
\begin{center}
        \includegraphics[scale=0.25]{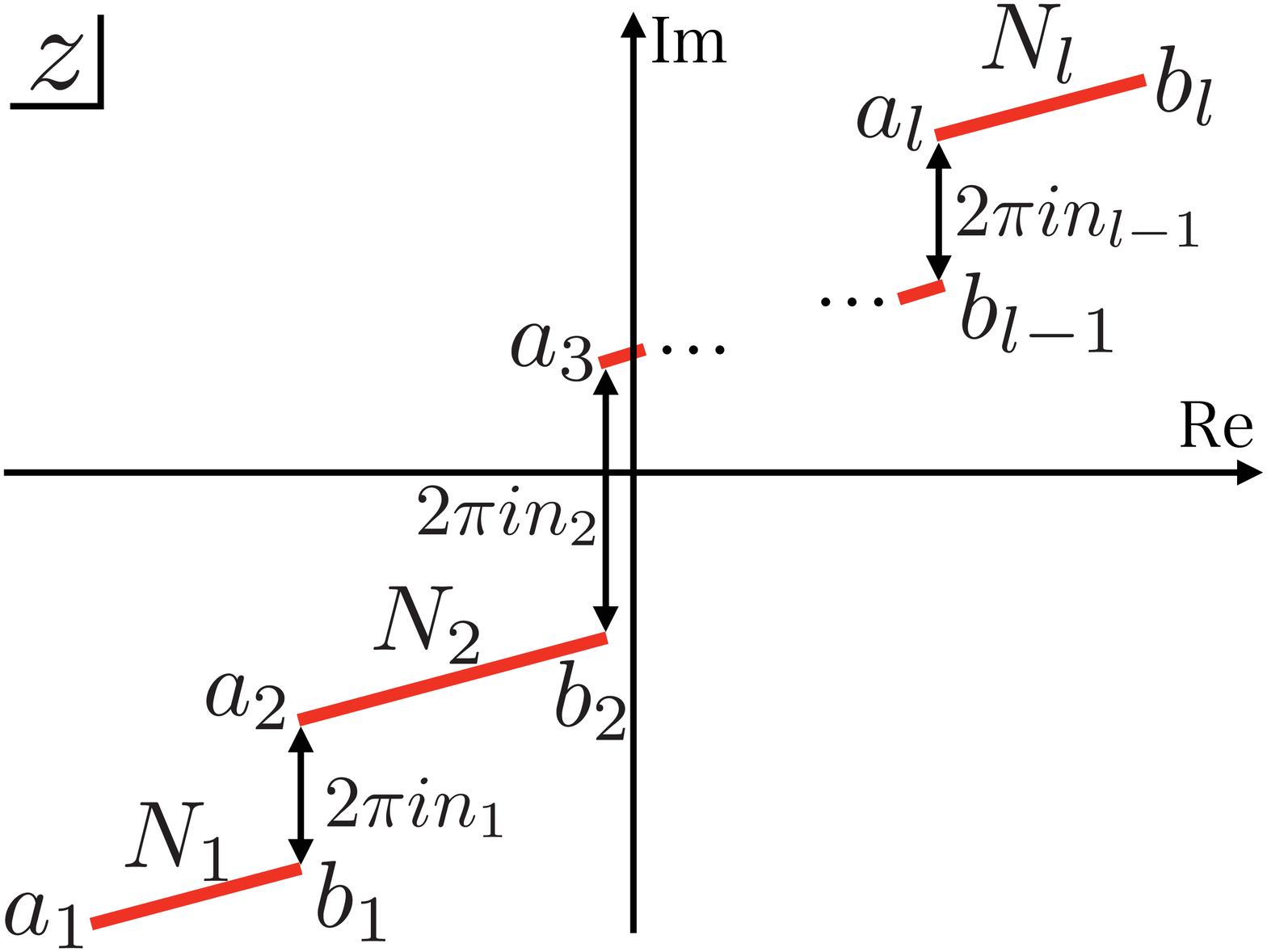}
  
\end{center}
\end{minipage}
\begin{minipage}{0.5\hsize}
\begin{center}
        \includegraphics[scale=0.25]{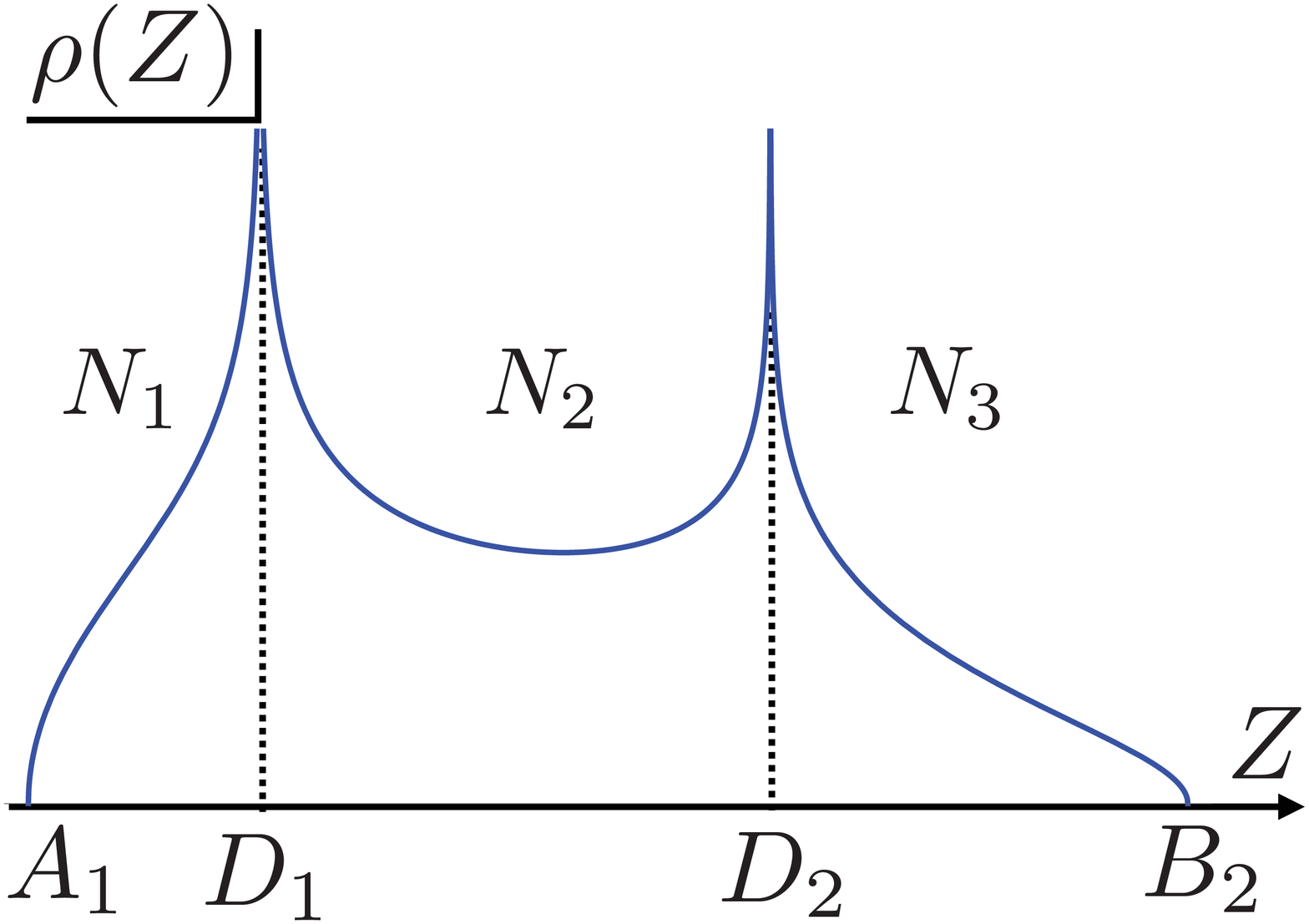}
\end{center}
\end{minipage}
\end{tabular}
       \caption{Sketch of the eigenvalue distribution and the eigenvalue density of the stepwise multi-cut solution. 
       In the eigenvalue density, we project the cuts on the different sheet to the same sheet and consider a small $\lambda$.} 
        \label{fig-g-CS}
\end{figure}

We develop the derivation of the stepwise two-cut solution in the previous section and consider the stepwise multi-cut solution in Figure \ref{fig-numerical-CS} (bottom-left).
For a stepwise $l$-cut solution, there would be cuts $[a_j, b_j]$ on the $z$-plane which satisfy ${\rm Re}(a_j) \le {\rm Re} (b_j) $ and ${\rm Im}(a_j) \le {\rm Im} (b_j) $, $(j=1, \cdots, l)$.
We assume that the $j$-th cut consists of $N_j$ eigenvalues ($\sum_{j=1}^l N_j=N$).
Similar to the stepwise two-cut solution, we impose that the end points of these cuts satisfy
\begin{equation}
a_{j+1}=b_j+2\pi i n_j,
\qquad
(j=1, \cdots, l-1),
\label{CSgansatz}
\end{equation}
where $\{ n_j \}$ are positive integers.
See Figure \ref{fig-g-CS}.
In terms of the $Z$ variable, this assumption implies the cut $[A_j, B_j]$ satisfying $A_{j+1}=e^{2\pi i n_j} B_j$.
Then, by generalizing (\ref{Migdal-CS2}) in the stepwise two-cut solution, we use the following ansatz for the resolvent 
\begin{equation}
v(Z)= \sum_{j=1}^l \oint_{ C_j} \frac{dW}{4\pi i} \frac{V'(W)}{Z-W} \sqrt{\frac{(Z-A_1)(Z-B_l)}{(W-A_1)(W-B_l)}},
\qquad
V'(Z)=\frac{1}{\pi i\lambda} \log{Z}.
\label{Migdal-CSg}
\end{equation}
Here the integral contour $C_j$  encircle the branch cut $[A_j,B_j]$ counterclockwise.
Again these contours are on the different sheets of $\log Z$, and we assume that the first contour $C_1$ is on the $n_0$-th sheet.

\begin{figure}
\begin{tabular}{cc}
\begin{minipage}{0.5\hsize}
\begin{center}
        \includegraphics[scale=1.1]{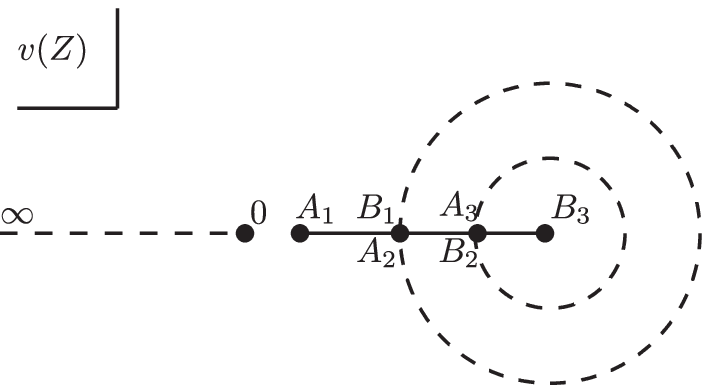}\\
\end{center}
\end{minipage}
\begin{minipage}{0.5\hsize}
\begin{center}
        \includegraphics[scale=0.25]{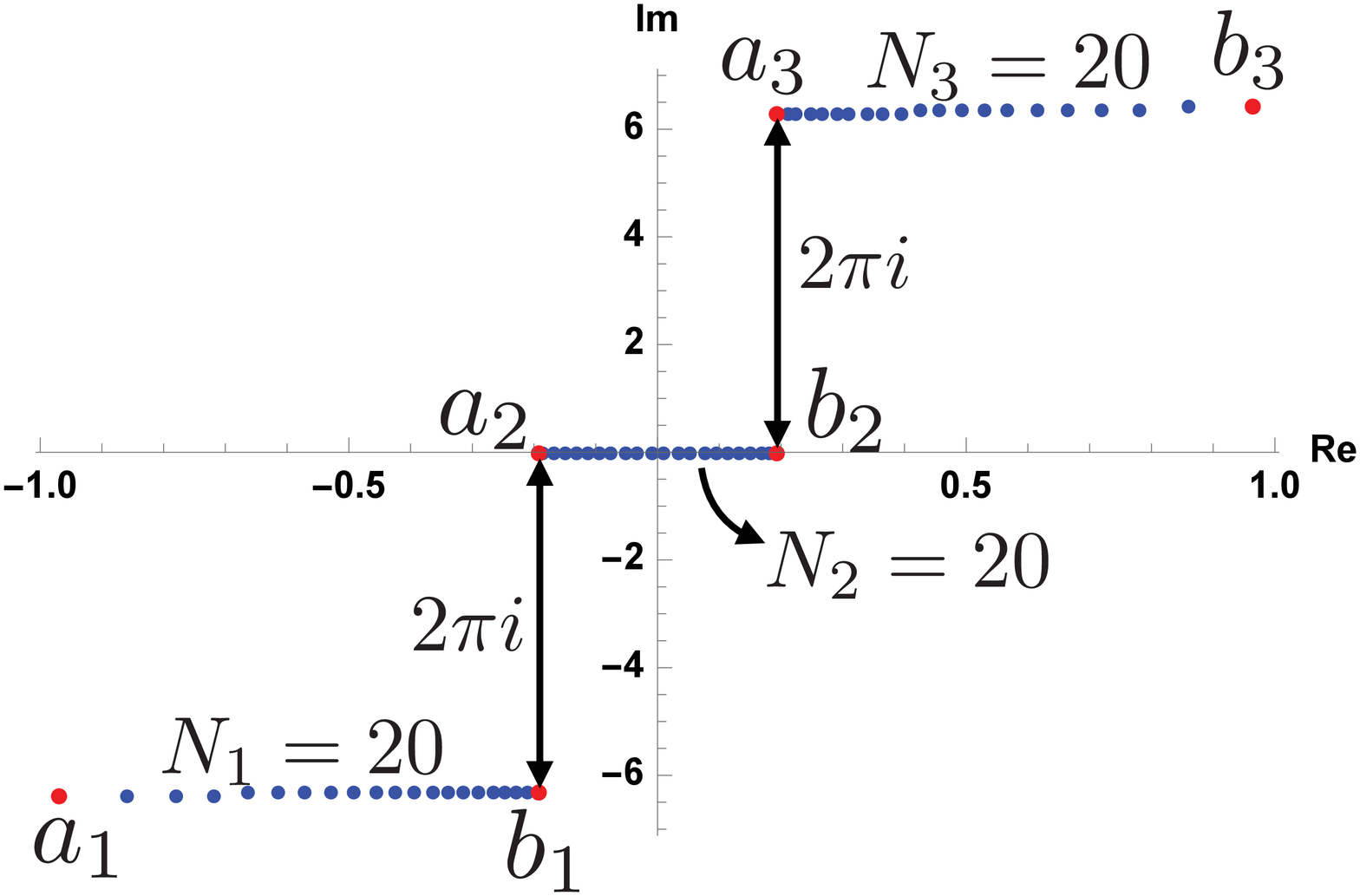}\\
\end{center}
\end{minipage}
\end{tabular}
       \caption{
       (Left) Branch cuts of the resolvent of the  stepwise three-cut solution (\ref{multi-cut-CS}) on the $Z$-plane.
       The solid lines denote the branch cuts of the square root: $[A_1,B_1]$, $[A_2,B_2]$ and $[A_3,B_3]$.
       The broken lines are the branch cuts of $\log$ which lie on the second sheet of the square root. 
       The logarithmic branch cut starting from $B_1(=D_1)$ on the first sheet
       immediately goes to the second sheet and terminates at $B_1$ on the second sheet.       
        (Right) Eigenvalue distributions through the Newton method (blue dots) and our method (red dots) for 
         the stepwise three-cut solution at $N_1/N=N_2/N=N_3/N=1$, $n_1=n_2=1$, $\lambda=0.3$. We take $N=60$ in the Newton method.
           }
        \label{fig-endpt-analysis-three-cut}
\end{figure}

We perform the integral in (\ref{Migdal-CSg}) through the similar calculations to the stepwise two-cut solution (\ref{two-cut-CS})  and obtain the resolvent of the stepwise $l$-cut solution
\begin{align}
&v(Z)=\frac{1}{\pi i\lambda} \log{\left( \frac{f(Z)-\sqrt{f^2(Z)-4Z}}{2} \right)}
+\sum_{j=1}^{l-1} \frac{n_j}{\pi i\lambda} \log{\left( \frac{q^{(j)}(Z)+\sqrt{\left( q^{(j)}(Z) \right)^2-4}}{2} \right)}+\frac{n_0}{\lambda}, \nonumber \\
&f(Z)=f_0+f_1 Z, \qquad\quad
f_0=\frac{2\sqrt{A_1B_l}}{\sqrt{A_1}+\sqrt{B_l}},
\quad
f_1=\frac{2}{\sqrt{A_1}+\sqrt{B_l}}, \nonumber
\\
&q^{(j)}(Z)=\frac{q_1^{(j)}Z-q_0^{(j)}D_j}{Z-D_j},
\quad
q_1^{(j)}=\frac{2(2D_j-A_1-B_l)}{B_l-A_1},
\quad
q_0^{(j)}=\frac{2(D_jB_l+D_jA_1-2A_1B_l)}{D_j(B_l-A_1)}.
\label{multi-cut-CS}
\end{align}
Here we have defined  $D_j:=B_j$ in order to emphasize the distinction between $B_l$ and other $B_j$'s.
Again $q^{(j)}(Z)$ has a pole at $Z=D_j$.
This pole causes a logarithmic singularity and we take the branch cut as sketched in Figure \ref{fig-endpt-analysis-three-cut} so that the equation (\ref{eom-CS3}) is satisfied correctly.

Then we obtain the eigenvalue density 
\begin{align}
\rho(Z) = & \frac{1}{4\pi^2 \lambda Z} \log{\left( \frac{Z+\sqrt{A_1B_l}-i \sqrt{(Z-A_1)(Z-B_l)}}{Z+\sqrt{A_1B_l}+i \sqrt{(Z-A_1)(Z-B_l)}} \right)} + \sum_{j=1}^{l-1} \rho_s^{(j)}(Z), \nonumber \\
&
\rho_{s}^{(j)}(Z)= \frac{n_j}{\pi^2 \lambda Z}
\left \{
\begin{array}{l}
{\rm arctanh} \left( \sqrt{\dfrac{Z-A_1}{B_l-Z}} \sqrt{\dfrac{B_l-D_j}{D_j-A_1}} \right),
\quad
\left( Z \in \underset{(k=1, \cdots, j)}{[A_k,B_k]} \right),
\\
-{\rm arctanh} \left( \sqrt{\dfrac{B_l-Z}{Z-A_1}} \sqrt{\dfrac{D_j-A_1}{B_l-D_j}} \right),
\quad
\left( \underset{(k=j+1, \cdots, l-1)}{Z \in [A_k,B_k]} \right).
\end{array}
\right.
\label{rho-general}
\end{align}
Again it shows the logarithmic divergence at each $D_j$, ($j=1,\cdots, l-1$).
We sketch the profile in Figure \ref{fig-g-CS}.

Lastly we have to fix the $l+1$ constant $A_1$, $B_l$ and $D_j$ ($j=1,\cdots, l-1$).
These are determined through the two boundary conditions at $Z=0$ and $Z=\infty$ (\ref{boundary-CS}) and $l-1$ normalization condition
\begin{equation}
\frac{N_i}{N}=\int^{B_i}_{A_i}  \rho(Z) dZ
=\frac{1}{4\pi i} \oint_{C_i} \frac{v(Z)}{Z} dZ
, \qquad
(i=1, \cdots, l).
\label{normalization-CS}
\end{equation}
(One of the normalization condition is not independent of the other conditions.)
We can solve these equations numerically.
The result for $\{\lambda, n_1,n_2,N_1/N,N_2/N\}=\{0.3,1,1,1/3,1/3 \}$ is shown in Figure \ref{fig-endpt-analysis-three-cut}. (In this case, the solution has a symmetry $z \to -z$ which fixes $b_3=-a_1$, $b_2=-a_2$, $b_1=-a_3$.)
This agrees with the result obtained from the Newton method.

\subsection{Composition of the stepwise multi-cut solutions}

\begin{figure}
\begin{tabular}{cc}
\begin{minipage}{0.5\hsize}
\begin{center}
        \includegraphics[scale=0.25]{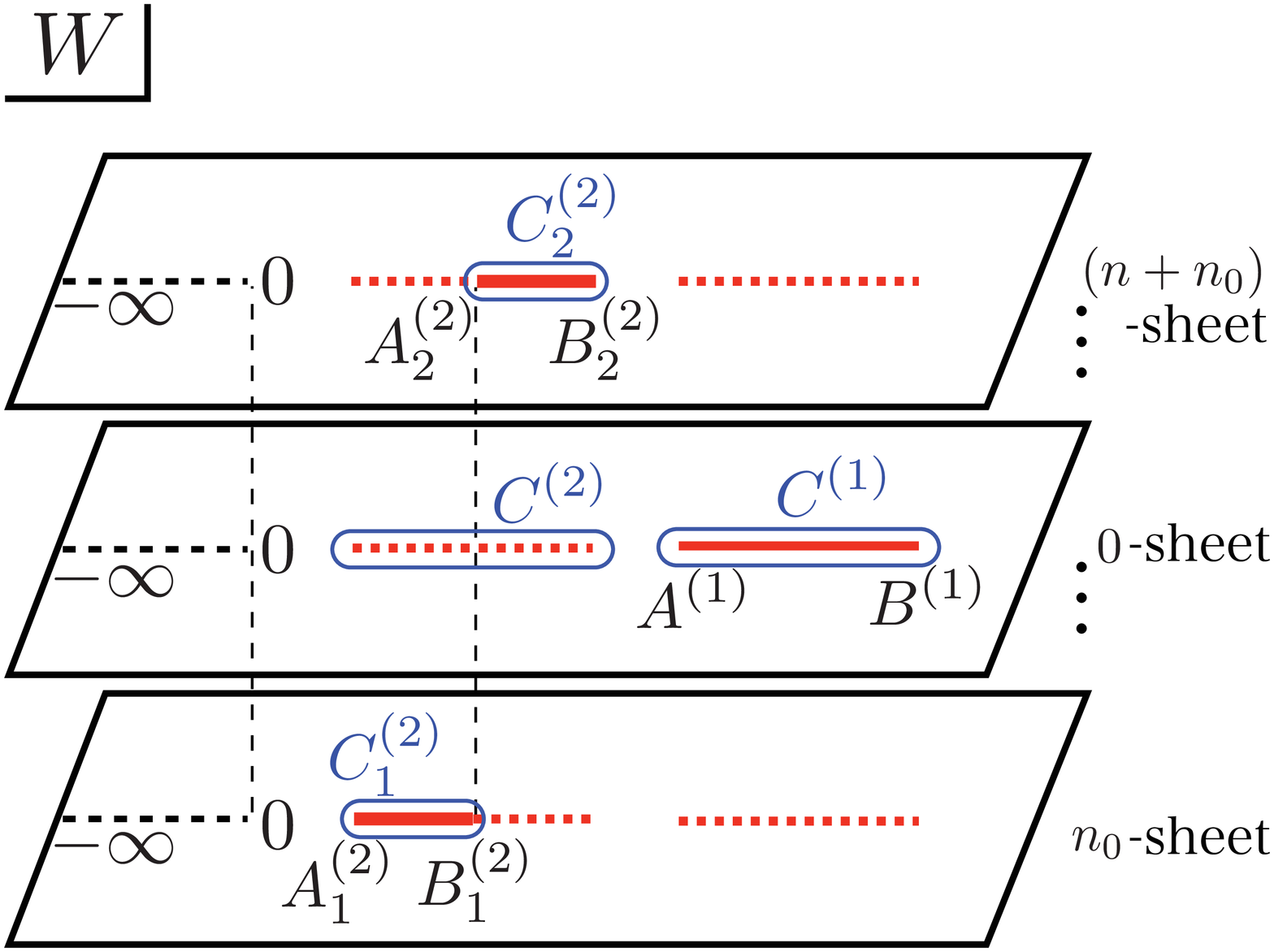}\\
\end{center}
\end{minipage}
\begin{minipage}{0.5\hsize}
\begin{center}
        \includegraphics[scale=0.25]{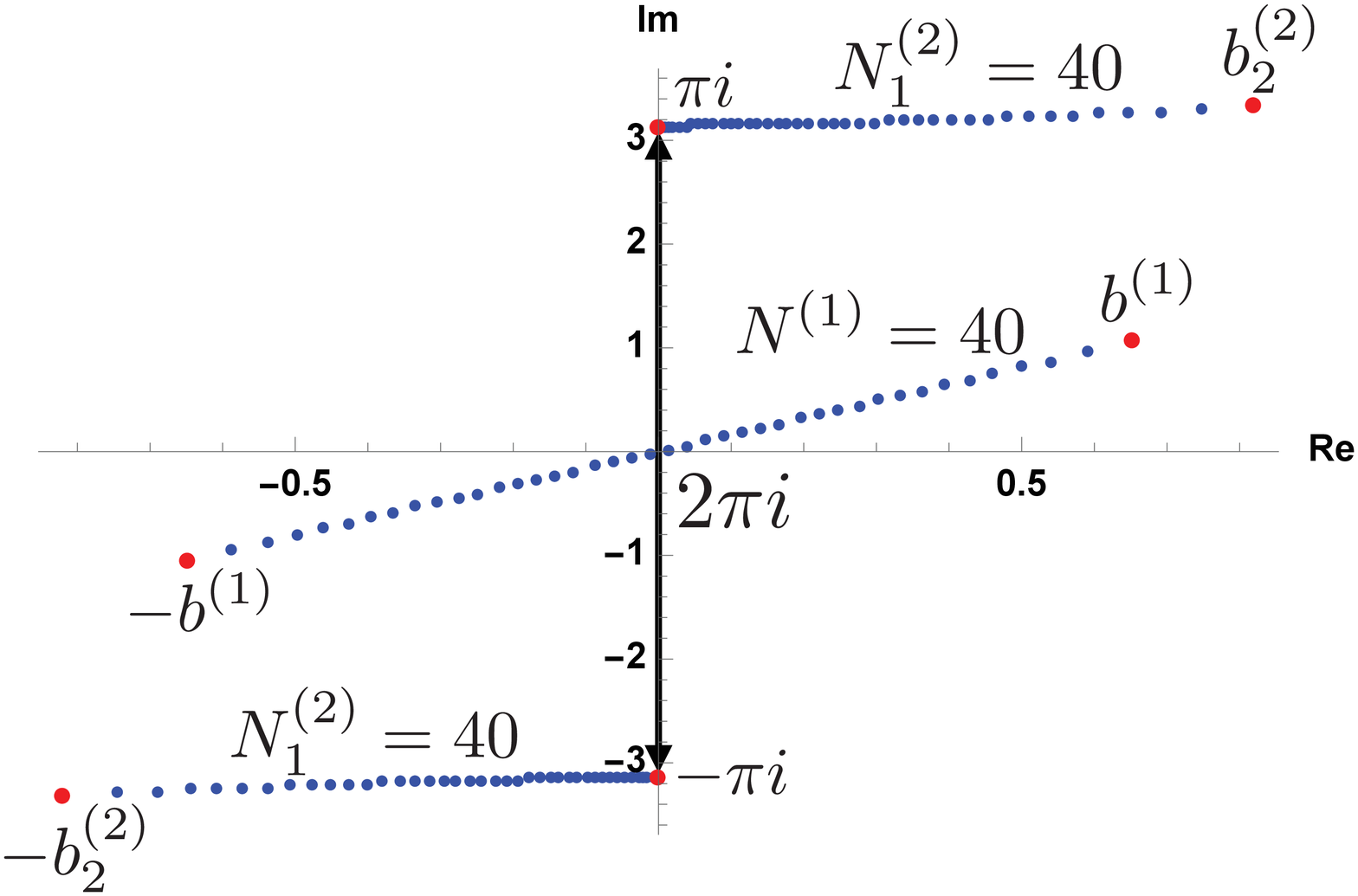}\\
\end{center}
\end{minipage}
\end{tabular}
       \caption{ (Left) Integral contours of the resolvent for the composite solution (one-cut + stepwise two-cut) in (\ref{1+2-CS}). 
        (Right) 
      Eigenvalue distributions through the Newton method (blue dots) and our method (red dots) for the  composite solution. We take $N^{(1)}/N=N_1^{(2)}/N=N_2^{(2)}/N=1/3$, $n=1$, $\lambda=0.2$.
$N=120$ is taken in the Newton method.           }
        \label{fig-1+2-cut}
\end{figure}

We explore the analytic solution for the last plot in Figure \ref{fig-numerical-CS} (bottom-right).
There, three cuts appear, and two of them are separated by $2\pi i$.
Thus they may be regarded as a composition of the one-cut solution (\ref{one-cut-CS})
 and the stepwise two-cut solution (\ref{two-cut-CS}).
 Hence we assume the three cuts as $[A^{(1)},B^{(1)}]$, $[A^{(2)}_1,B^{(2)}_1]$ and  
$[A^{(2)}_2,B^{(2)}_2]$.
Here $[A^{(1)},B^{(1)}]$ corresponds to the one-cut around the $z=0$ and $[A^{(2)}_i,B^{(2)}_i]$ ($i=1,2$) describe the stepwise two-cuts.
Hence we impose
\begin{align}
A^{(2)}_2=e^{2\pi in} B_1^{(2)},
\end{align}
where $n$ is a positive integer.
We also assume that the numbers of the eigenvalues on each cuts are $N^{(1)}$ and $N^{(2)}_i$ ($i=1,2$), respectively. 
Then the resolvent may be given as
\begin{equation}
v(Z)=\oint_{C^{(1)}  \cup \,  C^{(2)}_1  \cup \, C^{(2)}_2}  \frac{dW}{4\pi i} \frac{V'(W)}{Z-W} \sqrt{\frac{(Z-A^{(1)})(Z-B^{(1)})(Z-A^{(2)}_1)(Z-B^{(2)}_2)}{(W-A^{(1)})(W-B^{(1)})(W-A^{(2)}_1)(W-B^{(2)}_2)}},
\label{1+2-CS}
\end{equation}
where the contour $C^{(1)}$ encircles the cut $[A^{(1)},B^{(1)}]$ and $C_i^{(2)}$ encircles the cut $[A^{(2)}_i,B^{(2)}_i]$ ($i=1,2$). See Figure \ref{fig-1+2-cut}.
Note that we have the 5 constants:  $A^{(1)},B^{(1)},A^{(2)}_1,B^{(2)}_1$ and $B^{(2)}_2$, and these constants can be fixed by the 3 normalization conditions similar to (\ref{normalization-CS}) and the 3 boundary conditions (\ref{boundary-CS})\footnote{The resolvent (\ref{1+2-CS}) behaves as $v(Z) \to c_1 Z+ c_0 +O(1/Z) $, $(Z \to \infty)$ and the boundary condition (\ref{boundary-CS}) requires the two conditions: $c_1=0$ and $c_0=1$.}. (There are 6 conditions but only 5 of them are independent).
Therefore the consistent solution would exist.

%%%%%%%%%%%%%%%%%%%%%%%%%%%%%%%%%%%%
\paragraph{Symmetric solution}
%%%%%%%%%%%%%%%%%%%%%%%%%%%%%%%%%%%%
Performing the integral of the resolvent (\ref{1+2-CS}) is generally difficult.
However, if the solution is symmetric under $z \to -z$ ($Z \to 1/Z$), we can compute it as follows.
This symmetry requires the following conditions on the ansatz,
\begin{align}
A^{(1)}=1/B^{(1)}, \qquad  A_1^{(2)}=1/B_2^{(2)}, \qquad
A_2^{(2)}=1/B_1^{(2)}=e^{\pi in}, \qquad N_1^{(2)}= N_2^{(2)}.
\end{align}
Besides, the cut $[1/B^{(1)},B^{(1)}]$ should pass $Z=1$, and it demands $n$ to be odd so that the other two cuts do not hit this cut.
In this case, the cut $[e^{\pi in}, B_2^{(2)}]$ and $[ 1/B_2^{(2)}, e^{-\pi in}]$ are on the $(n+1)/2$-th sheet and $-(n+1)/2$-th sheet\footnote{This is because the cuts would tilt as we can see from the numerical result.
See footnote \ref{ftnt-tilt} also.}, respectively, and the integral  (\ref{1+2-CS}) can be written as
\begin{align}
v(Z)=& \oint_{C^{(1)}  \cup \,  C^{(2)}}  \frac{dW}{4\pi i} \frac{1}{\pi i\lambda} \frac{\log{W}}{Z-W} \sqrt{\frac{(Z-1/B^{(1)})(Z-B^{(1)})(Z-1/B^{(2)}_2)(Z-B^{(2)}_2)}{(W-1/B^{(1)})(W-B^{(1)})(W-1/B^{(2)}_2)(W-B^{(2)}_2)}}
\nonumber
\\
&+\oint_{C^{(2)}_1}  \frac{dW}{4\pi i} \frac{-(n+1)}{\lambda} \frac{1}{Z-W} \sqrt{\frac{(Z-1/B^{(1)})(Z-B^{(1)})(Z-1/B^{(2)}_2)(Z-B^{(2)}_2)}{(W-1/B^{(1)})(W-B^{(1)})(W-1/B^{(2)}_2)(W-B^{(2)}_2)}}
\nonumber
\\
&+\oint_{C^{(2)}_2}  \frac{dW}{4\pi i} \frac{n+1}{\lambda} \frac{1}{Z-W} \sqrt{\frac{(Z-1/B^{(1)})(Z-B^{(1)})(Z-1/B^{(2)}_2)(Z-B^{(2)}_2)}{(W-1/B^{(1)})(W-B^{(1)})(W-1/B^{(2)}_2)(W-B^{(2)}_2)}}.
\label{v-1+2-int}
\end{align}
We can compute this integral by using the technique developed in Ref.~\cite{Suyama:2010hr} and obtain
\begin{align}
v(Z)=&\frac{1}{2\pi i\lambda} \log{\left( \frac{f(Z)-\sqrt{f^2(Z)-4Z^2}}{2} \right)}-\frac{n}{2\pi i\lambda} \log{\left( \frac{q(Z)+\sqrt{q^2(Z)-4}}{2} \right)},
\nonumber
\\
&f(Z)=f_0+f_1Z+f_0Z^2,
\qquad
q(Z)=\frac{q_0+q_1Z+q_0Z^2}{(Z+1)^2},
\label{v-1+2}
\end{align}
where the constants are given as
\begin{align}
&f_0=\frac{4}{c^{(1)}-c^{(2)} },
\quad
f_1=-2 \frac{c^{(1)}+c^{(2)} }{c^{(1)}-c^{(2)}}, \quad
c^{(1)}=B^{(1)}+1/B^{(1)}, \quad c^{(2)}=B_2^{(2)}+1/B_2^{(2)},
\nonumber
\\
&q_0=2\frac{c^{(1)}+c^{(2)}+4}{c^{(1)}-c^{(2)}},
\quad
q_1=-\frac{4\left( c^{(1)}+c^{(2)} \right) +4 c^{(1)} c^{(2)} }{c^{(1)}-c^{(2)}}.
\end{align}
Remarkably, the first term of (\ref{v-1+2}) is similar to the resolvent of the $S^3/{\mathbf Z}^2$ Lens space matrix model \cite{Aganagic:2002wv, Halmagyi:2003ze} which is related to the ABJM matrix model (\ref{(1+1)-ABJM}).
The second term provides the logarithmic divergence at $Z= e^{\pm \pi i n}$ akin to the previous solutions (\ref{two-cut-CS}) and (\ref{multi-cut-CS}).
The appearance of the resolvent of the Lens space matrix model indicates that some geometrical interpretations of our multi-cut solutions might be possible.
We will consider it in a future research.

By numerically solving $B^{(1)}$ and $B^{(2)}_2$, we compare our solution with the one obtained via the Newton method.
We can see a good agreement as shown in Figure \ref{fig-1+2-cut}.
Note that we attempt to solve the equations (\ref{boundary-CS}) and the normalization condition like (\ref{normalization-CS}) directly by Mathematica and obtain a consistent result.
(Here we use FindRoot and NIntegral in (\ref{1+2-CS}).)
This is a good news.
Although it would be difficult to perform the integral such as (\ref{1+2-CS}) and obtain analytic expressions in general, this result indicates that we do not need the analytic expressions in order to evaluate the physical quantities.

%%%%%%%%%%%%%%%%%%%%%%%%%%%%%%%%
\subsection{Proposal for general solution in the pure CS matrix model}
\label{sec-CS-general}
%%%%%%%%%%%%%%%%%%%%%%%%%%%%%%%%

The generalization of the composite solution in the previous section is straightforward.
We can consider $p$-stepwise $l_q$-cuts: $[A^{(q)}_j,B^{(q)}_j]$ ($q=1,\cdots,p $ and $j=1,\cdots,l_q $)  satisfying
\begin{align}
A^{(q)}_j=e^{2\pi in^{(q)}_j} B_{j+1}^{(q)},
\end{align}
where $n^{(q)}_j$ are positive integers.
We also assign the numbers of the eigenvalues on the cut   $[A^{(q)}_j,B^{(q)}_j]$ as $N^{(q)}_j$.
Then the resolvent may be given by
\begin{equation}
v(Z)= \sum_{ r=1}^p\sum_{ j=1}^{l_r} \oint_{C^{(r)}_j}   \frac{dW}{4\pi i} \frac{V'(W)}{Z-W} \prod_{q=1}^{p} \sqrt{\frac{(Z-A^{(q)}_1)(Z-B^{(q)}_{l_q})}{(W-A^{(q)}_1)(W-B^{(q)}_{l_q})}},
\label{composite-CS}
\end{equation}
where the contour $C^{(r)}_j$ encircles the cut $[A^{(r)}_j,B^{(r)}_j]$ ($r=1,\cdots,p$ and $j=1,\cdots,l_r$).
This integral may be performed by using the genus $p-1$ generalizations of elliptic functions.
In this expression, we have the $p+\sum_{q=1}^{p} l_q $ undetermined constant $ \{ A^{(q)}_1, B^{(q)}_1, \cdots,  B^{(q)}_{l_q} \}  $, and these will be fixed by the $\sum_{q=1}^{p} l_q$ normalization condition (\ref{normalization-CS}) and $p$ boundary condition at $Z=\infty$ and one boundary condition at $Z=0$ (\ref{boundary-CS}).
(Again one of these conditions is not independent.)
As an example, we derive the solution shown in Figure \ref{fig-endpt-analysis-4cut} in Appendix \ref{app-negative-n} by using this ansatz.

In this way, our resolvent (\ref{composite-CS}) may describe all the solutions in Figure \ref{fig-numerical-CS} obtained through the Newton method.
Then one important question is whether any other solutions of the saddle point equation (\ref{eom-CS}) exist or not.
We explore the solutions through the Newton method, and it seems that all the solutions might be explained by our resolvent (\ref{composite-CS}).
Although it is hard to exclude the possibility of the existence of the other solutions, we presume that our solution (\ref{composite-CS}) may be the general solution of the saddle point equation of the pure CS matrix model.

Our method would be applicable to other CS matrix models.
As a demonstration, we consider the ABJM matrix model in the next section.

%%%%%%%%%%%%%%%%%%%%%%%%%%%%%%%%%%%%%%%%%%%%%%%%%%%%%%%%%
\section{Multi-cut solutions in the ABJM matrix model}
\label{sec-ABJM}
%%%%%%%%%%%%%%%%%%%%%%%%%%%%%%%%%%%%%%%%%%%%%%%%%%%%%%%%%

We will apply the technique for finding the multi-cut solutions developed in the previous section to the ABJM matrix model \cite{Kapustin:2009kz}.
The partition function of this model is given by
\begin{equation}
Z (k,N)
= \frac{1}{(N!)^2} \int  \prod_{i=1}^N \frac{d \mu_i}{2\pi}
e^{-\frac{N}{4\pi i \lambda}  \mu_i^2 }
 \prod_{j=1}^N \frac{d \nu_j}{2\pi}
e^{\frac{N}{4\pi i \lambda}  \nu_j^2 }
   \frac{\prod_{i<j}^{N} \left[ 2\sinh{\frac{\mu_i-\mu_j}{2}} \right]^2 \prod_{i<j}^{N} \left[ 2\sinh{\frac{\nu_i-\nu_j}{2}} \right]^2 }{\prod_{i,j=1}^{N} \left[ 2\cosh{\frac{\mu_i-\nu_j}{2}} \right]^2}.
\label{partition-ABJM}
\end{equation}
Here $k$ is the CS level and $\lambda := N/k$.
The saddle point equations of this model are 
\begin{eqnarray}
\mu_i &=& \frac{2\pi i \lambda}{N} \left[ \sum_{j \neq i}^{N} \coth{\frac{\mu_i-\mu_j}{2}}- \sum_{j=1}^{N} \tanh{\frac{\mu_i-\nu_j}{2}} \right],
\qquad
(i=1, \cdots, N), \nonumber
\\
-\nu_i &=& \frac{2 \pi i \lambda}{N} \left[ \sum_{j\neq i}^{N} \coth{\frac{\nu_i-\nu_j}{2}}-\sum_{j=1}^{N} \tanh{\frac{\nu_i-\mu_j}{2}} \right] ,
\qquad
(i=1, \cdots, N).
\label{eom-ABJM}
\end{eqnarray}

\begin{figure}[t]
\begin{tabular}{cc}
\begin{minipage}{0.5\hsize}
\begin{center}
        \includegraphics[scale=0.2]{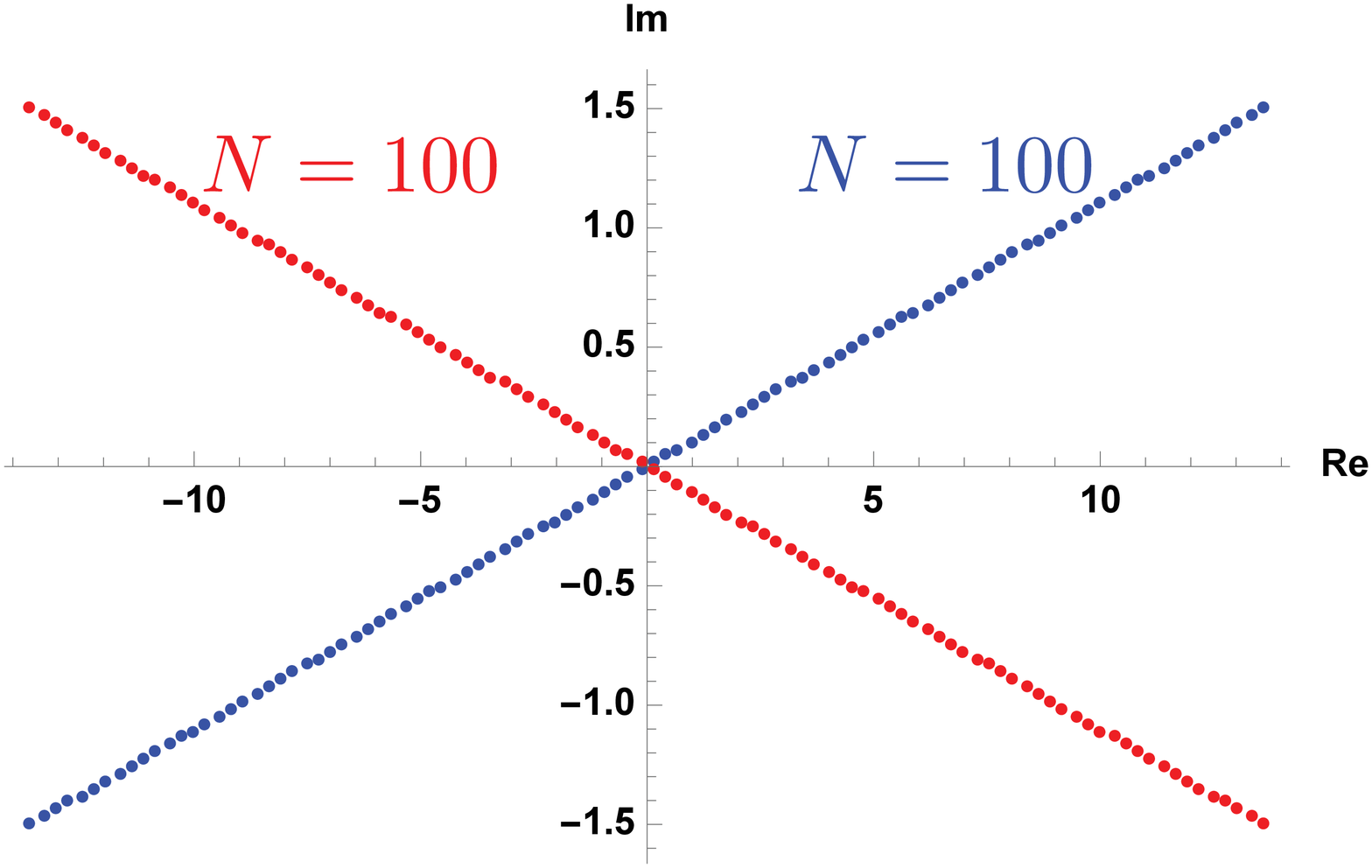}\\
    DMP solution
\end{center}
\end{minipage}
\begin{minipage}{0.5\hsize}
\begin{center}
        \includegraphics[scale=0.2]{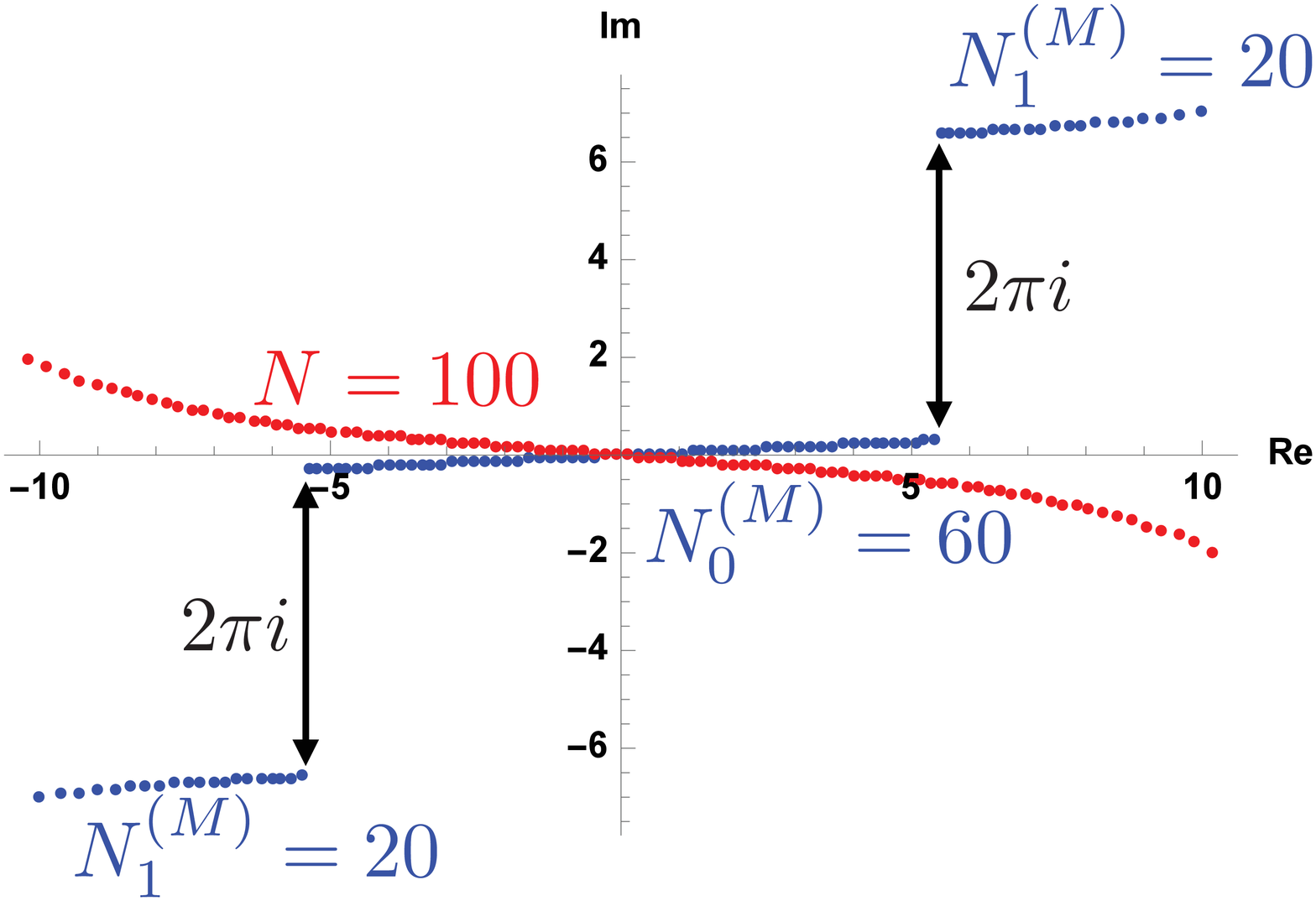}\\
  (Stepwise three + one)-cut solution
\end{center}
\end{minipage}\\
\begin{minipage}{0.5\hsize}
\begin{center}
        \includegraphics[scale=0.2]{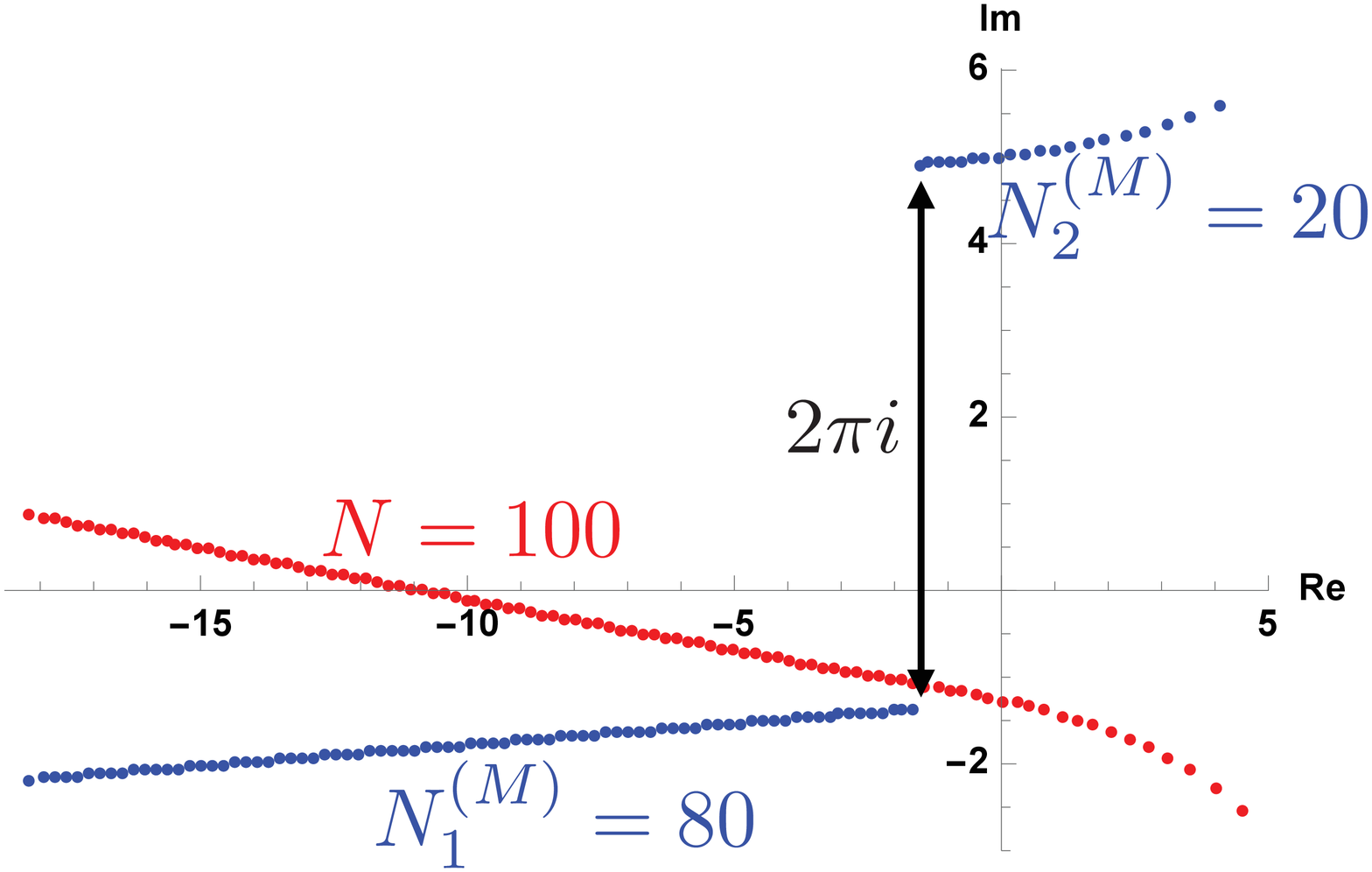}\\
   (Stepwise two+one)-cut solution
\end{center}
\end{minipage}
\begin{minipage}{0.5\hsize}
\begin{center}
        \includegraphics[scale=0.2]{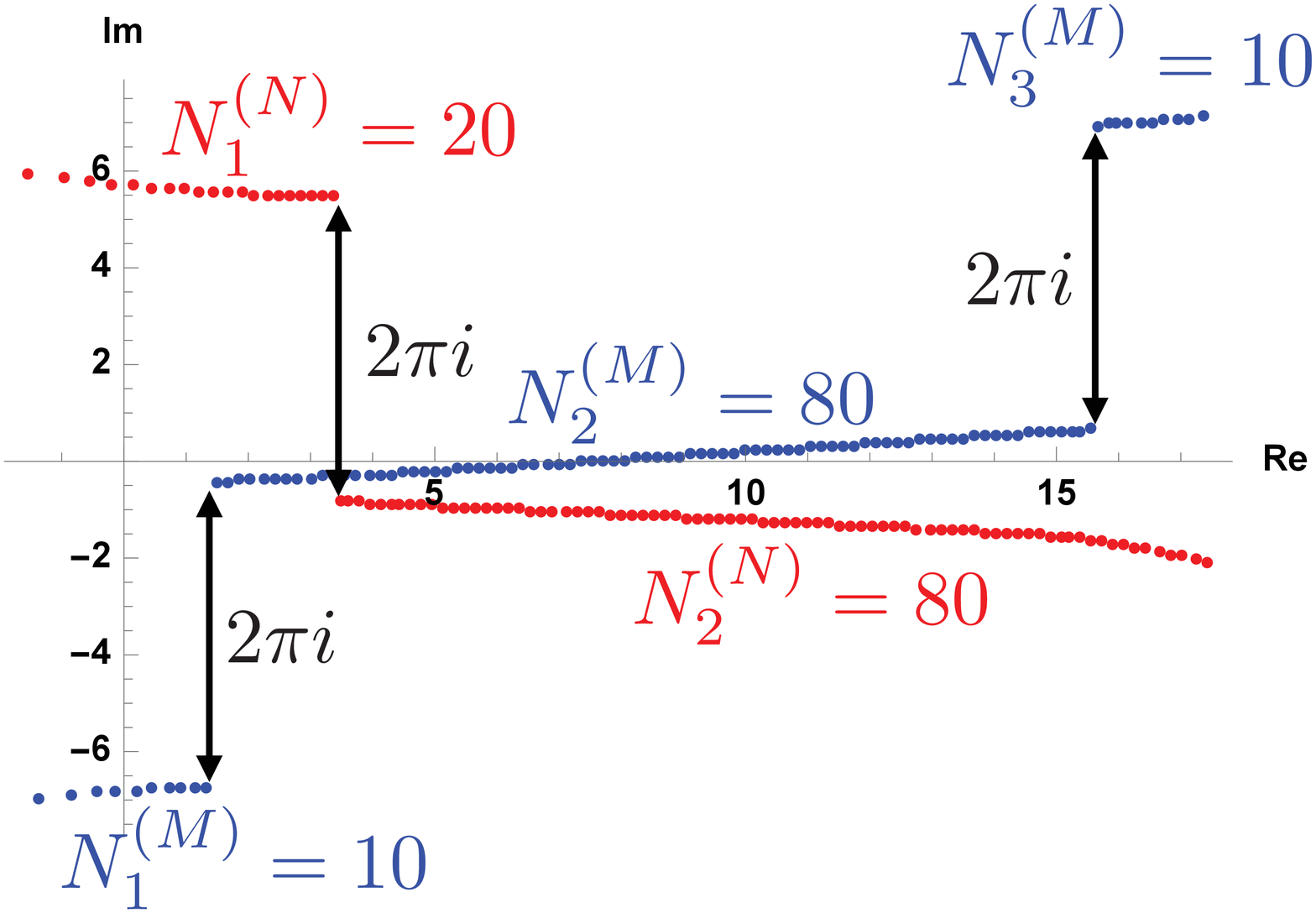}\\
        (Stepwise three+stepwise two)-cut solution
\end{center}
\end{minipage}
\end{tabular}
       \caption{Eigenvalue distributions of the numerical solutions of the saddle point equation (\ref{eom-ABJM}) in the ABJM matrix model. We take $N=100$ and $\lambda=10$.
  The blue and red dots denote the eigenvalues of  $\mu$ and $\nu$, respectively.
  The top-left plot corresponds to the DMP solution \cite{Drukker:2010nc}, and other various multi-cut solutions exist in this model. }
        \label{fig-numerical-ABJM}
\end{figure}

We explore the solutions of these equations in the 't Hooft limit $N \rightarrow \infty$ at finite $ \lambda$.
Again the resolvent is a convenient tool for solving these equations.
We define new variable $M_i := \exp{(\mu_i)}$ and $N_j := \exp{( \nu_j )}$, and rewrite the saddle point equations (\ref{eom-ABJM}) as
\begin{eqnarray}
\log{M_i} &=& \frac{2\pi i \lambda}{N} \left[ \sum_{j \neq i}^{N} \frac{M_i+M_j}{M_i-M_j}- \sum_{j=1}^{N} \frac{M_i-N_j}{M_i+N_j} \right],
\qquad
(i=1, \cdots, N), \nonumber
\\
-\log{N_i} &=& \frac{2 \pi i \lambda}{N} \left[ \sum_{j\neq i}^{N} \frac{N_i+N_j}{N_i-N_j}-\sum_{j=1}^{N} \frac{N_i-M_j}{N_i+M_j} \right] ,
\qquad
(i=1, \cdots, N).
\label{eom-ABJM2}
\end{eqnarray}
We introduce the eigenvalue densities of $M_i$ and $N_j$ and the resolvent as 
\begin{eqnarray}
&{}&
\rho_M(Z) := \frac{1}{N} \sum_{i=1}^N \delta (Z-M_i),
\qquad
\rho_N(Z) := \frac{1}{N} \sum_{i=1}^N \delta (Z-N_i),
\nonumber
\\
&{}&
w(Z) := \int_{\mathcal{C}_M}  \rho_M (W) \frac{Z+W}{Z-W} dW-\int_{\mathcal{C}_N} \rho_N(W) \frac{Z-W}{Z+W} dW,
\label{resolvent-ABJM}
\end{eqnarray}
where $\mathcal{C}_{M}$ and  $\mathcal{C}_{N}$ are the supports of $\rho_{M}(Z)$ and $\rho_{N}(Z)$, respectively.
Then the saddle point equations (\ref{eom-ABJM2}) become
\begin{eqnarray}
\frac{1}{\pi i\lambda} \log{Z}
&=&
\lim_{\epsilon \rightarrow 0} \left[ w(Z+i\epsilon)+w(Z-i\epsilon) \right],
\qquad
(Z \in \mathcal{C}_M),
\nonumber
\\
\frac{1}{\pi i\lambda} \log{Z}
&=&
\lim_{\epsilon \rightarrow 0} \left[ w(-Z+i\epsilon)+w(-Z-i\epsilon) \right],
\qquad
(Z \in \mathcal{C}_N),
\label{eom-ABJM3}
\end{eqnarray}
and the eigenvalue densities are described by 
\begin{eqnarray}
\rho_M(Z) &=& -\frac{1}{4\pi iZ} \lim_{\epsilon \rightarrow 0} \left[ w(Z+i \epsilon)-w(Z-i \epsilon) \right],
\qquad
(Z \in \mathcal{C}_M),
\nonumber
\\
\rho_N(Z) &=& +\frac{1}{4\pi iZ} \lim_{\epsilon \rightarrow 0} \left[ w(-Z+i \epsilon)-w(-Z-i \epsilon) \right],
\qquad
(Z \in \mathcal{C}_N).
\label{rho-ABJM}
\end{eqnarray}
Besides, the resolvent satisfies the boundary conditions
\begin{equation}
\lim_{Z \rightarrow \infty} w(Z)=0,
\qquad
\lim_{Z \rightarrow 0} w(Z)=0.
\label{boundary-ABJM}
\end{equation}
Therefore what we should do is finding the resolvent which satisfies the saddle point equations (\ref{eom-ABJM3}) and the boundary conditions (\ref{boundary-ABJM}).

Before considering the analytic solution, we attempt the numerical computations via the Newton method
in order to gain some insight.
Some of the obtained results are shown  in Figure \ref{fig-numerical-ABJM}.
The top-left panel corresponds to the well-known solution obtained by Drukker, Marino and Putrov \cite{Drukker:2010nc}.
We call this solution ``DMP" solution.
The top-right panel corresponds to the solution found in our previous study \cite{Morita:2017oev}.
  In addition, various multi-cut solutions exist.
These results indicate that the dynamics of the ABJM matrix model is similar to the pure CS matrix model.
While the eigenvalues tend to be around $z=0$,
 the strong interactions arise when the eigenvalues are separated by $2\pi i $, and they may cause   various solutions\footnote{Note that strong interactions work between $\mu_i$ and $\nu_j$ too in the saddle point equations (\ref{eom-ABJM}), if they are separated by $(2n+1) \pi i$.
However we could not find numerical solution in which $\mu_i$ and $\nu_j$ are separated by   $(2n+1) \pi i$. 
Since the sign of interactions (\ref{eom-ABJM}) between $\mu_i$ and $\nu_j $ in this case are opposite to those  between $\mu_i$ and $\mu_j$, we presume that the forces cannot balance and the solution could not exist. (It would be important to clarify this point rigorously.)
For this reason, we do not consider the analytic solutions for these configurations in this article.
}.
Therefore the technique in the pure CS matrix model would be useful in the ABJM matrix model too.

\subsection{Derivation of the DMP solution}
\begin{figure}
\begin{tabular}{cc}
\begin{minipage}{0.5\hsize}
\begin{center}
        \includegraphics[scale=0.25]{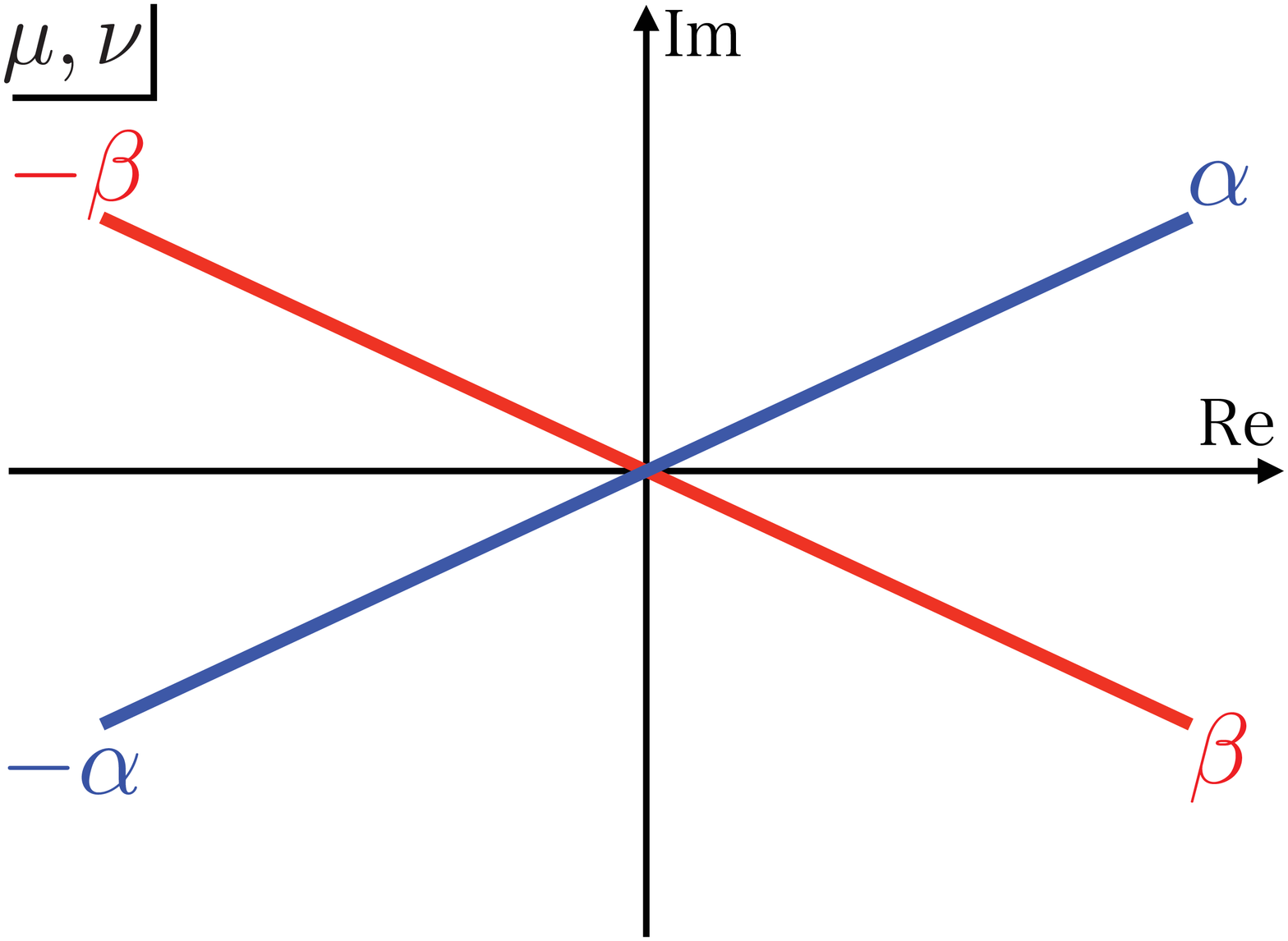}
        \end{center}
\end{minipage}
\begin{minipage}{0.5\hsize}
\begin{center}
        \includegraphics[scale=0.25]{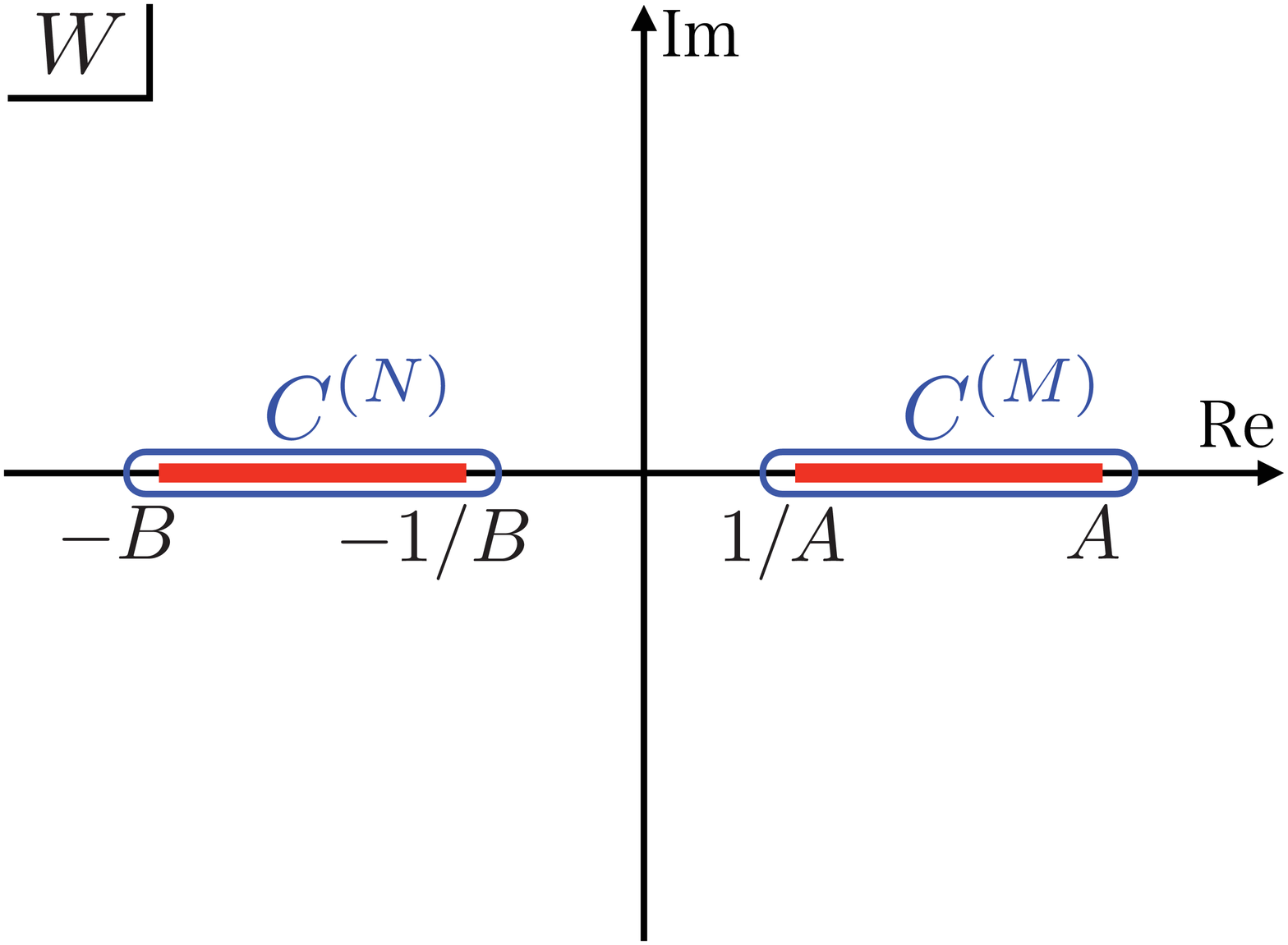}
 \end{center}
\end{minipage}
\end{tabular}
       \caption{(Left) Sketch of the cuts of the DMP solution. Here $\alpha=\log{A}$ and $\beta=\log{B}$. (Right) Integral contours of (\ref{Migdal-ABJM}).
       The contour $C^{(N)}$ encircles $[-B,-1/B]$ rather than $[1/B,B]$.
       }
        \label{fig-image-DMP}
\end{figure}

Before considering the multi-cut solutions, we first review the derivation of the DMP solution (Figure \ref{fig-numerical-ABJM} top-left) by using the technique in the previous section \cite{Suyama:2010hr}.
We assume the cut $[1/A,A]$ for $M_i$ and $[1/B,B]$ for $N_i$. (Here these cuts respect the symmetry $Z \to 1/Z$.)
We also assume $|A|,|B| \geq 1$.
See Figure \ref{fig-image-DMP}.
Then the resolvent which satisfies the saddle point equations (\ref{eom-ABJM3}) is given as
\begin{align}
w(Z)=&\oint_{C^{(M)}} \frac{dW}{4\pi i} \frac{V_M'(W)}{Z-W} \sqrt{\frac{(Z-A)(Z-1/A)(Z+1/B)(Z+B)}{(W-A)(W-1/A)(W+1/B)(W+B)}}
\nonumber
\\
&+\oint_{C^{(N)}} \frac{dW}{4\pi i} \frac{V_N'(W)}{Z-W} \sqrt{\frac{(Z-A)(Z-1/A)(Z+1/B)(Z+B)}{(W-A)(W-1/A)(W+1/B)(W+B)}}, \nonumber \\
&V_M'(Z):= \frac{1}{\pi i\lambda}\log{Z}, \qquad
V_N'(Z):= \frac{1}{\pi i\lambda}\log{\left( e^{\pi i}Z \right)}.
\label{Migdal-ABJM} 
\end{align}
Note that the cuts of this resolvent are on $[1/A,A]$ and $[-B,-1/B]$ rather than $[1/B,B]$.
This is because the saddle point equations (\ref{eom-ABJM2}) are singular when $M_i=-N_j$.  
Correspondingly the contour $C^{(M)}$ and $C^{(N)}$ encircle the cut $[1/A,A]$ and $[-B,-1/B]$, respectively.

We can perform this integral and obtain \cite{Suyama:2010hr}
\begin{align}
w(Z)=&\frac{1}{2\pi i\lambda} \log{\left( \frac{f(Z)-\sqrt{f^2(Z)-4Z^2}}{2} \right)},
\qquad
f(Z)=f_0+f_1Z+f_0Z^2,
\nonumber
\\
&f_0=\frac{4}{A+1/A+B+1/B},
\quad
f_1=\frac{2 \left( -A-1/A+B+1/B \right)}{A+1/A+B+1/B}.
\label{(1+1)-ABJM}
\end{align}
The parameter $A$ and $B$ are determined through the boundary conditions (\ref{boundary-ABJM}) and the normalization condition
\begin{equation}
1=\int_{1/A}^A \rho_M(Z) dZ=\frac{1}{4\pi i} \oint_{C^{(M)}} \frac{v(Z)}{Z} dZ.
\label{normalization-ABJM}
\end{equation}
Then we obtain the relations \cite{Drukker:2010nc}
\begin{equation}
A+\frac{1}{A}=2+i\kappa,
\qquad
B+\frac{1}{B}=2-i\kappa.
\label{endpt-ABJM}
\end{equation}
Here $\kappa$ is related to the 't~Hooft coupling $\lambda$ through
\begin{equation}
\lambda(\kappa) = \frac{\kappa}{8 \pi} {}_3F_2 \left( \frac12, \frac12, \frac12; 1, \frac32; - \frac{\kappa^2}{16} \right).
\label{kappa}
\end{equation}
Particularly, at the strong coupling $|\lambda| \gg 1$, we obtain
\begin{eqnarray}
&{}&A=e^{\alpha},
\qquad
\alpha=\pi  \sqrt{2 \hat{\lambda} }+ \frac{\pi}{2} i -2i e^{-\pi \sqrt{2 \hat{\lambda}}}+\cdots,
\nonumber
\\
&{}&B=e^{\beta},
\qquad
\beta=\pi  \sqrt{2 \hat{\lambda} }- \frac{\pi}{2} i +2i e^{-\pi \sqrt{2 \hat{\lambda}}}+\cdots,
\label{endpt-ABJM'}
\end{eqnarray}
where $\hat{\lambda} := \lambda-\frac{1}{24}$.

%%%%%%%%%%%%%%%%%%%%%%%%%%%%%%%%
\subsection{Proposal for general solution in the ABJM matrix model}
\label{sec-general-ABJM}
%%%%%%%%%%%%%%%%%%%%%%%%%%%%%%%%

We will apply the technique developed in the pure CS matrix model to the ABJM matrix model, and propose the general solution.
As the numerical computations shown in Figure \ref{fig-numerical-ABJM} suggest, there are various multi-cut solutions in which the eigenvalues of the same matrix are separated by $2\pi i$.
Thus each matrix can compose the stepwise multi-cuts. 
In addition, a composition of these stepwise multi-cuts would be a solution too as in the pure CS matrix model case. 
(Indeed we find these complicated solutions numerically, although we omit to show them in this article.)

By regarding these numerical results, we consider the following ansatz.
Suppose the eigenvalue $\{ M_i \}$ compose $p$ stepwise $l_r$-cuts ($r=1,\cdots,p $) and the eigenvalue 
$\{ N_i \}$ compose $q$ stepwise $m_r$-cuts ($r=1,\cdots,q $), and we define that the cut $[A^{(M,r)}_j,B^{(M,r)}_j]$ for  $\{ M_j \}$ and $[A^{(N,r)}_j,B^{(N,r)}_j]$ for  $\{ N_j \}$.
We assume that these cuts satisfy
\begin{align}
A^{(M,r)}_j=&e^{2\pi in^{(M,r)}_j} B_{j+1}^{(M,r)}, \qquad (j=1,\cdots,l_r,  \quad r=1,\cdots,p) ,\nonumber \\
A^{(N,r)}_j=&e^{-2\pi in^{(N,r)}_j} B_{j+1}^{(N,r)}, \qquad (j=1,\cdots, m_r,  \quad r=1,\cdots,q) ,
\end{align}
where $n^{(M,r)}_j$ and $n^{(N,r)}_j$ are positive integers.
We also assign the numbers of the eigenvalues on the cut $[A^{(M,r)}_j,B^{(M,r)}_j]$ and $[A^{(N,r)}_j,B^{(N,r)}_j]$ as $N^{(M,r)}_j$ and $N^{(N,r)}_j$, respectively. 
Then the resolvent may be given as
\begin{align}
w(Z)=&\sum_{ t=1}^p \sum_{ j=1}^{l_t} \oint_{C^{(M,t)}_j} \frac{dW}{4\pi i} \frac{V_M'(W)}{Z-W}
\prod_{r=1}^{p} \prod_{s=1}^{q}   \sqrt{\frac{(Z-A^{(M,r)}_1)(Z-B^{(M,r)}_{l_r})(Z+A^{(N,s)}_1)(Z+B^{(N,s)}_{m_s})}{(W-A^{(M,r)}_1)(W-B^{(M,r)}_{l_r})(W+A^{(N,s)}_1)(W+B^{(N,s)}_{m_s})}}
\nonumber
\\
&+\sum_{ t=1}^q \sum_{ j=1}^{m_t} \oint_{C^{(N,t)}_j} \frac{dW}{4\pi i} \frac{V_N'(W)}{Z-W}
\prod_{r=1}^{p} \prod_{s=1}^{q}   \sqrt{\frac{(Z-A^{(M,r)}_1)(Z-B^{(M,r)}_{l_r})(Z+A^{(N,s)}_1)(Z+B^{(N,s)}_{m_s})}{(W-A^{(M,r)}_1)(W-B^{(M,r)}_{l_r})(W+A^{(N,s)}_1)(W+B^{(N,s)}_{m_s})}}
, 
\label{general-ABJM}
\end{align}
where the contour $C^{(M,r)}_i$ and $C^{(N,s)}_j$ encircle the cut $[A^{(M,r)}_i,B^{(M,r)}_i]$ ($i=1,\cdots,l_r$ and $r=1,\cdots,p$) and $[-B^{(N,s)}_j,-A^{(N,s)}_j]$ ($j=1,\cdots,m_s$ and $s=1,\cdots,q$), respectively.
The end points of the cuts may be determined through the boundary conditions (\ref{boundary-ABJM}) 
and the normalization conditions akin to (\ref{normalization-ABJM}).

%%%%%%%%%%%%%%%%%%%%%%%%%%%%%%%%%%
\subsection{Symmetric stepwise multi-cut solution}
%%%%%%%%%%%%%%%%%%%%%%%%%%%%%%%%%%

\begin{figure}
\begin{center}
        \includegraphics[scale=0.25]{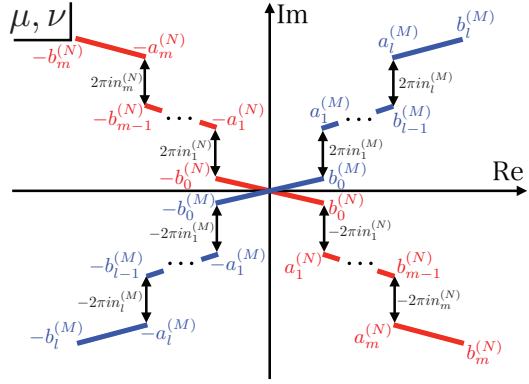}
\end{center}
\caption{Eigenvalue distribution of the symmetric stepwise $((2l+1)+(2m+1))$-cut solution (\ref{general-sym-ABJM}).}
        \label{fig-ABJMg}
\end{figure}
\par

Although the general solution (\ref{general-ABJM}) looks very complicated, if $p=q=1$ and the solution is symmetric under $Z \to 1/Z$, we will obtain a simple expression.
To see it, we consider a stepwise $2l+1$-cuts of $\mu_i$ and  stepwise $2m+1$-cuts of $\nu_i$ configuration as sketched in Figure \ref{fig-ABJMg}.
As we will see soon, the result depends on whether the number of each cut is odd or even, and we consider the both odd case first.
We assume that $\{ \mu_i \}$ are distributed between $[-b^{(M)}_0, b^{(M)}_0 ]$, $[a^{(M)}_j, b^{(M)}_j ]$ and  $[ -b^{(M)}_j, -a^{(M)}_j ]$, $(j=1, \cdots,l )$ and the number of the eigenvalues on each interval is $N^{(M)}_0$, $N^{(M)}_j$ and $N^{(M)}_j$, respectively,
so that the system is symmetric under $Z \to 1/Z$. 
Here $N^{(M)}_0+2 \sum_{j=1}^{l}N_j^{(M)}=N$ is imposed.
Similarly, for $\{\nu_i \}$, we take $[-b^{(N)}_0, b^{(N)}_0 ]$, $[a^{(N)}_j, b^{(N)}_j ]$ and  $[ -b^{(N)}_j, -a^{(N)}_j ]$, $(j=1, \cdots,m )$ and $N^{(N)}_j$ which satisfies $N^{(N)}_0+2 \sum_{j=1}^{m}N_j^{(N)}=N$.
Through the stepwise assumption, we impose the condition
\begin{align}
a_{j}^{(M)}&=b_{j-1}^{(M)}+ 2\pi i n_{j}^{(M)}, \qquad (j=1, \cdots, l ) ,\nonumber \\
a_{j}^{(N)}&=b_{j-1}^{(N)}- 2\pi i n_{j}^{(N)}, \qquad (j=1, \cdots, m ) ,
\end{align}
where $\{  n_{j}^{(M)} \}$ and $\{  n_{j}^{(N)} \}$ are positive integers.
On this set up, the resolvent (\ref{general-ABJM}) becomes
\begin{align}
w(Z)=&\sum_{ j=1}^{2l+1} \oint_{C^{(M)}_j} \frac{dW}{4\pi i} \frac{V_M'(W)}{Z-W}
   \sqrt{\frac{(Z-1/B^{(M)}_l)(Z-B^{(M)}_{l})(Z+1/B^{(N)}_m)(Z+B^{(N)}_{m})}{(W-1/B^{(M)}_l)(W-B^{(M)}_{l})(W+1/B^{(N)}_m)(W+B^{(N)}_{m})}}
\nonumber
\\
&+\sum_{ j=1}^{2m+1} \oint_{C^{(N)}_j} \frac{dW}{4\pi i} \frac{V_N'(W)}{Z-W}
   \sqrt{\frac{(Z-1/B^{(M)}_l)(Z-B^{(M)}_{l})(Z+1/B^{(N)}_m)(Z+B^{(N)}_{m})}{(W-1/B^{(M)}_l)(W-B^{(M)}_{l})(W+1/B^{(N)}_m)(W+B^{(N)}_{m})}}
, 
\label{general-sym-ABJM}
\end{align}
where $A^{(M)}_j= \exp\left(a^{(M)}_j\right) $, $B^{(M)}_j= \exp\left(b^{(M)}_j\right) $, 
 $A^{(N)}_j= \exp\left(a^{(N)}_j\right) $ and $B^{(N)}_j= \exp\left(b^{(N)}_j\right) $, and $C^{(M)}_j$ and $C^{(N)}_j$ are the contours which encircle the cuts as in (\ref{general-ABJM}).
 We will use $D_j^{(M)} := A_j^{(M)}  $ and $D_j^{(N)} :=  A_j^{(N)} $ when we emphasize the points of the steps. 
 Through calculations similar to section \ref{sec-2-cut-CS}, we can perform this integral and obtain
\begin{align}
&w(Z)=
\frac{1}{2\pi i\lambda} \log{\left( \frac{f(Z)-\sqrt{f^2(Z)-4Z^2}}{2} \right)}
\nonumber
\\
& \qquad \quad +\sum_{i=1}^l \frac{n_i^{(M)}}{\pi i\lambda} \log{\left( \frac{p^{(i)}(Z)+\sqrt{\left( p^{(i)}(Z) \right)^2 -4}}{2} \right)}
-\sum_{j=1}^m \frac{n_j^{(N)}}{\pi i\lambda} \log{\left( \frac{q^{(j)}(Z)+\sqrt{\left( q^{(j)}(Z) \right)^2 -4}}{2} \right)} .
\label{ABJMg}
\end{align}
Here $f(Z)$, $p^{(i)}(Z)$ and $q^{(j)}(Z)$ are rational functions
\begin{align}
&f(Z)=f_0+f_1Z+f_0Z^2,
\quad
p^{(i)}(Z)=\frac{p_0Z^2+p_1Z+p_0}{(Z-D_i^{(M)})(Z-1/D_i^{(M)})},
\quad
q^{(j)}(Z)=\frac{q_0Z^2+q_1Z+q_0}{(Z+D_j^{(N)})(Z+1/D_j^{(N)})},
\nonumber
\end{align}
where the coefficients are given by
\begin{align}
&f_0=\frac{4}{c^{(M)}+c^{(N)}},
\quad
f_1=\frac{2 \left( -c^{(M)}+c^{(N)} \right)}{c^{(M)}+c^{(N)}},
\nonumber
\\
&p^{(i)}_0=\frac{4\left( D_i^{(M)}+1/D_i^{(M)} \right)+2c^{(M)}-2c^{(N)} }{c^{(M)}+c^{(N)}},
\quad
p^{(i)}_1=\frac{2 \left( D_i^{(M)}+1/D_i^{(M)} \right) \left( c^{(M)} -c^{(N)} \right) -4c^{(M)} c^{(N)}}{c^{(M)}+c^{(N)}},
\nonumber
\\
&q^{(j)}_0=\frac{-4\left( D_j^{(N)}+1/D_j^{(N)} \right)+2c^{(M)}-2c^{(N)} }{c^{(M)}+c^{(N)}},
\quad
q^{(j)}_1=\frac{-2 \left( D_j^{(N)}+1/D_j^{(N)} \right) \left( c^{(M)} -c^{(N)} \right) -4c^{(M)} c^{(N)}}{c^{(M)}+c^{(N)}},
\nonumber
\\
&c^{(M)}=B^{(M)}_l+1/B^{(M)}_l, \qquad c^{(N)}=B^{(N)}_m+1/B^{(N)}_m.
\end{align}
In (\ref{ABJMg}),  the first term is identical to the DMP solution (\ref{(1+1)-ABJM}) and the rest of the terms resemble the terms in the stepwise multi-cut solutions in the pure CS matrix model (\ref{multi-cut-CS}) and (\ref{v-1+2}).
Particularly, the resolvent shows the logarithmic singularities at $Z=D_i^{(M)}$, $1/D_i^{(M)}$, $-D_j^{(N)}$ and $-1/D_j^{(N)}$ ($i=1,\cdots, l $ and $j=1,\cdots, m $ ).

If the number of the cuts is even, the result should be modified, since the cut at the origin disappears.
Suppose the number of the cuts of $\mu_i$ is even, we should remove the cut $[1/B_0^{(M)}, B_0^{(M)}]$ and fix $D_1^{(M)}=\exp \left( \pi i n^{M}_1 \right)$.
Similarly, if the number of the cuts of $\nu_i$ is even, the cut $[1/B_0^{(N)}, B_0^{(N)}]$ is removed and $D_1^{(N)}=\exp \left( -\pi i n^{N}_1 \right)$.
With these modifications, the expression (\ref{ABJMg}) works in these cases.

%%%%%%%%%%%%%%%%%%%%%%%%%%%%%%%%%%%%%%%%%%%%%%%
\subsection{Connection to the large-$N$ instantons}
\label{sec-instanton}
%%%%%%%%%%%%%%%%%%%%%%%%%%%%%%%%%%%%%%%%%%%%%%%

Once we obtain the multi-cut solutions, we may obtain the large-$N$ instantons which are the ``tunneling" of the eigenvalues between two solutions \cite{David:1990sk, David:1992za}.
Particularly the instantons in the DMP solution which corresponds to the AdS${}_4\times$CP${}^3$ vacuum of the string theory might be related to non-perturbative objects of strings.
In this section, we argue that some of the instantons may be related to the so-called D2-brane instantons \cite{Drukker:2011zy, Grassi:2014cla}.

We consider the stepwise two+one-cut solution plotted in Figure \ref{fig-numerical-ABJM} (bottom-left).
If we take $N_2^{(M)} \to 0$ limit, this solution reduces to the DMP solution.
Thus $N_2^{(M)} \to 1$ limit of this solution may correspond to the instanton of the single eigenvalue tunneling in the DMP solution.
We can rudely estimate the instanton action of this instanton as follows \cite{Morita:2017oev}.
We consider the effective potential for the $N$-th eigenvalue, say $\mu_N$, in the DMP solution.
From (\ref{partition-ABJM}), the effective potential for $\mu_N$ is given by 
\begin{align}
V_{\rm eff} (\mu_N)& =  \frac{N}{4\pi i \lambda}  \mu^2_N + V_{\rm int} (\mu_N),
\nonumber
\\
V_{\rm int} (\mu_N) & :=
-\sum_{j =1 }^{N-1}
\log \left[ 2\sinh{\frac{\mu_N-\mu_j}{2}} \right]^2+\sum_{j=1}^{N} \log \left[ 2\cosh{\frac{\mu_N-\nu_j}{2}} \right]^2 .
\label{eff-ABJM}
\end{align}
Here we fix $\{ \mu_i \}$ ($i \neq N$) and  $\{ \nu_j \}$  to be the DMP solution and we ignore  the back-reaction of $\mu_N$ to the other eigenvalues. (We will soon see that ignoring the back-reaction is too rude.)
If $\mu_N=\alpha$ where $\alpha$ is the location of the right end point of the cut defined in (\ref{endpt-ABJM'}), it corresponds to the DMP solution.
Then the instanton action is estimated as the difference of the values of the effective potentials
\begin{align}
S_{\rm inst}(\mu) = V_{\rm eff}(\mu) -V_{\rm eff}(\alpha).
\end{align}
By using this equation, we can estimate the instanton action of the $N_2^{(M)} \to 1$ limit of the stepwise two-cut solution (Figure \ref{fig-numerical-ABJM}) by taking $\mu=\alpha+ 2\pi  i $,
\begin{align}
S_{\rm inst} (\alpha+ 2\pi  i )& =  \frac{N}{4\pi i \lambda}  (\alpha+ 2\pi i  )^2 + V_{\rm int} (\alpha+ 2\pi  i ) - \left( \frac{N}{4\pi i \lambda}  \alpha ^2 + V_{\rm int} (\alpha ) \right) \nonumber \\
&=  \frac{  N  \alpha}{\lambda}  + i  \frac{N \pi  }{\lambda} \nonumber \\
&=   N  \pi \sqrt{2 /\lambda}+ \cdots, \qquad (|\lambda| \gg 1). 
\label{S-inst}
\end{align}
Here we have used the periodicity of the interaction $ V_{\rm int} (\alpha+ 2\pi  i n)=V_{\rm int} (\alpha)$, and equation (\ref{endpt-ABJM'}).
Remarkably, the obtained value at strong coupling $(|\lambda| \gg 1)$ agrees with the D2-brane instanton which was obtained through a sophisticated cycle integral of the spectral curve \cite{Drukker:2011zy, Grassi:2014cla}
\begin{eqnarray}
S^{\rm D2}_{\rm inst}=\pi N \sqrt{2/\lambda} , \qquad (|\lambda| \gg 1).
\label{F1D2}
\end{eqnarray}
This quantitative agreement indicates that our multi-cut solutions might be interpreted as the condensations of the D2-brane instantons\footnote{Although the real part of the instanton action (\ref{S-inst}) at the leading order of the strong coupling agrees with the result of \cite{Drukker:2011zy, Grassi:2014cla}, the additional imaginary factor $iN \pi/ \lambda= i \pi k $ in (\ref{S-inst}) does not appear in \cite{Drukker:2011zy, Grassi:2014cla}. 
This contributes to the phase factor of the instanton action.
However, since we have merely considered the value of the effective action, 
we cannot evaluate the additional phase factor coming from the deformation of the contour of the path-integral.
Hence we cannot ask the precise relation between our multi-cut solution and the D2-brane instanton of \cite{Drukker:2011zy, Grassi:2014cla}.
In order to evaluate this phase factor, we may need to consider the path integral including the back-reaction, and it is a challenging problem.
}.

However our evaluation of the instanton action (\ref{S-inst}) is too rude, since $\mu= \alpha+ 2\pi i $ does not satisfy the equation of motion  $0=V'_{\rm eff}(\mu)=N \mu/2 \pi i \lambda +V'_{\rm int}(\mu) $.
We can see it as follows.
Since we have assumed that $\mu= \alpha$ is the DMP solution,  it should satisfy  $0=N \alpha/2 \pi i \lambda +V'_{\rm int}(\alpha) $.
However it immediately means that $\mu=\alpha+2\pi i   $ is not a solution due to the periodicity $V'_{\rm int}(\mu +2\pi  i n )=V'_{\rm int}(\mu  )$.
It implies that the back-reaction to the other eigenvalues is crucial to construct the instanton solution\footnote{Indeed if we do not  consider the back reaction, the classical equation of motion derived from the effective action $V_{\rm eff}(\mu)$ is given by $y=0$ where $y$ is the spectral curve of the DMP solution \cite{Drukker:2011zy}. 
We can easily see that it allows only the trivial solutions $\mu= \pm \alpha$.}.

\begin{figure}
\begin{tabular}{cc}
\begin{minipage}{0.5\hsize}
\begin{center}
        \includegraphics[scale=0.35]{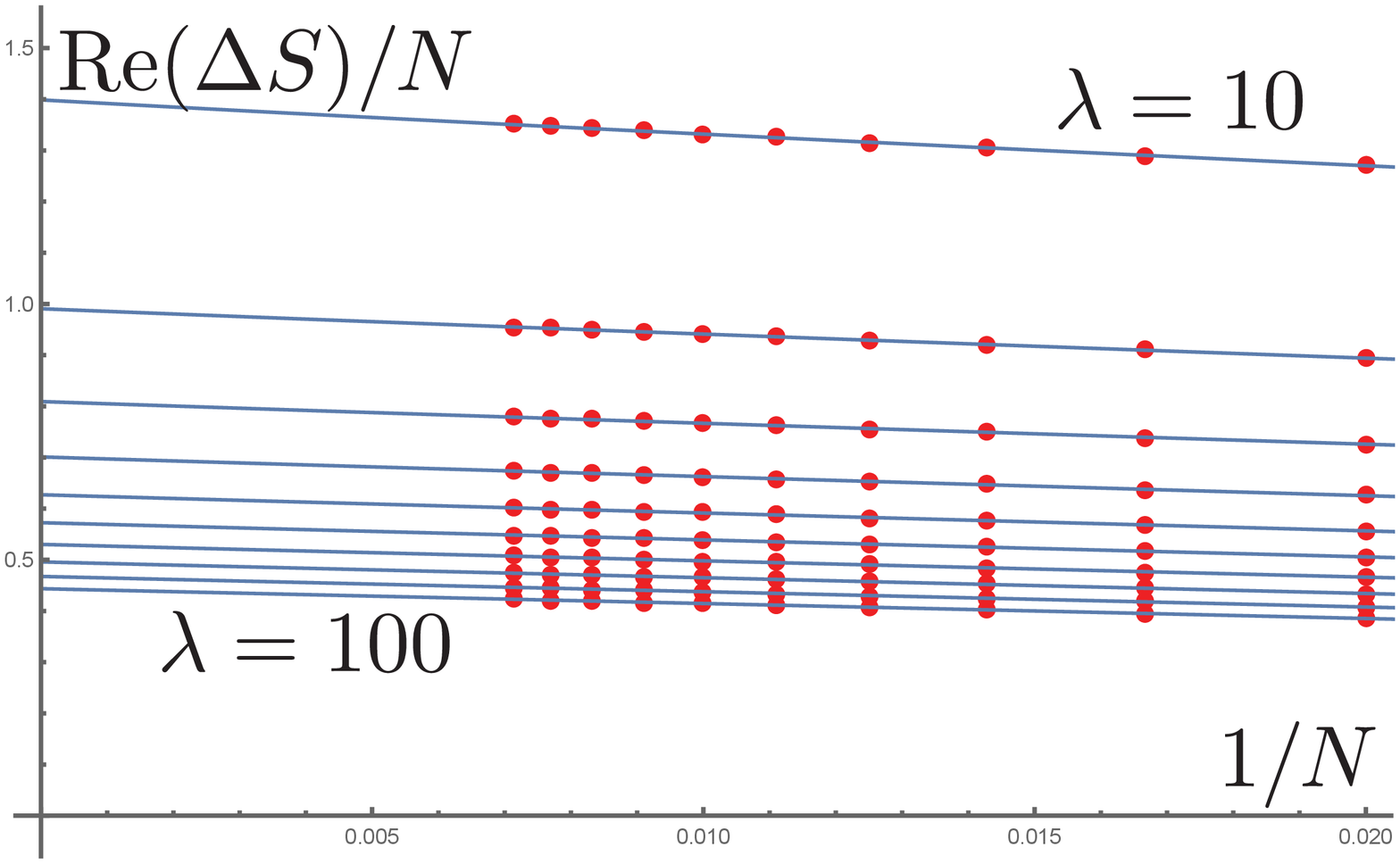}\\
    ${\rm Re} (\Delta S)/N $ vs. $1/N$
\end{center}
\end{minipage}
\begin{minipage}{0.5\hsize}
\begin{center}
        \includegraphics[scale=0.35]{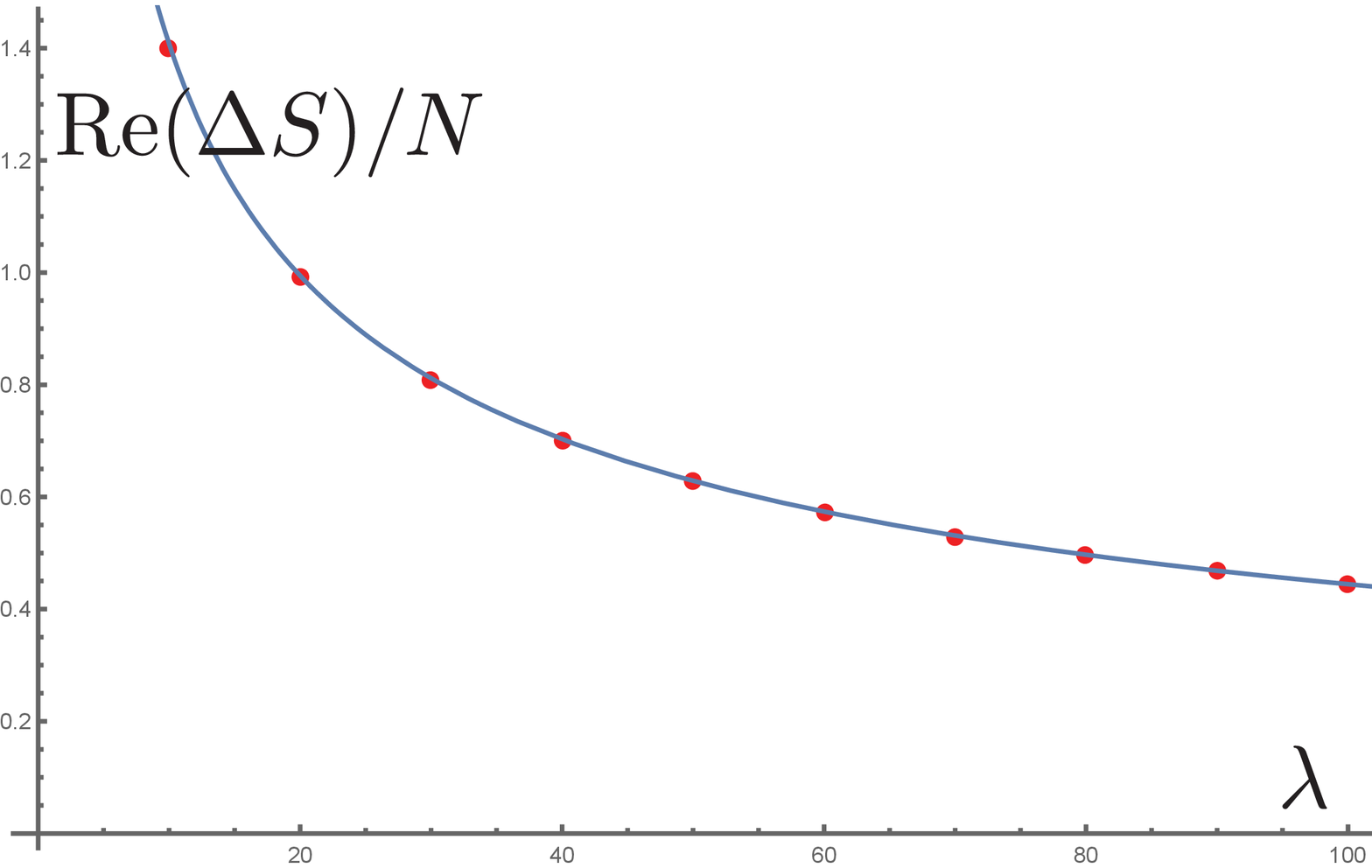}\\
        $ {\rm Re} (\Delta S)/N $ vs. $\lambda$
\end{center}
\end{minipage}
\end{tabular}
       \caption{(Left)  $N$ and $\lambda$ dependence of the real part of the instanton action through the Newton method.
  We compute the classical action of the DMP solution and the instanton solution and evaluate their differences $\Delta S$ at various $N$ and $\lambda$ (the red dots).
Then we fit these data at each fixed $\lambda$ (solid lines) and extrapolate  $\Delta S(\lambda) |_{N \to \infty} $.
(Right) Plot of $\Delta S(\lambda) |_{N \to \infty}/N $ (the red dots). The solid line is analytic prediction of the D2-brane instanton action $\pi \sqrt{2/\hat{\lambda}}$ (\ref{F1D2}).
We can see a good agreement between them.     }
        \label{fig-instanton}
\end{figure}

In principle, we can evaluate the  back-reaction by using the stepwise two+one-cut solution  (\ref{ABJMg}).
Starting from this solution, by taking $N_2^{(M)} \to 1$ in the free energy, we would obtain the instanton action including the back-reaction.
However the computation of the free energy of the stepwise two+one-cut solution is technically difficult, and we instead evaluate the instanton action numerically by employing the Newton method.
The result is summarized in Figure \ref{fig-instanton}.
It indicates that somehow the contributions of the back-reaction to the instanton action is suppressed and the rude estimation (\ref{S-inst}) works well\footnote{
We can confirm that the imaginary part of the instanton action in the numerical calculation also agrees with the estimation (\ref{S-inst}).
We can also check that the results in the $N_2^{(M)}=2$ case are consistent with the $N_2^{(M)}=1$ case.
These results are omitted in this article.
}.
This result supports our conjecture that the stepwise multi-cut solutions are related to the D2-brane instantons in the ABJM theory.

\paragraph{Large-$N$ instantons in the pure CS matrix model}

We can apply the estimation of the instanton action in the ABJM matrix model (\ref{S-inst}) to other CS matrix models if the model allows the stepwise multi-cut solutions.
For example, in the case of the pure CS matrix model (\ref{partition-CS}), we can estimate the instanton action as
\begin{align}
S_{\rm inst} (b+ 2\pi  i )& = V_{\rm eff}(b+2\pi i) -V_{\rm eff}(b) \nonumber \\
&=  \frac{N}{4\pi i \lambda}  (b+ 2\pi i  )^2 + V_{\rm int} (b+ 2\pi  i ) -\left(\frac{N}{4\pi i \lambda}  b^2 + V_{\rm int} (b )\right) \nonumber \\
&=  \frac{  N  b}{\lambda}  + i  \frac{N \pi  }{\lambda} =  2 \pi i  N + \cdots, \qquad (|\lambda| \gg 1). 
\end{align}
Here $b$ is the end point of the one-cut solution (\ref{endpt-CS}), and $V_{\rm eff}$ and
  $V_{\rm int}$ are defined similar to \eqref{eff-ABJM}.
Again we have ignored the back-reaction in this estimation without any justification.
However, the obtained value of the instanton action agrees with the membrane instanton of the pure CS matrix model argued in \cite{Pasquetti:2009jg, Hatsuda:2015owa},
\begin{eqnarray}
&{}&\text{membrane instanton:}
\quad
S^{\rm M2}_{\rm inst}= \frac{2\pi t}{g_s}=2\pi iN,
\qquad (|t| \gg 1),
\label{CSinstanton}
\end{eqnarray}
where $t:=ig_sN=2\pi i \lambda$.
This agreement suggests that the stepwise multi-cut solutions might be regarded as the condensations of the membrane instantons.

 \paragraph{Other instantons?}
 So far we have discussed the instanton limit of the stepwise multi-cut solutions.
 As we have seen in section \ref{sec-CS-general} and \ref{sec-general-ABJM}, the composite solutions also exist in the CS matrix models.
 However, they are composed of at least three cuts, and we cannot take the ordinary instanton limit, namely taking the configuration of the stable solution plus single tunnelling eigenvalue.
 At least two tunnelling eigenvalues are required, and, in this sense, the composite solution might provide a novel type of large-$N$ instantons. (The interaction between the tunneling eigenvalues is crucial similar to the $N=2$ analysis in \eqref{N=2}.)
 However, we have not found any simple estimation of the instanton action for these solutions so far, and the quantitative comparison to D-branes and the known non-perturbative effects in the CS matrix models \cite{Pasquetti:2009jg, Hatsuda:2015owa, Honda:2016vmv, Honda:2017qdb, Drukker:2011zy, Grassi:2014cla} have not been done.
 We leave this issue for future work.

%%%%%%%%%%%%%%%%%%%%%%%%%%%%%%%%%%%%%%%%%%%%%%%
\section{Conclusions and Discussions}
\label{sec-discussion}
%%%%%%%%%%%%%%%%%%%%%%%%%%%%%%%%%%%%%%%%%%%%%%%

In this article, we proposed the ansatz (\ref{composite-CS}) and (\ref{general-ABJM}) for the general solutions of the pure CS matrix model and ABJM matrix model, respectively.
By solving these ansatz, we obtained the multi-cut solutions which quantitatively agree with the Newton method.
Besides, these solutions exhibit the various curious properties: the two types of the multi-cuts (the composite and stepwise), the logarithmic divergences of the eigenvalue densities and the instanton limit. 
Since the multi-cut solutions may describe the various vacua of the systems, these solutions may be crucial to reveal the non-perturbative structures of the CS matrix models.
Indeed we have found the quantitatively evidences that our multi-cut solutions are related to the membrane instantons \cite{Pasquetti:2009jg, Hatsuda:2015owa} and the D2-brane instantons \cite{Drukker:2011zy, Grassi:2014cla}.

One important future direction is the analytic computations of the integral (\ref{composite-CS}) and (\ref{general-ABJM}) in  the general situations.
They might provide us further curious structures of the CS matrix models.
The holomorphy might also help us to find the general solutions as discussed in Appendix \ref{app-holomorphy}.

Another interesting future direction is exploring the gravity duals of our multi-cut solutions.
Since we have considered the 't Hooft limit of the CS gauge theories, the dual gravity description in superstring theory may work.
Particularly the existence of the infinite number of the solutions in the gauge theories reminds us the story of  the bubbling geometries \cite{Lin:2004nb}.
If the corresponding infinite number of the gravity solutions were found, it would be very important in the  supergravities.
The researches on the Lens space matrix models \cite{Aganagic:2002wv, Halmagyi:2003ze, Halmagyi:2003mm, Okuda:2004mb, Halmagyi:2007rw} may give us some insight about the connection between the geometries and the eigenvalue distributions of the CS matrix models.

\paragraph{Acknowledgements}
The authors would like to thank  Tomoki Nosaka,  Kazumi Okuyama, Takao Suyama and Asato Tsuchiya for valuable discussions and comments.
The authors would also like to thank the referee for her/his careful reading of this manuscript and helpful comments.
The authors would also like to thank YITP for financially supporting {\it ``Chube summer school 2017''}, where they had the opportunity to present and develop this work.
The work of T.~M. is supported in part by Grant-in-Aid for Scientific Research (No. 15K17643) from JSPS.

\appendix
%%%%%%%%%%%%%%%%%%%%%%%%%%%%%%%%%%%%
\section{Derivation of the stepwise multi-cut solution via holomorphy}
\label{app-holomorphy}
%%%%%%%%%%%%%%%%%%%%%%%%%%%%%%%%%%%%

We will show that the stepwise multi-cut solution can be derived by using holomorphy\footnote{The advantage of the derivation of the resolvent via holomorphy is that we do not need to perform the  integral of the ansatz \eqref{composite-CS} if we found a suitable holomorphic function.
However, in the case of the CS matrix models, we do not have a guidance principle to find such a holomorphic function and we have to do it through trial and error. 
We mention related issues in footnote \ref{ftnt-holo}
.} too which have been employed in the CS matrix models \cite{Drukker:2010nc, Aganagic:2002wv, Halmagyi:2003ze, Drukker:2011zy}.

\subsection{One-cut solution}
We review the derivation of the one-cut solution (\ref{one-cut-CS}) via holomorphy \cite{Aganagic:2002wv, Halmagyi:2003ze}.
We assume that the resolvent $v(Z)$ has the branch cuts on $\mathcal{C}:$ $[A,B]$.
On this cut, the resolvent should satisfy the saddle point equation (\ref{eom-CS3}).
Then we can define a holomorphic function 
\begin{align}
f(Z)=e^{\pi i\lambda v (Z)}+Z e^{-\pi i\lambda v(Z)}.
\label{f(Z)}
\end{align}
From the boundary conditions (\ref{boundary-CS}), $f(Z)$ satisfies $f(Z) \to  e^{-\pi i\lambda } Z$ ($Z\to \infty$) and $f(Z) \to  e^{-\pi i\lambda } $ ($Z\to 0$).
Then such a holomorphic function is uniquely determined as
\begin{align}
f(Z)=f_0+f_1Z,
\qquad
f_0= e^{-\pi i \lambda},
 \qquad
f_1= e^{-\pi i \lambda}.
\end{align}
On the other hand, by solving  (\ref{f(Z)}) with respect to $v(Z)$, we obtain
\begin{align}
v(Z)=\frac{1}{\pi i\lambda} \log{\left( \frac{f(Z)-\sqrt{f^2(Z)-4Z}}{2} \right)}.
\label{v_1-soln}
\end{align}
This result agrees with (\ref{one-cut-CS}).
One can confirm that this resolvent correctly satisfies the saddle point equation (\ref{eom-CS3}) 
\begin{align}
\lim_{\epsilon \rightarrow 0} \left[ v(Z+i\epsilon)+v(Z-i\epsilon) \right]
&=
\frac{1}{\pi i\lambda} \left[ \log{\left( \frac{f(Z)-i \sqrt{4Z-f^2(Z)}}{2} \right)}+\log{\left( \frac{f(Z)+i \sqrt{4Z-f^2(Z)}}{2} \right)} \right]
\nonumber
\\
&=
\frac{1}{\pi i\lambda} \log{Z},
\quad
\left( Z \in \mathcal{C} \right).
\label{check-eom-v0}
\end{align}

Note that the end points of the cut $A$ and $B$ are determined through the relation $\sqrt{f^2(Z)-4Z}  \propto \sqrt{(Z-A)(Z-B)}$, and they are given as the solution of
\begin{align}
\frac{2f_0f_1-4}{f_1^2}=\left( A+B \right) ,
\qquad
\frac{f_0^2}{f_1^2}=AB
\label{f_0f_1-AB}.
\end{align}

\subsection{Stepwise two-cut solution}
We consider the derivation of the stepwise two-cut solution (\ref{two-cut-CS}) by developing the argument in the previous section.
We assume that the resolvent $v(Z)$ has the branch cuts on $\mathcal{C}_1$: $[A_1,B_1]$ and $\mathcal{C}_2$: $[A_2,B_2]$ where $A_2= e^{2\pi i n}B_1$ with a positive integer $n$ as in (\ref{CS-step}).
On these cuts, the resolvent should satisfy the saddle point equation  (\ref{eom-CS3}).
As sketched in Figure \ref{fig-two-CS}, we assume that $\mathcal{C}_1$ locates on the $n_0$-th sheet and $\mathcal{C}_2$ locates on the $n_0+n$-th sheet.

We will see that the resolvent of the one-cut solution (\ref{v_1-soln}) plays a key role in this problem.
As a trial, let us rotate $Z \to e^{2\pi i n_0} Z$ around $Z=0$ in the saddle point equation (\ref{check-eom-v0}) of the one-cut solution (\ref{v_1-soln}) and see what happens.
On the left hand side of (\ref{check-eom-v0}), since $v(Z)$ is non-singular at $Z=0$, the rotation does not change the value\footnote{$v(Z)$ has the logarithmic singularity at $Z=0$ only on the second sheet of the square root.}. (We rotate $Z $ so that it avoids the branch cut of the square root of $v(Z)$.) 
On the right hand side, since $\log Z$ has the branch cut, the additional constant $2n_0/\lambda$ appears.
Thus, the resolvent of the one-cut solution almost satisfies the saddle point equation (\ref{eom-CS3}) on $\mathcal{C}_1$ except the constant term $ 2n_0/ 
\lambda$.
Similarly, on $\mathcal{C}_2$, $ 2(n_0+n)/ \lambda$ arises.

Therefore, if we find a function $v_1(Z)$ which satisfies\footnote{If the cut $\mathcal{C}_1$ or $\mathcal{C}_2$ crosses the branch cut of $\log Z$, (\ref{eom-v1}) should be modified. 
A simple way is rotating the branch cut so that it avoids the cuts $\mathcal{C}_1$ and $\mathcal{C}_2$. }
\begin{align}
\lim_{\epsilon \rightarrow 0} \left[ v_1(Z+i\epsilon)+v_1(Z-i\epsilon) \right]=&
\frac{2n_0}{\lambda},
\qquad
(Z \in \mathcal{C}_1), \nonumber 
\nonumber
\\
\lim_{\epsilon \rightarrow 0} \left[ v_1(Z+i\epsilon)+v_1(Z-i\epsilon) \right]=&
\frac{2(n+n_0)}{\lambda} ,
\qquad
(Z \in \mathcal{C}_2 ),
\label{eom-v1}
\end{align}
the resolvent of the stepwise two-cut solution may be given as
\begin{align} 
v(Z)=v_0(Z)+v_1(Z),
\label{v0-v1}
\end{align}
where $v_0(Z)$ denotes the one-cut solution (\ref{v_1-soln}) which satisfies
\begin{align}
\lim_{\epsilon \rightarrow 0} \left[ v_0(Z+i\epsilon)+v_0(Z-i\epsilon) \right]=&
\frac{1}{\pi i\lambda} \log Z,
\qquad
(Z \in \mathcal{C}).
\label{eom-v0}
\end{align}
Here we have defined  the cut $\mathcal{C} =\mathcal{C}_1  \cup \mathcal{C}_2$ on the 0-th sheet of the branch cut of $\log Z$. 

However $v_0(Z)$ is not exactly identical to (\ref{v_1-soln}).
This is because, through the boundary conditions (\ref{boundary-CS}), $v_0(Z)$ and $v_1(Z)$ should satisfy
\begin{align}
&\lim_{Z \rightarrow \infty} v_0(Z) =s,
\qquad
\lim_{Z \rightarrow \infty} v_1(Z)=1-s, \nonumber
\\
&\lim_{Z \rightarrow 0} v_0(Z) =-\tilde{s},
\qquad
\lim_{Z \rightarrow 0} v_1(Z)=-(1-\tilde{s}),
\label{decomposition-boundary}
\end{align}
where $s$ and $\tilde{s}$ are constants. (We have assumed that $v_0(Z)$ and $v_1(Z)$ are finite at $Z=0$ and $Z=\infty$.) Hence $f(Z)$ is modified,
\begin{align}
f(Z)=f_0+f_1Z, \qquad
f_0=e^{-\pi i\lambda \tilde{s}},
\qquad
f_1=e^{-\pi i\lambda s}.
\label{f(Z)-boundary-2cut}
\end{align}
Thus $v_0(Z)$ is given by (\ref{v_1-soln}) with this $f(Z)$.

Next we consider $v_1(Z)$.
Similar to the $v_0(Z)$, we define a function, 
\begin{align}
q(Z)=e^{-\frac{n_0 \pi i}{n}} e^{\frac{\pi i\lambda}{n} v_1(Z)}+ e^{\frac{n_0 \pi i}{n}} e^{-\frac{\pi i \lambda}{n} v_1(Z)}.
\label{q(Z)}
\end{align}
We can see that $q(Z)$ is smooth on the cut  $\mathcal{C}_1$ and $\mathcal{C}_2$ through (\ref{eom-v1})\footnote{\label{ftnt-holo}$q_m(Z)=e^{\frac{\pi i \lambda m}{n} (v_1(Z)-n/\lambda)}+e^{-\frac{\pi i \lambda m}{n} (v_1(Z)-n/\lambda)}$ is also holomorphic on the cuts $\mathcal{C}_1$ and $\mathcal{C}_2$, if $m$ is an integer. 
However, the resolvent obtained through $q_m(Z)$ may involve constants which cannot be determined through the boundary conditions unless $m = \pm 1$, and we do not consider these cases. 
Similar ambiguity exists in $f(Z)$ of (\ref{f(Z)}) and other cases too.
For the composite type multi-cut solutions, we may need general $m$.
For example, $q(Z)$ with $m=2$ appears in (\ref{v-1+2}).
}.
(Note that we will soon see that $q(Z)$ has to have a pole.)
By solving (\ref{q(Z)}) with respect to $v_1(Z)$, we obtain
\begin{align}
v_1(Z)=\frac{n}{\pi i\lambda} \log{\left( \frac{q(Z)+\sqrt{q^2(Z)-4}}{2} \right)} +\frac{n_0}{\lambda}.
\label{v_2-soln}
\end{align}

Now we determine the function $q(Z)$.
Through the boundary conditions (\ref{decomposition-boundary}), $q(Z)$ satisfies
\begin{align}
q_1:=&\lim_{Z \rightarrow \infty} q(Z)=e^{-\frac{n_0 \pi i}{n}} e^{\frac{\pi i\lambda}{n} (1-s)}+e^{\frac{n_0 \pi i}{n}}e^{-\frac{\pi i\lambda}{n} (1-s)} ,
\nonumber \\
q_0:=&\lim_{Z \rightarrow 0} q(Z)=e^{-\frac{n_0 \pi i}{n}}e^{-\frac{\pi i\lambda}{n} (1-\tilde{s})}+e^{\frac{n_0 \pi i}{n}}e^{\frac{\pi i\lambda}{n} (1-\tilde{s})} .
\label{q(Z)-boundary}
\end{align}
Besides, since $v_1(Z)$ should have the branch cut between $A_1$ and $B_2$, we demand
\begin{align}
\sqrt{q^2(Z)-4} \propto \sqrt{(Z-A_1)(Z-B_2)}.
\label{q(Z)-cut}
\end{align}
However we can easily see that this condition and the boundary conditions (\ref{q(Z)-boundary}) are inconsistent if $q(Z)$ is a holomorphic function on the entire complex plane.
Hence we relax holomorphy and allow $q(Z)$ to have poles.
A natural candidate of the location of the pole is $Z=D_1:=B_1$ where the value of the right hand side of (\ref{eom-v1}) changes.
Then the conditions (\ref{q(Z)-boundary}) and (\ref{q(Z)-cut}) are satisfied, if 
\begin{equation}
q(Z)=\frac{q_1Z-q_0D_1}{Z-D_1},
\label{q(Z)-laurent}
\end{equation}
where $q_0$ and $q_1$ are related to $A_1$, $D_1$ and $B_2$ via
 \begin{align}
 \frac{2(4-q_0q_1)D_1}{q_1^2-4}=- \left( A_1+B_2 \right) , \qquad \frac{(q_0^2-4)D_1^2}{q_1^2-4}=A_1B_2.
\label{q_0q_1-AB}
\end{align}

It will be instructive to see how the resolvent $v_1(Z)$ (\ref{v_2-soln}) satisfies the equation (\ref{eom-v1}).
On $Z \in \mathcal{C}_1$, (\ref{eom-v1}) is satisfied because
\begin{align}
&\lim_{\epsilon \rightarrow 0} \left[ v_1(Z+i\epsilon)+v_1(Z-i\epsilon) \right]
\nonumber
\\
=&
\frac{2n_0}{\lambda}+\frac{n}{\pi i\lambda} \left[ \log{\left( \frac{q(Z)-i \sqrt{4-q^2(Z)}}{2} \right)}+\log{\left( \frac{q(Z)+i \sqrt{4-q^2(Z)}}{2} \right)} \right]
\nonumber
\\
=&
\frac{2n_0}{\lambda},
\quad (Z \in \mathcal{C}_1).
\label{check-eom-v1}
\end{align}
At $Z=B_1$, the imaginary part of $v_1(Z)$ diverges logarithmically and the real part of $v_1(Z)$ (\ref{v_2-soln}) changes by $n/\lambda$.
Thus the right hand side of (\ref{check-eom-v1}) becomes $2(n_0+n)/\lambda$ on $ \mathcal{C}_2$, and it satisfies (\ref{eom-v1}) correctly. 
The configuration of the branch cut corresponding to this divergence can be seen in Figure \ref{fig-endpt-analysis-three-cut}. 
Note that, although the cuts $[A_1,B_2]$ exist on the every sheet of $\log Z$, $v(Z)$ (\ref{v0-v1}) satisfies  the equation (\ref{eom-CS3}) only on $ \mathcal{C}_1$ on the $n_0$-th sheet and on $ \mathcal{C}_2$ on the $n_0+n$-th sheet.

By using the obtained $v_0(Z)$ and $v_1(Z)$, the stepwise two-cut solution is given by
\begin{align}
v(Z)=&\frac{1}{\pi i\lambda} \log{\left( \frac{f(Z)-\sqrt{f^2(Z)-4Z}}{2} \right)}
+\frac{n}{\pi i\lambda} \log{\left( \frac{q(Z)+\sqrt{q^2(Z)-4}}{2} \right)}-\frac{n_0}{\lambda}, \nonumber \\
&f(Z)=f_0+f_1 Z, \qquad q(Z)=\frac{q_1Z-q_0D_1}{Z-D_1} .
\label{CS-2cut-sol}
\end{align}
This expression involves five constants: $f_0$, $f_1$, $q_0$, $q_1$ and $D_1$, and we can rewrite $f_0$, $f_1$, $q_0$ and $q_1$ by $A_1$, $B_2$ and $D_1$ through (\ref{f_0f_1-AB}) and  (\ref{q_0q_1-AB}).
Also we can fix $A_1$, $B_2$ and $D_1$ by imposing the normalization condition (\ref{cycle-CS2}) and the boundary conditions (\ref{decomposition-boundary}), (\ref{f(Z)-boundary-2cut}) and (\ref{q(Z)-boundary}), and will obtain the solution consistently.
This agrees with the stepwise two-cut solution via the integral formula (\ref{two-cut-CS}).

The generalization of such a derivation via holomorphy to the stepwise multi-cut solution (\ref{multi-cut-CS}) is straightforward.
However the generalization to the composite type solution (\ref{composite-CS}) would be difficult.
As we can see in (\ref{v0soln}), the holomorphic function $f(Z)$ has a pole at $Z=-1$, which is not the location of any step.
Such additional poles may appear in the composite solution generally and, we have not understood the correct rule for the assumptions on $f(Z)$ and $q(Z)$ in these cases yet.

%%%%%%%%%%%%%%%%%%%%%%%%%%%%%%%%%%%%
\section{Comments on the positivity of $\{n_i \}$. }
\label{app-negative-n}
%%%%%%%%%%%%%%%%%%%%%%%%%%%%%%%%%%%%
\begin{figure}
\begin{tabular}{cc}
\begin{minipage}{0.5\hsize}
\begin{center}
        \includegraphics[scale=0.25]{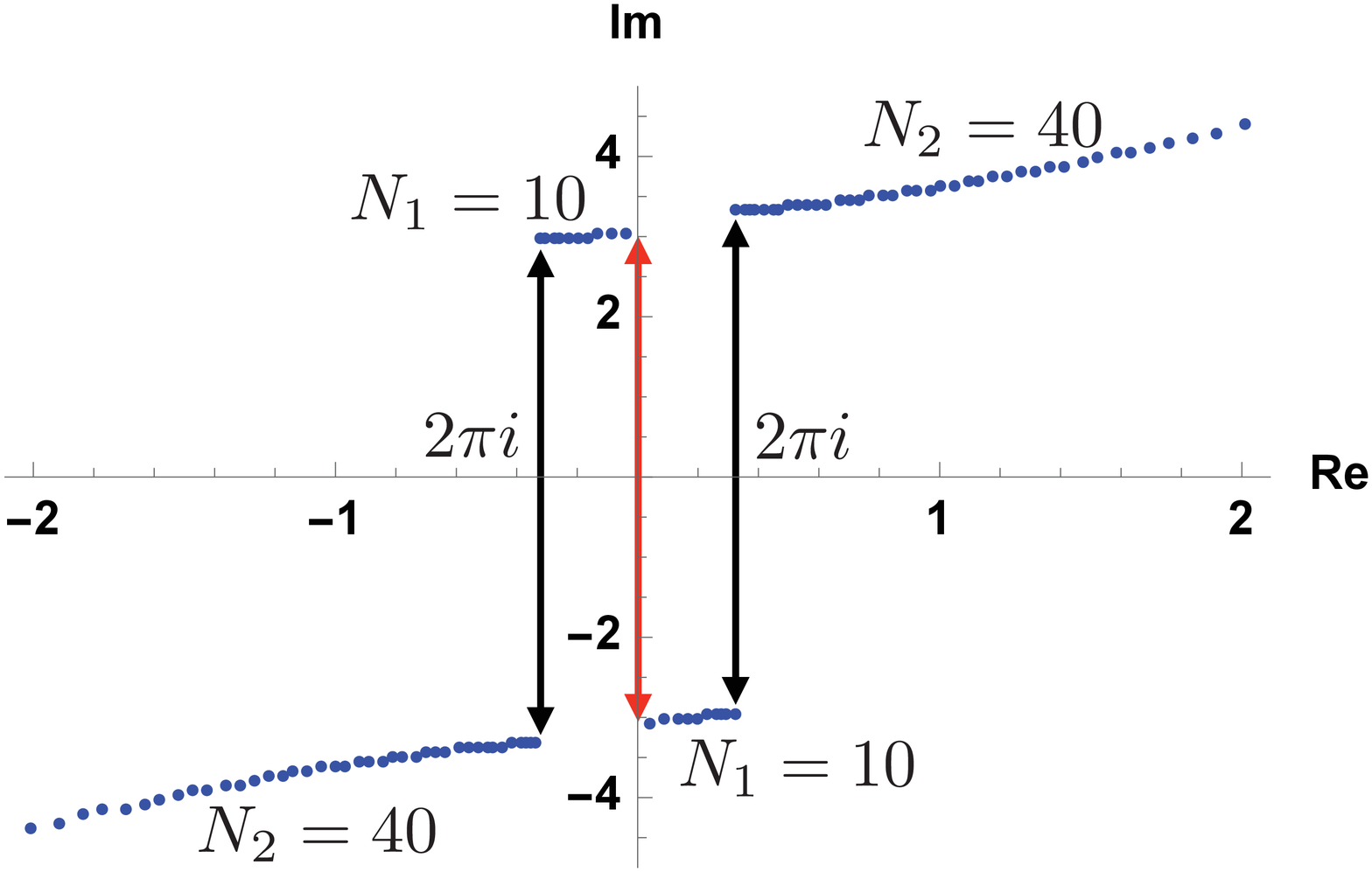}
        \\
        Symmetric four-cut solution
\end{center}
\end{minipage}
\begin{minipage}{0.5\hsize}
\begin{center}
        \includegraphics[scale=0.25]{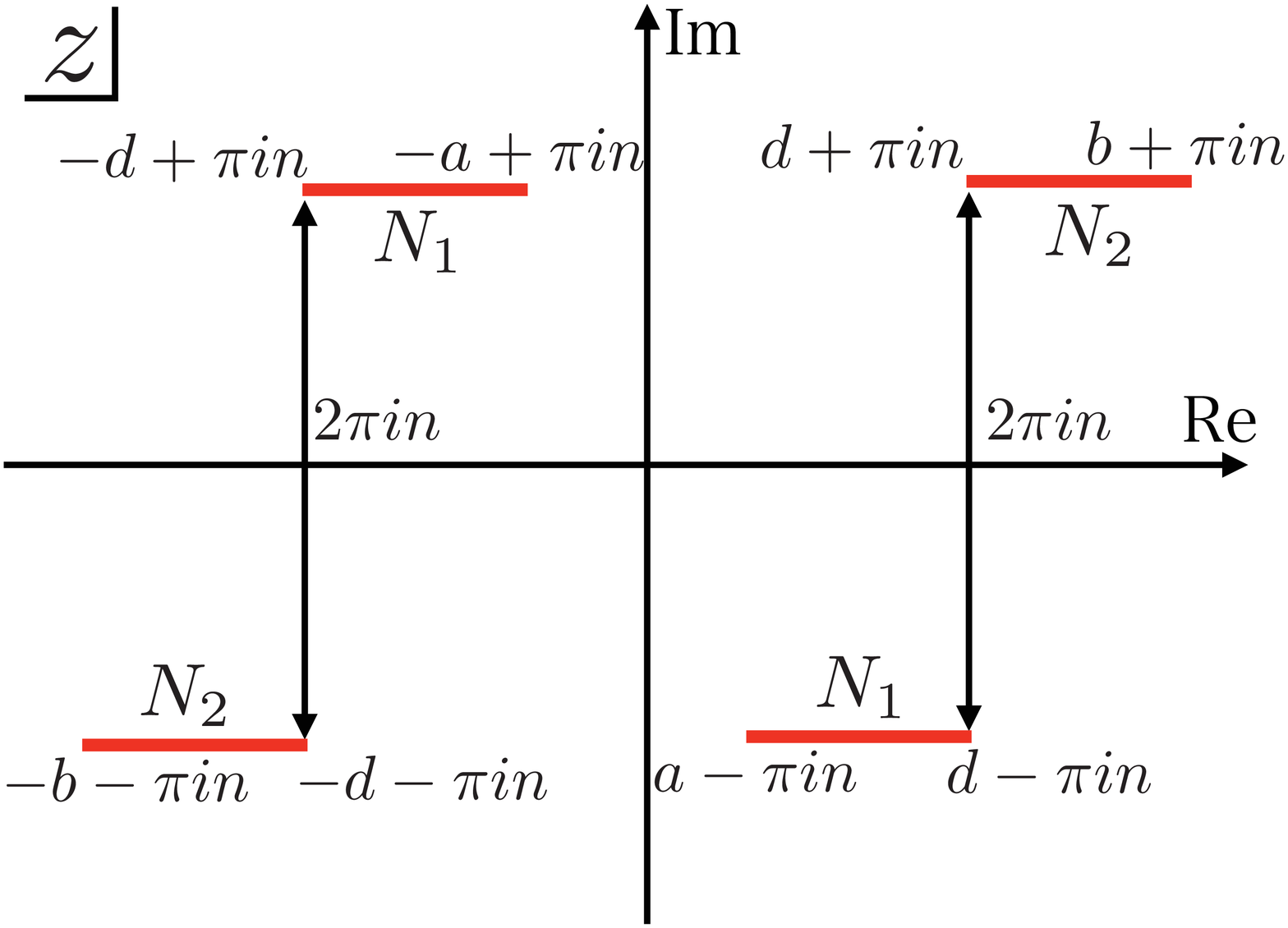}
        \\
        Composite (two+two)-cut solution
\end{center}
\end{minipage}
\end{tabular}
       \caption{ (Left) Symmetric four-cut solution through the Newton method ($\lambda=0.5, N=100$). The question is whether the red arrow interval is ``a negative step" or ``a small gap".
  (Right) Sketch of the composite (two+two)-cut solution.
  This type of the solution may describe the symmetric four-cut solution if $a$ is sufficiently small.}
        \label{fig-four}
\end{figure}

When we considered the stepwise multi-cut solutions, we assumed that $\{ n_i \}$ in (\ref{CSgansatz}) are positive integers.
In this appendix, we discuss why we imposed this assumption.

Actually we can find a numerical solution of the saddle point equation (\ref{eom-CS}) plotted in Figure \ref{fig-four} (left) through the Newton method.
This solution seems to have ``a negative step'' against our assumption.
(Here ``a negative step'' means a negative $n_j$ in (\ref{CSgansatz}).)
However we cannot distinguish a negative step and a small gap through the numerical calculation.
(Here ``a gap'' means $a_{j+1} \neq b_{j}+2 \pi i n_j$ in (\ref{CSgansatz}).)

If it was a negative step, we would naively expect that this solution may be described by our stepwise multi-cut solution (\ref{multi-cut-CS}) with a negative $n$.
However, if we set $n$ negative, the eigenvalue density (\ref{rho-general}) near the negative step may become negative\footnote{ The argument of the appearance of the negative eigenvalue density is subtle, since it is generally difficult to find how the eigenvalues are distributed between the end points $A_1$ and $\{B_i\}$ of the cuts. 
However for a small real $\lambda$, the eigenvalues are distributed parallel to the real axis as we can read off from (\ref{weak-ab}), and we can indeed see that a negative $n$ always causes negative eigenvalue density.} as $\rho(Z) \sim n \log (Z-D)$.
Since negative eigenvalue densities are not allowed physically, our stepwise multi-cut solution (\ref{multi-cut-CS}) may not be applied to the solution in Figure  \ref{fig-four}.
This is one reason that we restrict $\{ n_i \}$ to be positive.

In addition, we can indeed find a solution which has a gap rather than the negative step by composing two stepwise two-cut solutions.
See the sketch in Figure  \ref{fig-four} (right).
In the rest of this appendix, we will derive this solution and show another evidence that the negative step solution may not be allowed. 
Besides we will see that this solution itself has several interesting properties.

We assume that the solution is symmetric under $z \to -z$ and the four cuts locate on $[-b-\pi in, -d-\pi in]$, $[-d+\pi in, -a+\pi in]$, $[a-\pi in, d-\pi in]$ and $[d+\pi in, b+\pi in]$ as in Figure \ref{fig-four} (right).
We take $n$ positive even for simplicity\footnote{
In the case of an odd $n$, the branch cuts $C_i^{(j)}$ may locate near the branch cut of the $\log{Z}$ in the saddle point equation, and it makes the analysis a bit complicated.}.

\begin{figure}
\begin{tabular}{cc}
\begin{minipage}{0.5\hsize}
\begin{center}
        \includegraphics[scale=0.25]{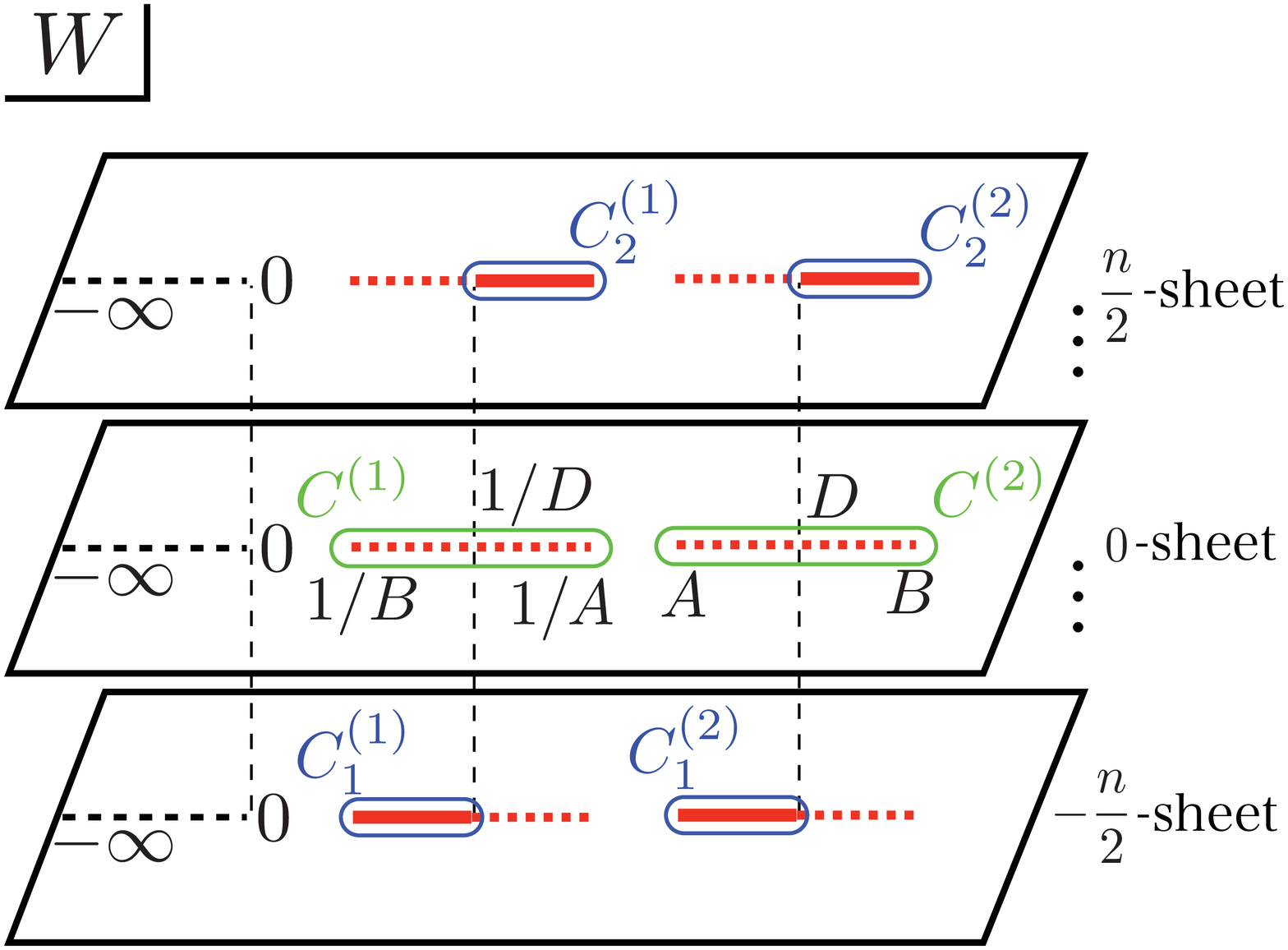}
\end{center}
\end{minipage}
\begin{minipage}{0.5\hsize}
\begin{center}
        \includegraphics[scale=1.1]{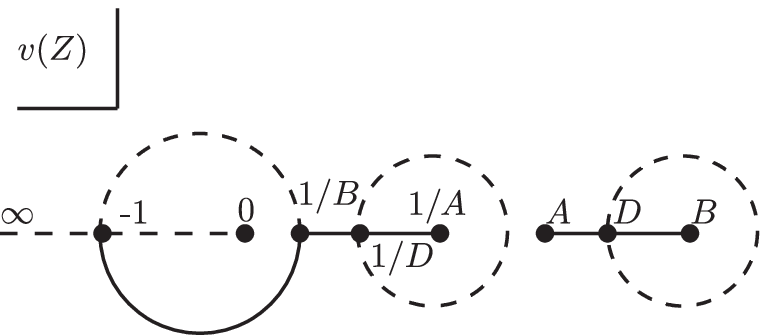}
\end{center}
\end{minipage}
\end{tabular}
       \caption{ (Left) Sketch of the integral contours of the composite (two+two)-cut solution (\ref{Migdal-4cut}).
  (Right)  Branch cuts of the resolvent of the  (two+two)-cut solution (\ref{v0soln}) plus (\ref{v1-soln}) on the $Z$-plane.
       The solid lines denote the branch cuts on the first sheet of the square root.
       The broken lines denote the branch cuts on the second sheet of the square root. }
        \label{fig-cycle-4cut}
\end{figure}

Through the formula for the general multi-cut solution (\ref{composite-CS}), we obtain the resolvent 
\begin{align}
v(Z)=\oint_{C_1^{(1)}  \cup \, C_2^{(1)}  \cup \,  C_1^{(2)}  \cup \, C_2^{(2)}} \frac{dW}{4\pi i} \frac{1}{\pi i\lambda} \frac{\log{W}}{Z-W} \sqrt{\frac{(Z-A)(Z-1/A)(Z-B)(Z-1/B)}{(W-A)(W-1/A)(W-B)(W-1/B)} },
\label{Migdal-4cut}
\end{align}
where the contour $C_1^{(1)}$, $C_2^{(1)}$, $C_1^{(2)}$ and $C_2^{(2)}$ encircle $[1/B, 1/D]$, $[1/D, 1/A]$, $[A, D]$ and $[D,B]$ respectively as shown in Figure \ref{fig-cycle-4cut}.
By regarding the value of $\log W$ on the cuts, 
this integral becomes
\begin{align}
v(Z)=&v_0(Z)+v_1(Z),
\label{v0+v1}
\\
v_0(Z)=&\oint_{C^{(1)}  \cup \, C^{(2)}} \frac{dW}{4\pi i} \frac{1}{\pi i\lambda} \frac{\log{W}}{Z-W} \sqrt{\frac{(Z-A)(Z-1/A)(Z-B)(Z-1/B)}{(W-A)(W-1/A)(W-B)(W-1/B)} },
\label{v0int}
\\
v_1(Z)=&\oint_{C_1^{(1)}  \cup \, C_1^{(2)}} \frac{dW}{4\pi i} \frac{-n/\lambda}{Z-W} \sqrt{\frac{(Z-A)(Z-1/A)(Z-B)(Z-1/B)}{(W-A)(W-1/A)(W-B)(W-1/B)} },
\nonumber
\\
&+\oint_{C_2^{(1)}  \cup \, C_2^{(2)}} \frac{dW}{4\pi i} \frac{+n/\lambda}{Z-W} \sqrt{\frac{(Z-A)(Z-1/A)(Z-B)(Z-1/B)}{(W-A)(W-1/A)(W-B)(W-1/B)} },
\label{v1int}
\end{align}
where the contour $C^{(1)}$ and $C^{(2)}$ encircle the cut $[1/B,1/A]$ and $ [A,B]$ respectively as shown in Figure \ref{fig-cycle-4cut}.

First we evaluate  $v_0(Z)$.
Through a similar computation to (\ref{v-1+2-int}) and (\ref{Migdal-ABJM}), we obtain
\begin{align}
v_0(Z)=& \frac{1}{\pi i\lambda} \log{\left( \frac{f(Z)-\sqrt{f^2(Z)-4Z}}{2} \right)},
\qquad
f(Z)=\frac{f_0+f_1Z+f_0Z^2}{Z+1},
\nonumber
\\
f_0=&\frac{2 \sqrt{AB}}{\left( \sqrt{A}+\sqrt{B} \right) \left( 1+\sqrt{AB} \right)},
\quad
f_1=\frac{2\left( \sqrt{B} (1+A)(B+1/B)- \sqrt{A} (1+B)(A+1/A) \right)}{\sqrt{AB} (B+1/B-A-1/A)}.
\label{v0soln}
\end{align}
Here we can show that $\sqrt{f^2-4Z} \propto \sqrt{(Z-A)(Z-1/A)(Z-B)(Z-1/B)}$.
This term resembles (\ref{v-1+2}) and the DMP solution, while it shows a logarithmic divergence at $Z=-1$. 
The branch cut from $Z=-1$ may terminate at $Z=-1$ on the second sheet of the square root.
See Figure \ref{fig-cycle-4cut}.

Next we consider $v_1(Z)$.
Since this integral is complicated, we employ holomorphy discussed in Appendix \ref{app-holomorphy} to derive $v_1(Z)$.
Similar to (\ref{eom-v1}), $v_1(Z)$ satisfies
\begin{align}
-\frac{n}{\lambda}
&=
\lim_{\epsilon \rightarrow 0} \left[ v_1(Z+i\epsilon)+v_1(Z-i\epsilon) \right] ,
\qquad
(Z \in [1/B,1/D], [A,D]), 
\nonumber
\\
+\frac{n}{\lambda}
&=
\lim_{\epsilon \rightarrow 0} \left[ v_1(Z+i\epsilon)+v_1(Z-i\epsilon) \right] ,
\qquad
(Z \in [1/D, 1/A], [D,B]).
\label{eom-v1-4cut}
\end{align}
Besides $v_1(Z)$ is symmetric under
\begin{align}
v_1(1/Z)=-v_1(Z),
\label{v-sym}
\end{align}
which can be seen from the definition of $v_1(Z)$ (\ref{v1int}).
We also assume that $v_1(Z)$ satisfies the boundary conditions
\begin{align}
\lim_{Z \rightarrow \infty} v_1(Z)=s,
\qquad
\lim_{Z \rightarrow 0} v_1(Z)=-s,
\label{v1-boundary}
\end{align}
where $s$ is a constant and the symmetry (\ref{v-sym}) has been taken into account.

From (\ref{eom-v1-4cut}), we can find a function which is holomorphic on the cuts as
\begin{align}
g(Z)=e^{\frac{\pi i \lambda}{n} v_1(Z)}-e^{-\frac{\pi i \lambda}{n} v_1(Z)}.
\label{g(Z)}
\end{align}
Besides, through the boundary conditions (\ref{v1-boundary}),  $g(Z)$ satisfies
\begin{align}
g_2:=\lim_{Z \rightarrow \infty} g(Z)=e^{\frac{\pi i \lambda}{n} s}-e^{-\frac{\pi i \lambda}{n} s},
\quad
\lim_{Z \rightarrow 0} g(Z)=e^{-\frac{\pi i \lambda}{n} s}-e^{\frac{\pi i \lambda}{n} s} = -g_2.
\label{g(Z)-boundary}
\end{align}
By solving (\ref{g(Z)}), we obtain
\begin{align}
v_1(Z)=\frac{n}{\pi i\lambda} \log{\left( \frac{g(Z)+\sqrt{g^2(Z)+4} }{2} \right)}.
\label{v1-soln}
\end{align}
Here we impose the following relation
\begin{align}
\sqrt{g^2(Z)+4} \propto \sqrt{(Z-A)(Z-1/A)(Z-B)(Z-1/B)},
\label{g(Z)-cut}
\end{align}
so that $v_1(Z)$ has the suitable cuts.
Similar to $q(Z)$ in appendix \ref{app-holomorphy}, $g(Z)$ has to have some singularities in order to satisfy both this relation and the boundary conditions (\ref{g(Z)-boundary}).
Since the left hand side of the relations (\ref{eom-v1-4cut}) is discontinuous at $Z=D$ and $1/D$, we assume that $g(Z)$ has poles there.
Also $g(Z)$ satisfies $g(1/Z)=-g(Z)$ through (\ref{v-sym}) and (\ref{g(Z)}). 
Then we can find $g(Z)$ which satisfies (\ref{g(Z)-boundary}) and (\ref{g(Z)-cut}) as
\begin{align}
g(Z)=\frac{g_2 (Z^2-1)}{(Z-D)(Z-1/D)},
\label{g(Z)-laurent}
\end{align}
where the following relations have been imposed on the constants
\begin{align}
-8 \frac{D+1/D}{g_2^2+4}=-\left( A+\frac{1}{A}+B+\frac{1}{B} \right) ,
\quad
\frac{-2g_2^2+4D^2+4/D^2+16}{g_2^2+4}=2+\left( A+\frac{1}{A} \right) \left( B+\frac{1}{B} \right).
\label{g2-AB}
\end{align}
These relations can be written as
\begin{align}
&g_2=2 \sqrt{\frac{2(D+1/D)-(A+1/A+B+1/B)}{A+1/A+B+1/B}},
\label{g2-ABZ0} \\
&\left( D+1/D \right)^2-4 \kappa \left( D+1/D \right)+4=0,
\quad
\kappa := \frac{4+(A+1/A)(B+1/B)}{2(A+1/A+B+1/B)},
\label{Z0-AB}
\end{align}
and the second equation leads to
\begin{align}
D=\exp{\left[ {\rm arccosh} \left( \kappa+\sqrt{\kappa^2-1} \right) \right]}.
\label{Z0-soln}
\end{align}
In this way, $g_2$ and $D$ are determined by $A$ and $B$.

\begin{figure}
\begin{tabular}{cc}
\begin{minipage}{0.5\hsize}
\begin{center}
        \includegraphics[scale=0.25]{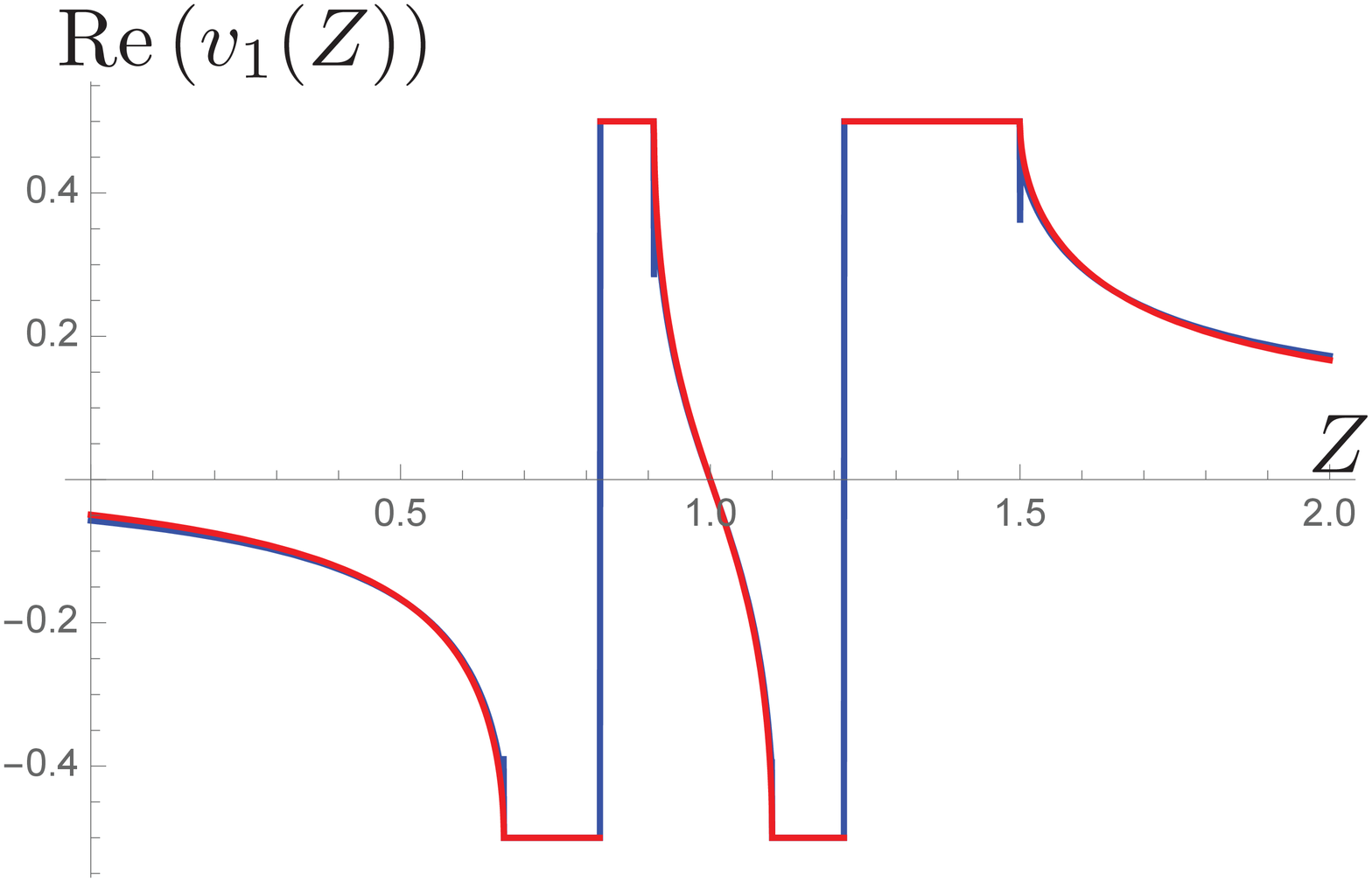}
        \\
        Real part of $v_1(Z)$
\end{center}
\end{minipage}
\begin{minipage}{0.5\hsize}
\begin{center}
        \includegraphics[scale=0.25]{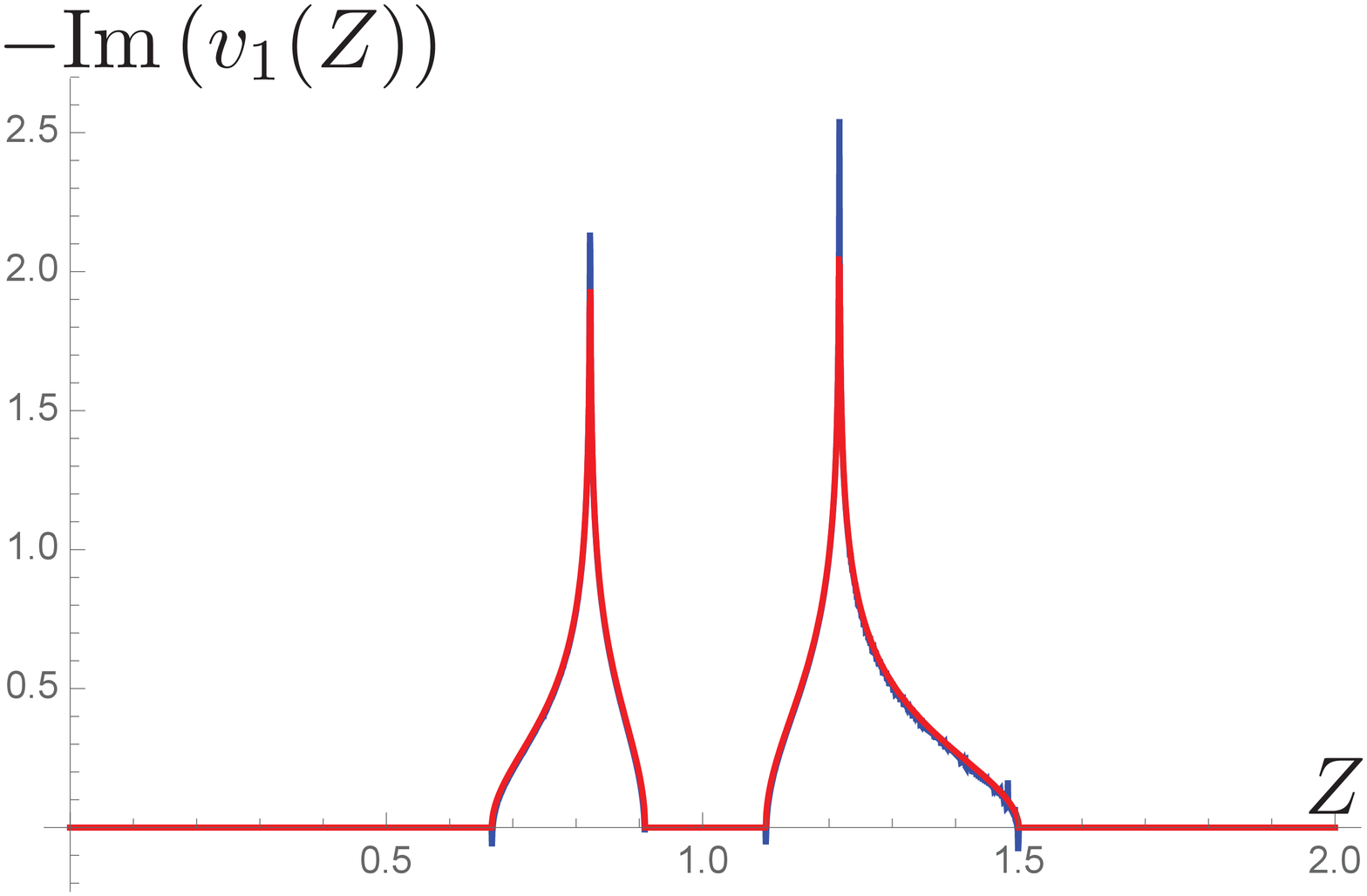}
        \\
        Imaginary part of $v_1(Z)$
\end{center}
\end{minipage}
\end{tabular}
       \caption{ Plots of $v_1(Z)$ via holomorphy (\ref{v1-soln}) (red curves) and the integral formula (\ref{v1int}) (blue curves).
       We take $A=1.1$ and $B=1.5$, and $D$ is fixed via (\ref{Z0-soln}). (See footnote \ref{ftnt-Z0} about $D$ in (\ref{v1int}).)
This agreement indicates that holomorphy provides the answer of the integral (\ref{v1int}).
Note that the plateaus in the real part correspond to the left hand side of (\ref{eom-v1-4cut}).
Besides, the imaginary parts of $v_1(Z)$ may provide the eigenvalue densities.
}
        \label{fig-v1}
\end{figure}

We can confirm that the obtained $v_1(Z)$ is consistent with the integral formula (\ref{v1int}) by comparing them numerically\footnote{
\label{ftnt-Z0}
If we take $Z \rightarrow \infty$ in the integral formula (\ref{v1int}), $v_1(Z)$ linearly grows as
\begin{align}
\lim_{Z \rightarrow \infty} v_1(Z)=& Z \left[ \oint_{C_1^{(1)} \cup \, C_1^{(2)}} \frac{dW}{4\pi i} \frac{-n/\lambda}{\sqrt{(W-A)(W-1/A)(W-B)(W-1/B)}} \right.
\nonumber
\\
& \left. +\oint_{C_2^{(1)}  \cup \, C_2^{(2)}} \frac{dW}{4\pi i} \frac{+n/\lambda}{\sqrt{(W-A)(W-1/A)(W-B)(W-1/B)}} \right]+ \mathcal{O}(Z^0),
\end{align}
and the boundary condition (\ref{boundary-CS}) demands vanishing this term.
We can numerically check that, if $D$ is given by (\ref{Z0-soln}) which has been derived via holomorphy, this term becomes 0.
This coincidence supports the consistency of the integral formula (\ref{v1int}) and holomorphy.}.
 See Figure \ref{fig-v1}.

\begin{figure}
\begin{center}
        \includegraphics[scale=0.25]{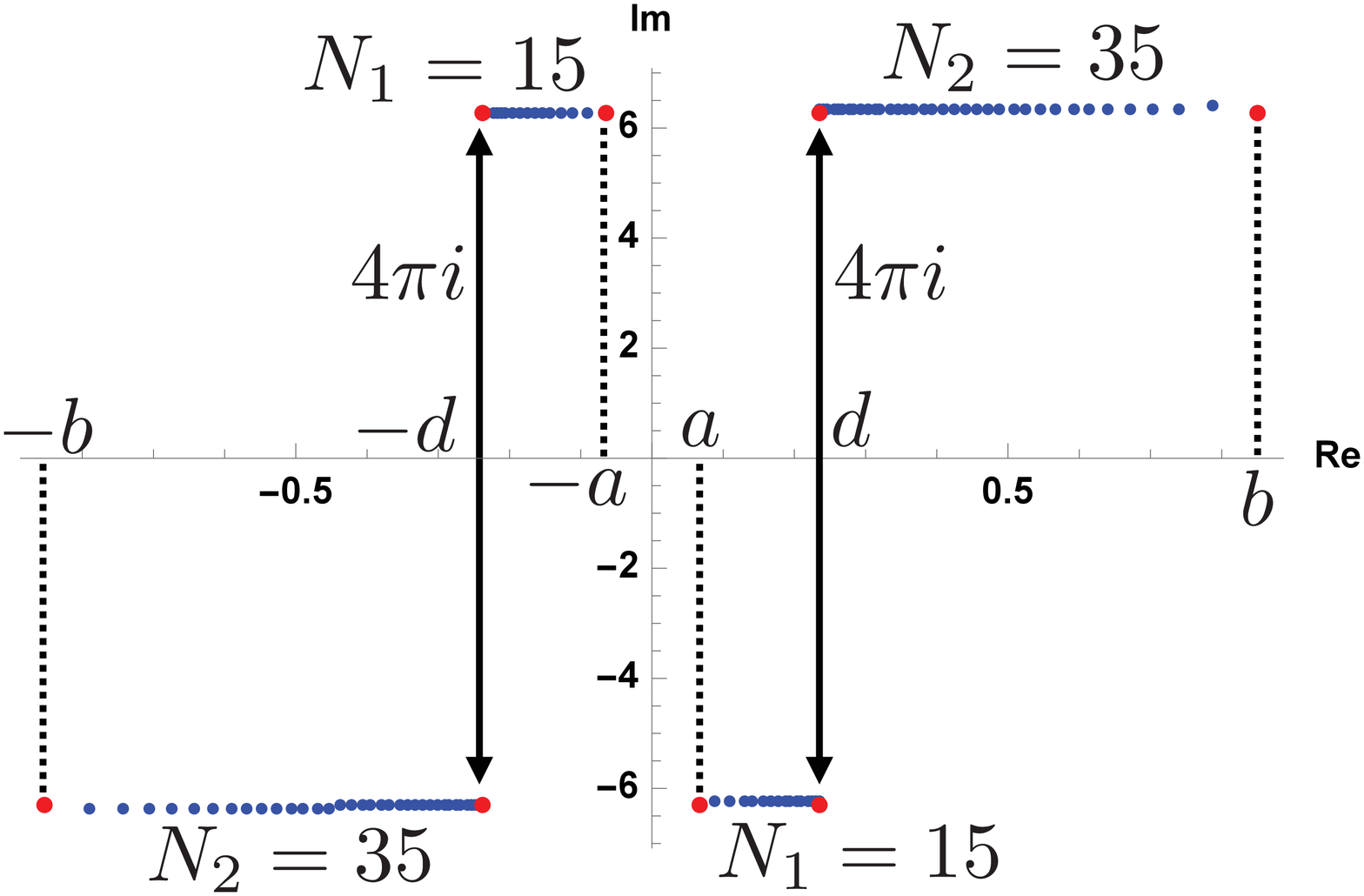}\\
\end{center}
       \caption{Composite (two+two)-cut solution via the Newton method (blue) and our result (\ref{v1-soln}) (red).
       We take $n=2$, $\lambda=0.25$, $N_1=15$ and $N_2=35$ in the Newton method.
       In our method, we ignore $v_0$ by regarding small  $\lambda$, and consider the contribution of $v_1$ only.
We solve the conditions (\ref{cycle-4cut}) and (\ref{boundary-CS}) numerically, and find $A$ and $B$.
These two results agree very well.
       }
        \label{fig-endpt-analysis-4cut}
\end{figure}

Now we obtain the resolvent $v=v_0+v_1$ via (\ref{v0soln}) and (\ref{v1-soln}).
It involves the undetermined constant $A$ and $B$.
They can be fixed by the boundary condition (\ref{boundary-CS}) and the normalization condition
\begin{align}
\frac{N_1}{N} =\int_{A}^{D} \rho(Z) dZ=\frac{1}{4\pi i} \oint_{C_1^{(2)}} \frac{w(Z)}{Z} dZ
\label{cycle-4cut}.
\end{align}
We can numerically solve these conditions for given $n,\lambda,N_1/N$.
See Figure \ref{fig-endpt-analysis-4cut} for the result at a weak coupling. 
It correctly reproduces the solution obtained through the Newton method.

Lastly we discuss the properties of the solution.
One question is whether it continues to a negative step solution.
To answer it, we regard $A$ as the input parameter of the solution instead of $N_1/N$.
As we take $A \to 1$ ($a \to 0$), if the solution continues to a negative step solution, the cut $[a,d]$ should remains finite.
However the relation (\ref{Z0-soln}) tells us that 
\begin{align}
\lim_{a \rightarrow 0} d= \sqrt{\frac{2a(B-1)}{B+1}}+O(a^{3/2}).
\label{Z0-lim}
\end{align}
Thus the cut shrinks as $a \to 0$, and the solution rather continues to the stepwise two-cut solution with the cuts $[-b,0]$ and $[0,b]$.
This result may indicate that the negative step is dynamically not allowed, and the numerical result plotted in Figure \ref{fig-four} is our composite type solution\footnote{By using the assumption about the four cuts and the change of the variables $u_j= \pm \pi i n +x_j $ where $\pm$ depends on which cut $u_j$ belongs to, we can show that $\{ x_j \}$ feel an effective potential 
\begin{align}
V(x)=n[-x-d+2(x+d)\theta(x+d)-2x\theta(x)+2(x-d)\theta(x-d) ] 
\end{align}
at a weak coupling ($|\lambda| \ll 1 $)  \cite{Morita:2017oev}.
Then the non-existence of the negative step solution implies the non-existence of  ``one-cut solution" in this potential.
We presume that the potential at $x=0$ is so sharp that the one-cut solution is not allowed.
See \cite{Okuyama:2017feo} for a related problem.
}.

 \bibliographystyle{JHEP} 
 \bibliography{CS2} 
  \end{document}